\documentclass[12pt]{article}

\usepackage{scicite}

\usepackage{times}

\usepackage{amsmath}
\usepackage{amsfonts}
\usepackage{amssymb}
\usepackage{graphicx}
\usepackage{ulem}

\usepackage{bm,color,amsmath,amssymb,latexsym,graphicx,psfrag,accents,float}
\usepackage{amscd,dsfont,wasysym,mathrsfs,mathtools}
\usepackage{multirow}
\usepackage{dcolumn}
\usepackage{xcolor}
\usepackage{comment}

\usepackage{booktabs} 

\usepackage{lineno}

\newenvironment{sciabstract}{%
\begin{quote} \bf}
{\end{quote}}

        \definecolor{AAcolor}{rgb}{0.7,0.1,0.4}

\title{Quasi-symmetry protected topology in a semi-metal}

\author
{Chunyu Guo,${}^{1\ast}$ Lunhui Hu,${}^{2\ast}$  Carsten Putzke,${}^{1,3}$ Jonas Diaz,${}^{1}$  Xiangwei \\ Huang,${}^{1}$ Kaustuv Manna,${}^{4,5}$ Feng-Ren Fan,${}^{4}$ Chandra Shekhar,${}^{4}$ Yan Sun,${}^{4}$\\ Claudia Felser,${}^{4}$ Chaoxing Liu,${}^{2,6\dagger}$  B. Andrei Bernevig,${}^{6,7,8\dagger}$ Philip J. W. Moll${}^{1\dagger}$\\
  \normalsize{${}^{1}$Laboratory of Quantum Materials (QMAT), Institute of Materials (IMX),}\\ \normalsize{\'{E}cole Polytechnique F\'{e}d\'{e}rale de Lausanne (EPFL), CH-1015 Lausanne, Switzerland}\\
\normalsize{${}^{2}$Department of Physics, The Pennsylvania State University, University Park, PA, USA}\\
\normalsize{${}^{3}$Max Planck Institute for the Structure and Dynamics of Matter, 22761 Hamburg, Germany}\\
 \normalsize{${}^{4}$Max Planck Institute for Chemical Physics of Solids, 01187 Dresden, Germany}\\
  \normalsize{${}^{5}$Department of Physics, Indian Institute of Technology Delhi, New Delhi 110016, India}\\
  \normalsize{${}^{6}$Department of Physics, Princeton University, Princeton, New Jersey 08544, USA}\\
 \normalsize{${}^{7}$Donostia International Physics Center,P. Manuel de Lardizabal 4, 20018}
 \\ \normalsize{ Donostia-San Sebastian, Spain}\\
 \normalsize{${}^{8}$IKERBASQUE, Basque Foundation for Science, Bilbao, Spain}\\
 \\
  \normalsize{$^\ast$These authors contributed equally to this work.}\\
 \normalsize{$^\dagger$Corresponding authors: 
 cxl56@psu.edu(C.X.L.); bernevig@princeton.edu(B.A.B.);}
 \\
 \normalsize{philip.moll@epfl.ch(P.J.W.M.)}\\}

\date{}

\begin{document}

\baselineskip24pt

\maketitle

\begin{sciabstract}

The crystal symmetry of a material dictates the type of topological band structures it may host, and therefore symmetry is the guiding principle to find topological materials. Here we introduce an alternative guiding principle, which we call ‘quasi-symmetry’. This is the situation where a Hamiltonian has an exact symmetry at lower-order that is broken by higher-order perturbation terms. This enforces finite but parametrically small gaps at some low-symmetry points in momentum space. Untethered from the restraints of symmetry, quasi-symmetries eliminate the need for fine-tuning as they enforce that sources of large Berry curvature will occur at arbitrary chemical potentials. We demonstrate that a quasi-symmetry in the semi-metal CoSi stabilizes gaps below 2 meV over a large near-degenerate plane that can be measured in the quantum oscillation spectrum. The application of in-plane strain breaks the crystal symmetry and gaps the degenerate point, observable by new magnetic breakdown orbits. The quasi-symmetry, however, does not depend on spatial symmetries and hence transmission remains fully coherent. These results demonstrate a class of topological materials with increased resilience to perturbations such as strain-induced crystalline symmetry breaking, which may lead to robust topological applications as well as unexpected topology beyond the usual space group classifications.

\end{sciabstract}


The introduction of concepts of topology in the past years into the field of electronic dispersions has captured the imagination of condensed matter physics and sparked a flurry of new research directions. In topologically non-trivial metals, the wavefunctions are inherently non-local, and novel electronic states are expected and observed at the crystal surface terminating them \cite{DiracAndWeyl_Armitage_RMP,ARPES_RMP,FermiArc1st,Hourglass,Bradlyn_newfermions,zhijun_na3bi,zhijun_cd3as2,Drumhead,CoSi_chiralFermion,FermiArc}. Beyond these surface properties, interesting anomalies in the bulk dispersion lead to novel physical phenomena, such as the emergence of quasiparticles mimicking ultra-relativistic Weyl- and Dirac-Fermions in 3D; a solid-state analogon of the Adler-Bell-Jackiw anomaly; a planar Hall effect; topological piezoelectric effect; second harmonic generation in Weyl semimetals; or strong Berry curvature that impacts the semi-classical and quantum dynamics of quasiparticles \cite{	ChiralAnomaly_Burkov1,ChiralAnomaly_Burkov2,ChiralAnomaly_Na3Bi,Piezo,SHG1,SHG2,PHE,Co3Sn2S2_AHE,Mn3Si_AHE}.


Given the wealth of new phenomena and potential applications, it is important to identify such "ideal" Weyl-, Dirac- or nodal-line semi-metals in which the a priori unrelated chemical potential and the topological anomaly coincide. That such a coincidence is rather rare can be understood from the Wigner-Von Neumann theorem, which states that a two-level crossing generally requires three tuning parameters\cite{VonNeumann,TPT}. To stabilize crossings of more than two bands (nodal points) or achieve higher dimensional degeneracies (nodal lines, planes), additional crystalline symmetries must provide further constraints on the form of the Hamiltonian, such as rotational symmetries for the fourfold Dirac points in the 3D Dirac semi-metals Cd$_3$As$_2$ and Na$_3$Bi\cite{zhijun_na3bi,zhijun_cd3as2} or two-fold screw axis for the nodal plane in MnSi \cite{MnSi_Plef}.



\begin{figure}
\includegraphics[width=0.8\columnwidth]{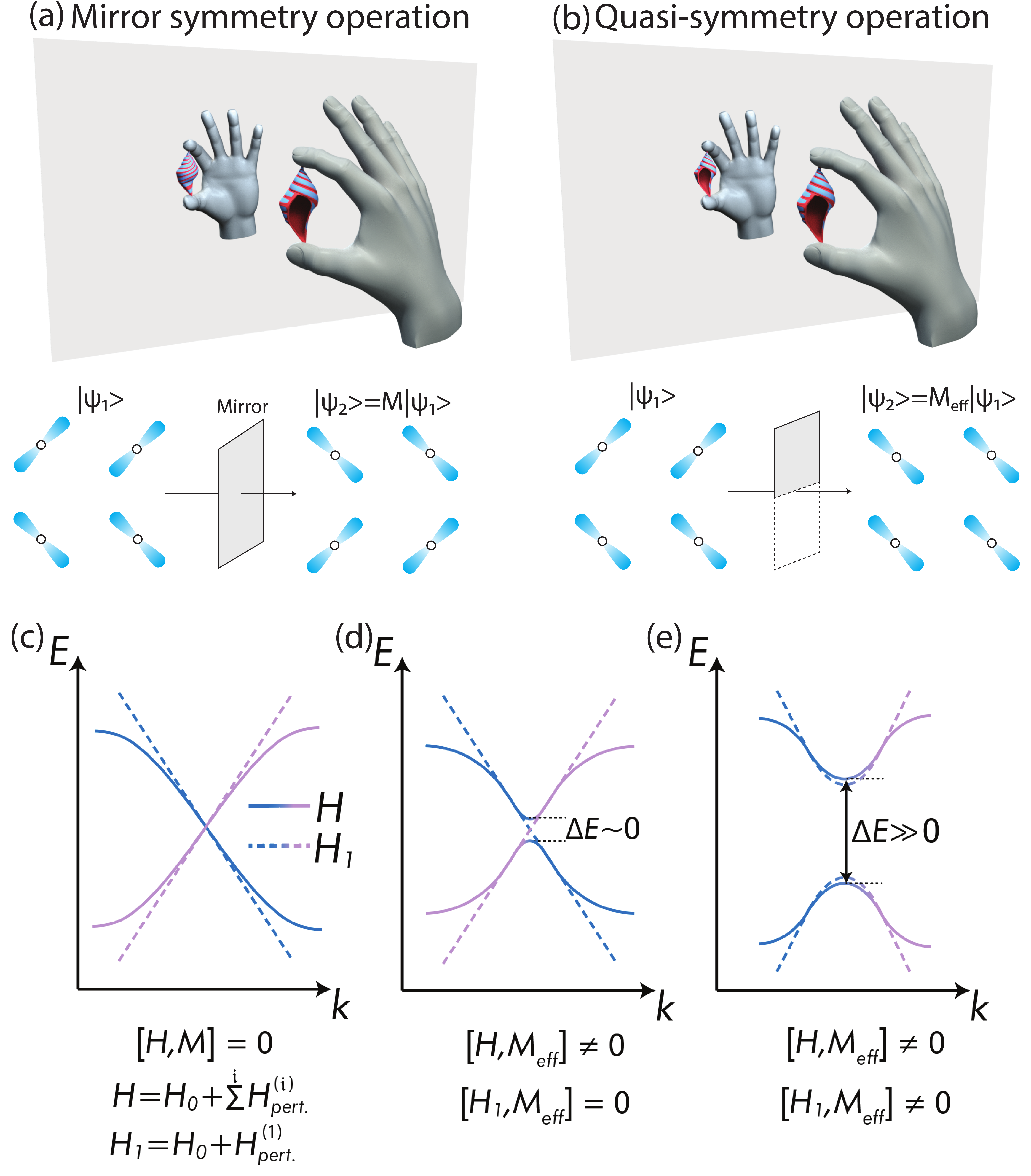}
\centering
		\caption{(a) Illustration of a mirror symmetry operation, which acts at the whole object consistently. (b) In contrast to the mirror symmetry, quasi-symmetry operation acts differently on different parts of the system. (c) If a crystalline symmetry operation commutes with the Hamiltonian system, the band crossing point is therefore symmetry-protected. (d) In this case the Hamiltonian itself does not commute with the symmetry operation yet its first order perturbation does. This lead to the situation where though the band crossing is numerically avoided but the resulted gap size is negligible for the physical properties of the system. This type of symmetry is called quasi-symmetry. (e) If the symmetry operation does not commute with the Hamiltonian to any order of perturbation of the system, the crossing is not protected by any symmetry. This will result in a sizeable gap and lost of topological character.  }
	\label{Intro}
\end{figure}

Hence crystalline symmetry and band topology are inseparably intertwined, and it is natural that symmetry should guide our search for materials. Here we refine this picture, and argue for the importance of approximate symmetries as an additional concept (Fig.~\ref{Intro}) with clearly observable experimental consequences. A crystalline symmetry operation, $M$, commutes with the Hamiltonian of the system, $[H(k),M]=0$. The well-known representation theory dictates that low-symmetry points usually support only one-dimensional representations, and hence bands anti-cross at these points. It may be possible, however, that at these low symmetry points {\it approximate} symmetries emerge. They are not exact symmetries of the crystal yet still commute with a dominant part of the Hamiltonian. Such approximate symmetries would stabilize almost linear band crossings by enforcing finite yet perturbatively small energy gaps. A trivial example is a weak lattice distortion that removes a spatial symmetry that otherwise stabilizes a band crossing. Graphene could be considered as an alternative example in the spin channel, as its Dirac point is weakly gapped when spin-orbit coupling is considered, turning spin-rotation into an approximate symmetry\cite{TB_graphene,Gap_graphene}. However, as approximate symmetries leave the bounds set by crystalline point-groups, more exciting approximate symmetries may emerge that have no counterpart in the crystal symmetry. 



\begin{figure}
\includegraphics[width=0.95\columnwidth]{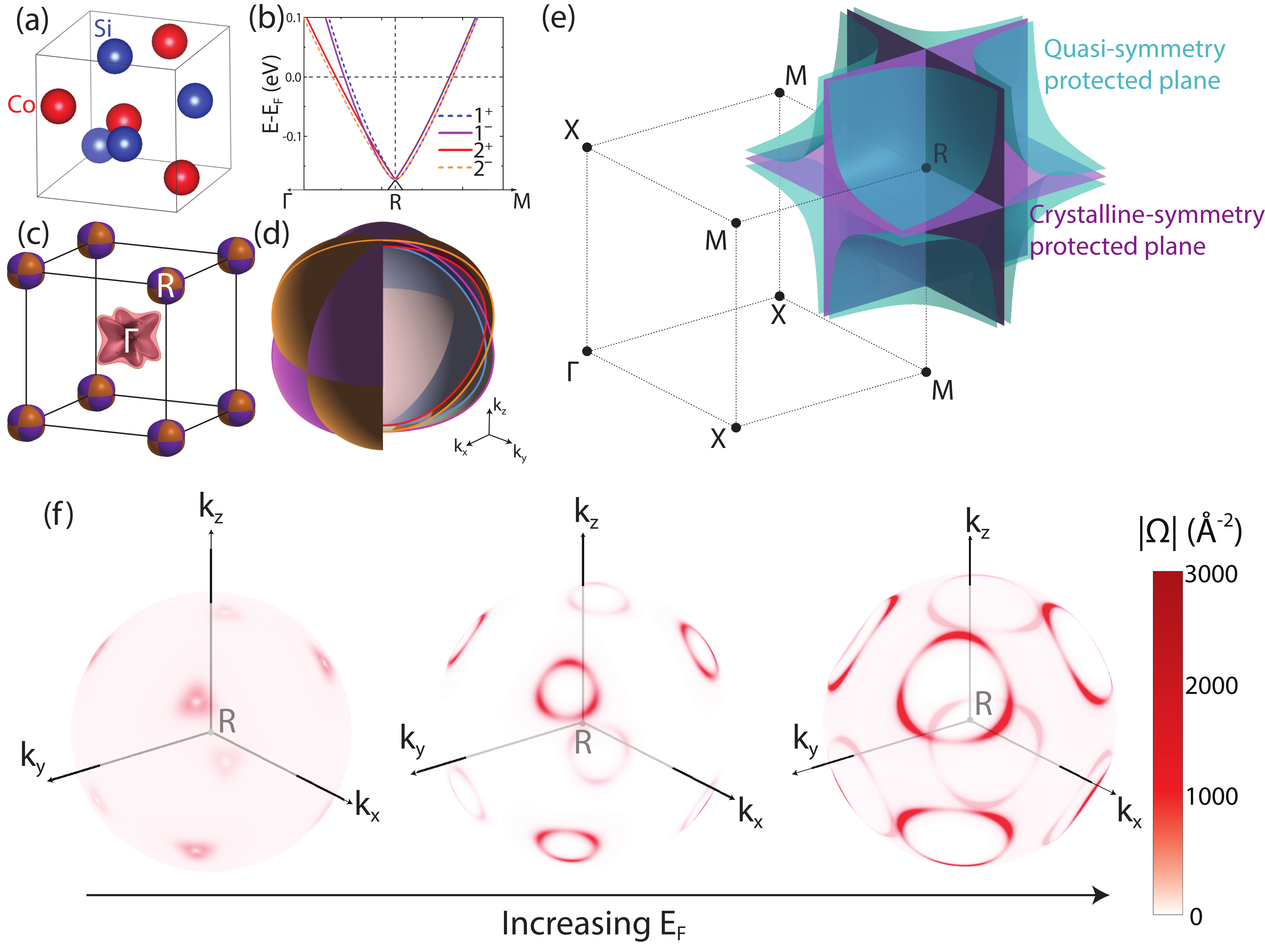}
\centering
	\caption{(a) Crystal structure of CoSi. Co and Si atoms are presented as red and blue spheres respectively. (b) Ab-initio-calculated band structure of CoSi around R-point. Here 1/2 denotes orbital character while +/- stands for the spin character of the band. (c) 3D view of all Fermi surfaces, which are centered around either R- or $\Gamma$- point of the Brilloiun zone. (d) 3D view of Fermi surfaces centered around R-point with a quadrant cut. (e) Quasi-symmetry and crystalline-symmetry protected degenerate planes. (f) The Berry curvature distribution of the $1^{-}$-band defined in (b) at different Fermi energy calculated from the model Hamiltonian~Eq.~\eqref{eq:HR}.}
	\label{Crystal}
\end{figure}

We adopt the term ``quasi-symmetry'' to denote a special type of approximate symmetry of electronic band structures, that emerges from the hierarchy of a $k\cdot p$ type of perturbation expansion around a high-symmetry point. Quasi-symmetries are operators $M_{eff}$ that are the {\it exact} symmetries of the lower-order expansion but not of higher-order perturbation terms. The main goal of this paper is to show such a quasi-symmetry exists in the semi-metal CoSi as well as a large set of similar materials, and that it enforces a low-symmetry plane of near-degeneracies spanning the Brillouin zone which pin a source of Berry curvature to the Fermi level (Fig.~\ref{Crystal}). Excitingly, it falls into the latter category of quasi-symmetries that are unrelated to approximate point-group symmetries, but can be treated as a symmetry of internal degrees of freedom, akin to spin in graphene. Unlike graphene, however, the form of quasi-symmetry is non-trivial as the corresponding symmetry operator is generally k-dependent and forms a part of the Hamiltonian. The key distinction is that while a proper crystal symmetry acts on all atomic coordinates uniformly, the quasi-symmetry operation exchanges 3d orbitals on subsets of inequivalent Co-sites selectively. A simplified example may be an incomplete mirror image [Fig.~\ref{Intro}(a) and (b)], in which the mirror operation acts only on parts of the object but not the whole consistently.



CoSi crystallizes in a chiral cubic structure of space group $P2_13$ without an inversion center (Fig.~\ref{Crystal}). Recent studies revealed the existence of exceptionally long Fermi arcs which connect the sixfold and fourfold band degenerate points\cite{CoSi_chiralFermion,CoSi_QPI,CoSi_longFermiArc,MultiWeyl}. These two degenerate points correspond to two branches of Fermi surfaces located at at the $\Gamma$- and R-point, respectively, yet only the R-point pockets are observed in quantum oscillation experiments. Band structure calculations reveal four split bands due to spin-orbit coupling which correspond to four Fermi surfaces at R, which we label as $1^+,1^-,2^+,2^-$, representing the orbital and spin degree of freedom respectively. States at the boundary of the Brillouin zone are doubly degenerate due to crystalline symmetry, hence the non-degenerate Fermi surfaces are enforced to touch there \cite{Bradlyn_newfermions}. The spin-orbit coupling and the orbital gap are both small yet similar in magnitude, which explains how four slightly degenerate, interpenetrating spheroids make up a surprisingly rich internal structure of the Fermi surfaces. 

\begin{figure}
\includegraphics[width=0.97\columnwidth]{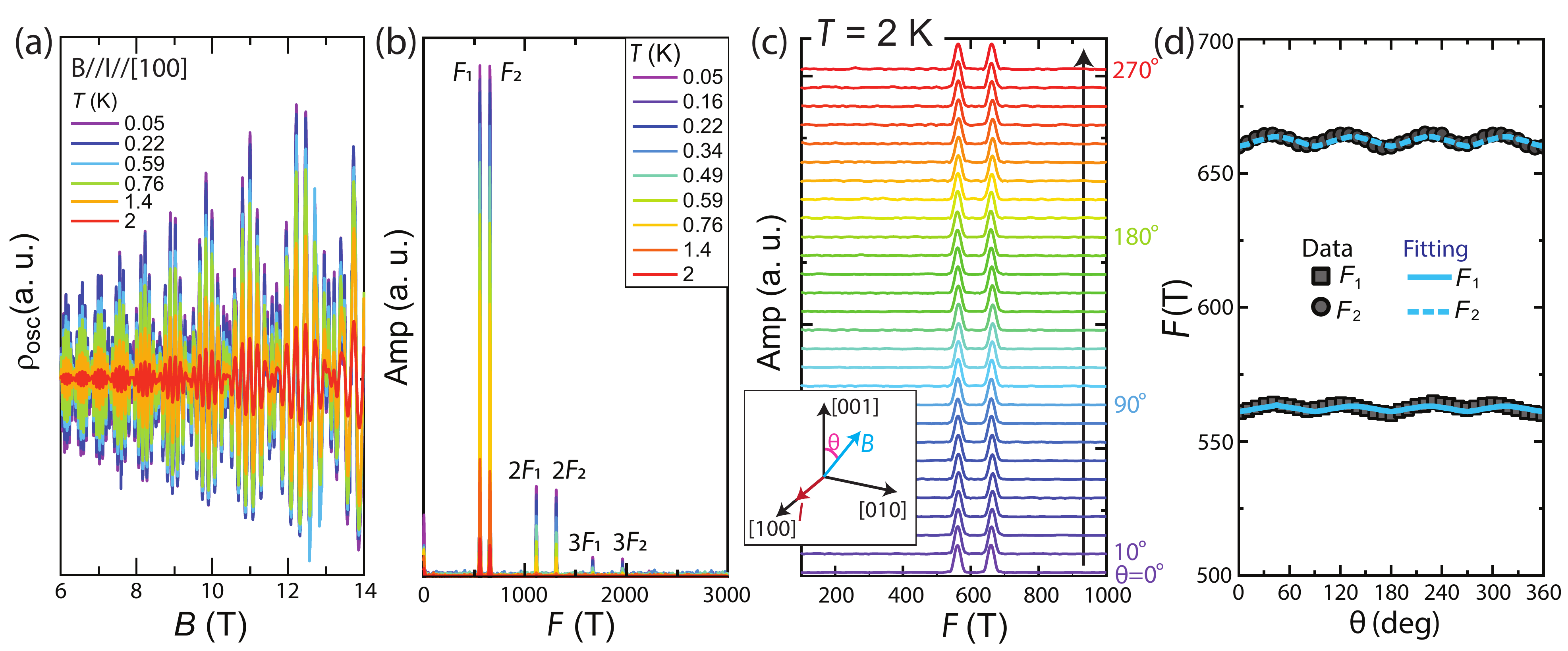}
\centering
	\caption{(a) Temperature-dependent SdH oscillations with field and current applied along [100] axis. Here $\rho_{osc}=\Delta \rho / \rho_{BG}$, with $\Delta \rho$ the oscillatory part of the magnetoresistivity, and $\rho_{BG}$ the background obtained from a 3$^{rd}$-order polynomial fit to the magnetoresistivity. (b) Fast-Fourier-transformation spectrum of the SdH oscillations presented in (a) with the field window of 3 to 14 T. Two main peaks, as well as their higher-harmonic components, can be clearly observed. The suppression of peak amplitude with increasing temperature is due to the thermal damping effect. (c) Fast-fourier-transformation (FFT) spectrum of angle-dependent quantum oscillations measured at $T$ = 2 K. Here the magnetic field is rotated within (100) plane and the angle is defined between the field direction and [001] axis. (d) Summary of angular dependence of oscillation frequencies. Here the fitting is genereated by calculating the orbital area based on band structure calculations with taking extrem magnetic breakdown due to quasi-symmetry into account (see supplement). The near-perfect fitting clearly demonstrates that quasi-symmetry is the only option to explain the experimental results of nearly angle-independent oscillation frequencies.}
	\label{Data}
\end{figure}

CoSi single crystals resemble octahedra, indicating a dominant growth along the [111] direction (see supplement Sec.I.A. for growth details). These crystals were fabricated into microstructures using focused-ion-beam machining \cite{moll2018focused} (see supplement Sec.I.B.) to increase the current path homogeneity and signal of Shubnikov-de Haas oscillations. This microfabrication will later be the key to study the effects of unidirectional strain. The resistivity as well as the large residual resistivity ratio in the samples agrees well with previous bulk measurements \cite{CoSi_JS}, indicating an unchanged material quality during fabrication (see supplement Fig. S1). Hence not surprisingly, large quantum oscillations of the magnetoresistance are readily observed at low fields (Fig.~\ref{Data}). Subtracting a polynomial background uncovers strong oscillations that resemble a beating pattern of two frequencies around $F_1\sim$ 550 T and $F_2 \sim$ 660 T, in good agreement with existing literature\cite{CoSi_CPL,CoSi_JS,CoSi_dHvA,CoSi_Plef}. The temperature dependence of the oscillations follows well the Lifshitz-Kosevich form, leading to a cyclotron effective mass of $m_c\sim 0.84~m_e$ (see supplement Fig. S14). 

\begin{figure}
\includegraphics[width=0.55\columnwidth]{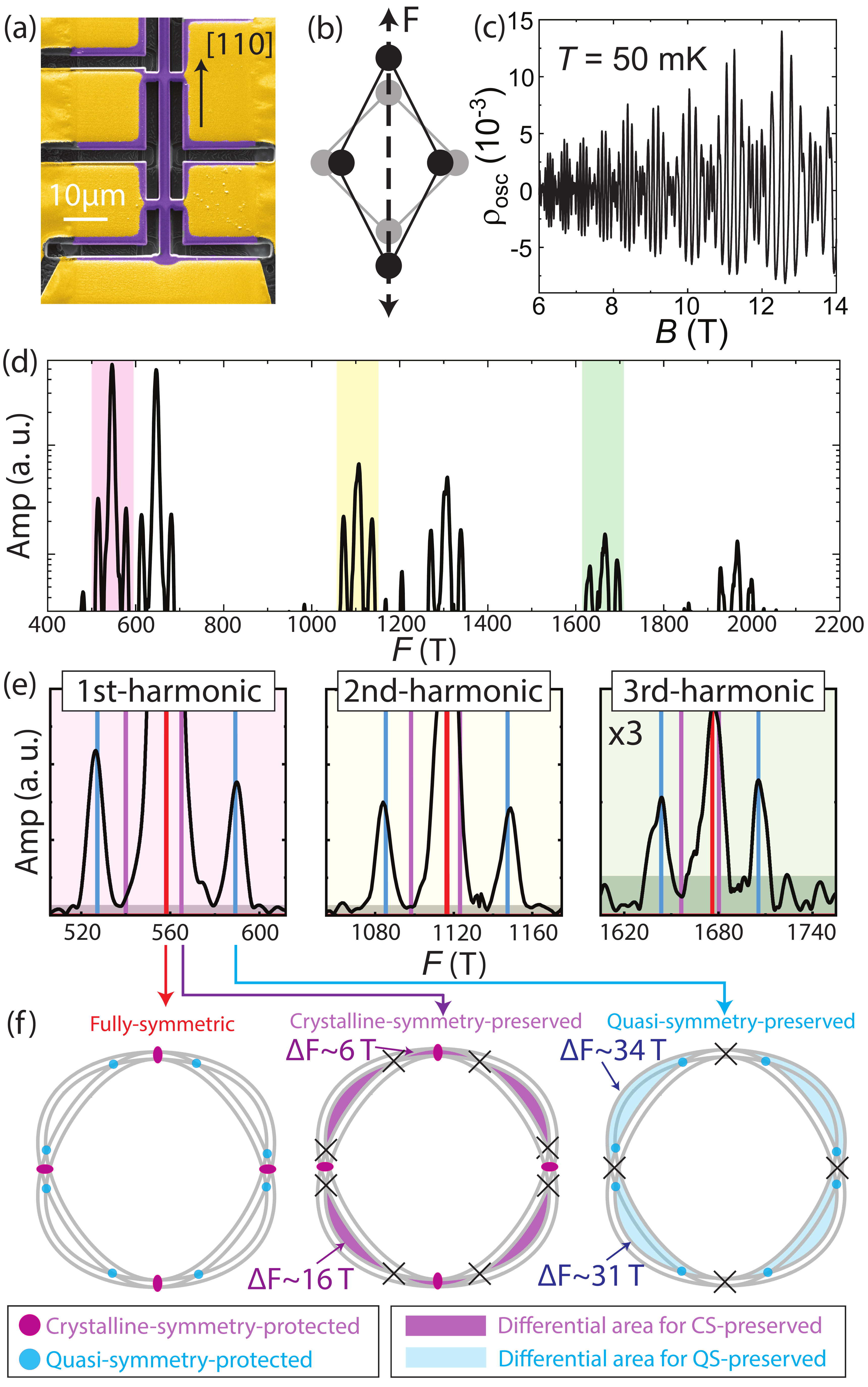}
\centering
	\caption{(a) Scanning electron microscope image of CoSi microdevice. A long bar roughly along [110] direction with a 2.5 by 2.4 $\mu m^2$ cross section is fabricated by FIB. (b) Illustration of tensile strain approximately along [110] which breaks the $C_2$ rotational symmetry of the crystal structure. (c) SdH oscillations with both field and current applied along the fabricated bar direction at $T = 50$ mK. (d) Logarithmic-scaled FFT spectrum of SdH oscillations displayed in (c). The main peaks are always accompanied with satellite peaks up to the third harmonics. (e) Enlarged view of satellite peaks correspond to the 1$^{st}$ to 3$^{rd}$ harmonic oscillations. The red, purple and blue vertical lines correspond to the FFT spectrum produced by the fully symmetric, crystalline-symmetry-preserved and quasi-symmetry-preserved scenarios respectively. (f) Corresponding Landau orbits for three different scenarios. Here the colored area illustrates the orbital area difference compared to the fully-symmetric case, and the black crosses represent the degeneracies that are lifted in different scenarios. Only the quasi-symmetry-preserved scenario reproduces FFT peaks that match perfectly well with the experimental data.}
	\label{Main}
\end{figure}

In light of the complexity of self-intersecting Fermi surfaces it appears at first surprising that the quantum oscillation spectrum is tantalizingly simple, with only two frequences that, most importantly, only weakly depend on the direction of the applied magnetic field (Fig.~\ref{Data}). When magnetic quantum oscillations arise from such highly degenerate Fermi surfaces, the main question is which, if any, possible  trajectories become quantum coherent. As the Fermi-surface is centered around the R-point, any orbit necessarily crosses the symmetry-enforced degeneracies at the Brillouin zone boundary multiple times. This has been argued to stabilize band degeneracies pinned to the Fermi level in this crystal structure\cite{CoSi_Plef,MnSi_Plef} on the planes of Brilloiun zone boundary [marked purple in Fig.~\ref{Main}(f)].

Yet the orbits further intersect at low-symmetry points at which a gap must open [marked blue in Fig.~\ref{Main}(f)]. Neither the frequencies nor their angle dependence obtained from DFT calculations match the data if this gap is considered to separate the orbits via significant avoided-crossing. As the gap is small, the quasiparticle can tunnel across it continuing on the original trajectory, a phenomenon known as magnetic breakdown\cite{MB,Cuprates,Organics,ZrSiS}. Based on estimations of the gap from DFT and further analytical calculations, vanishingly small breakdown fields fall below {0.11~T} for any field orientation, and hence the experimental data is always obtained in an extreme breakdown regime (see supplement Sec. V. for details). Thus the wavefunctions propagate through these points at near-perfect transmission, and indeed an excellent match of frequencies and their angle dispersion between theory and experiment is found (Fig.~\ref{Data}).

These observations raise a key question about the electronic structure, namely why in a metal with eV bandwidth two bands anti-cross with a gap no larger than 2~meV in extended regions of the Brillouin zone. To obtain such a parametrically small gap accidentally, an unreasonable degree of fine-tuning is required that acts simultaneously at many k-points, as well as at a wide range of energies as we will show. Instead, a quasi-symmetry enforces the uniform smallness of this gap, and thus explains why magnetic breakdown at full transparency occurs at any arbitrary angle.

To understand this quasi-symmetry, we next consider an effective model around the R-point. Without spin-orbit coupling (SOC), all the bands are eight-fold degenerate at R due to the combination of two-fold screw axis symmetries along the x and y direction and time reversal symmetry. Including spin doubles the degeneracy and the SOC splits the eight-fold degenerate states into six-fold and two-fold degenerate states at the R-point. An effective model for these eight bands around the R-point can be constructed as
\begin{equation}\label{eq:HR}
\mathcal{H}_{R}=\mathcal{H}_{0}(\mathbf{k})+\mathcal{H}_{soc}+\mathcal{H}_{k^2}(\mathbf{k}),
\end{equation}
where $\mathcal{H}_{0}(\mathbf{k})=  C_0 + 2A_1 ({\bf k\cdot L})$ is the lower-order expansion of the spin-independent Hamiltonian, and three 4-by-4 matrices ${\bf L}$ form an emergent angular momentum algebra $[L_i,L_j]=i \varepsilon_{ijk} L_k$ with Levi-Civita symbol $\varepsilon_{ijk}$ and $i,j,k=x,y,z$. The SOC term is given by $\mathcal{H}_{soc}=2\lambda_0 ( {\bf s\cdot L})$, where ${\bf s}$ is the spin operator. $\mathcal{H}_{k^2}(\mathbf{k})$ is the higher-order spin-independent term with its form given in the supplement. Here $\mathcal{H}_0(k)+\mathcal{H}_{k^2}(k)$ is the SOC-free expanded Hamiltonian to second order. By choosing appropriate parameters such as $C_0, A_1, \lambda_0$, the energy dispersion of $\mathcal{H}_{R}$ well reproduces that from the DFT calculations (see supplement Sec. VII.). 

A striking feature in the Hamiltonian $\mathcal{H}_{R}$ is that the SOC term $\mathcal{H}_{soc}$ takes a similar form as the linear-momentum term, just by replacing the momentum ${\bf k}$ by spin ${\bf s}$. This is because spin, as a pseudo-vector, behaves exactly the same as a vector due to the lack of inversion, mirror or other roto-inversion symmetries for a chiral crystal. Due to this similarity, up to the first-order perturbation, we find that the spin of the eigen-states is parallel or anti-parallel to the momentum ${\bf k}$ and thus we can label the spin states of these bands by $\pm$ in Fig.~\ref{Crystal}(b). We derive an effective 
Hamiltonian for the near degenerate bands. This can be done by first projecting the full Hamiltonian $\mathcal{H}_{R}$ into the 4 closest bands ($1^\pm$ and $2^\pm$ bands) to get a 4-band model and then constructing a 2-band model for the $1^-$ and $2^+$ bands, as discussed in supplement. Up to the first-order perturbation, the 2-band Hamiltonian takes the form $H_{eff}=\epsilon_0+d_z({\bf k})\sigma_z$, where $d_z({\bf k})=\lambda_0-\sqrt{3}\tilde{C}\frac{|k_x k_y k_z|}{k}$ and the form of the function $\epsilon_0$ is given in supplement. Here $\sigma_z$ is the Pauli matrix on the basis of $1^-$ and $2^+$ bands, $\lambda_0$ term comes from the SOC Hamiltonian $\mathcal{H}_{soc}$ and the $\tilde{C}$ term from $\mathcal{H}_{k^2}(\mathbf{k})$. It is striking to see that the operator $\mathcal{M}_{eff}=\sigma_z$ commutes with $H_{eff}$ and thus serves as an {\it exact} symmetry at the first-order expansion. The existence of $\mathcal{M}_{eff}$ forbids any terms that are coupled to $\sigma_{x,y}$ in the effective Hamiltonian, thus reducing the codimension of a two-level crossing from 3 to 1 and stabilizing a nodal plane, which is defined by the equation $\lambda_0= \sqrt{3}\tilde{C}\frac{|k_x k_y k_z|}{k}$ in this two-band model. A more accurate and complete description around the Fermi energy is the 4-band model, which shares a similar k-dependent hidden quasi-symmetry and is essential in deriving the 2-band model (see supplement Sec. VII.), and the whole perturbation theory reveals a striking hierarchy structure. Higher-order perturbation expansion breaks $\mathcal{M}_{eff}$ through inducing additional terms coupled to $\sigma_{x,y}$ in the effective model and thus leads to a small gap opening of this nodal plane (see supplement). Therefore, $\mathcal{M}_{eff}$ is not a symmetry of the system, but an emergent quasi-symmetry in the sense of low-energy effective theory. Importantly, this curved nodal plane spans low-symmetry regions of the Brillouin zone, in contrast to symmetry-enforced exact degeneracies. It is this combined structure of symmetry and quasi-symmetry induced degeneracies that forms the basis of the complex structure of the Fermi surface centered at R, and reduces the complexity of the quantum orbits in magnetic fields to two, angle-independent frequencies. 

Physically, quasi-symmetry is linked to the orbital structure of the wavefunctions, and its operation only interchanges the orbitals at different atomic sites (internal degrees). Thus, unlike crystalline symmetries, they remain robust against symmetry-breaking lattice distortions (Fig. 4). This is probed experimentally by the application of tensile strain to a bar cut approximately along the [110] direction. The low-symmetry distortion removes the critical screw symmetries at the zone boundaries, and gaps the otherwise symmetry protected crossing points. Experimentally, such strain can be easily applied by thinning the transport bar by FIB-milling, thus reducing its effective spring constant and increasing the total distortion under equal force. The microstructures are coupled to a sapphire substrate and their differential thermal contraction generates the desired tensile strain along the bar [see supplement and \cite{Maja_Ce115}]. As such strain lifts the screw symmetry protecting the degeneracy, one naturally expects the sample to gradually leave the magnetic breakdown regime as the strain in increased and a zoo of sum frequencies to emerge that correspond to all possible area differences between the orbits\cite{Cuprates,Organics,ZrSiS}. Indeed, the strained bar exhibits pronounced satellite peaks that are symmetrically offset from the main peaks by $\Delta F \approx 32~T$ for all harmonics. This matches exactly with the calculated area difference between the crystalline symmetry protected points, fully explaining the observed spectrum. However, those areas associated with a quasiparticle tunneling at a quasi-symmetry-protected point are not observed, implying that the magnetic breakdown field associated with that gap remains much smaller than the lowest field at which quantum oscillations are observed.  This is a direct experimental proof that the quasi-symmetry protection is insensitive to breaking of crystalline symmetries, a notion that can be well corroborated by calculations (Fig. 4). This resilience clearly distinguishes symmetry and quasi-symmetry enforced topological band anomalies.

Their unique implications in the Berry curvature distribution in momentum space are another key characteristic of the topological character of near-degeneracies enforced by quasi-symmetry, compared to the exact degeneracies due to symmetry [Fig.~\ref{Crystal}(f)]. The Berry curvature almost vanishes in the vicinity of the exact degenerate plane, yet it is concentrated to a ring on the near nodal plane at any fixed energy. This large Berry curvature around the near nodal lines, which always occurs at the Fermi energy, may strongly affect physical phenomena that are related to local Berry curvature, including intrinsic spin Hall effect\cite{NodalLine,SHE} and quantum non-linear Hall effect\cite{NonLinearHall}. 



The quasi-symmetry is the mathematical and physical framework behind the unexpectedly simple and strain-resistant quantum oscillation spectrum of CoSi. It is, however, just one example of a large class of materials in which quasi-symmetries can exist. Clearly, this new class of near degenerate manifolds at low-symmetry k-points is not found by current search programs for topological materials based on crystalline symmetry without further guideline. To search for materials with quasi-symmetries, a systematic expansion of the ${\bf k\cdot p}$-type effective Hamiltonian order by order is required for all space groups, which is in principle feasible \cite{Encyclopedia}. Typically, lower-order terms with higher symmetries than the crystalline symmetry itself possess larger prefactors than the higher-order terms with lower symmetries\cite{SOCbook}, and a symmetry hierarchy is expected for the terms of different orders in the ${\bf k\cdot p}$-type Hamiltonian. Further interesting flavors of quasi-symmetry may exist which emerge when higher-order terms with additional symmetries become dominant. Experimentally, we propose that quantum oscillations may provide guidance by pointing to materials with similarly extreme magnetic breakdown on extended regions on the Fermi surface, which a quasi-symmetry would naturally explain. While the here described search methods need to be implemented in the future, it is natural to anticipate quasi-symmetries in related compounds crystallizing in the same structure, such as e.g. PtGa, PtAl, RhSi and the magnetic MnSi where the influence of magnetism can be explored.\cite{	PtGa_fermiArc,PtGa_NodalWall,PtGa_QO,RhSi_hasan,PtAl,PdGa_Hand,PtGa_CN,RhSi_optic,MnSi_Plef,MnSi_skr}. Our DFT calculations indeed reveal similar energy dispersions and Fermi surface shapes with quasi-symmetry-induced near degeneracies in PtAl, PtGa and RhSi (see supplement Sec. VI.). The concept of the quasi-symmetry raises the interesting prospect that even those materials in which crystalline symmetries do not allow stable band crossings may still host exotic quasiparticles and strong Berry curvature at the Fermi level. If such features at the Fermi level are desired, the current approach of automated symmetry classifications may not prove to lead to the most practical materials. Without the necessity of a crystalline symmetry protecting higher-order band crossing points, quasi-symmetries hold an advantage for topological applications. The topological anomalies in real materials often are far away from the Fermi level, and in thin-film form they are exposed to mismatch strains inherent to heterostructures. These key bottlenecks of topological materials do not restrict quasi-symmetric materials. A consensus of high-throughput efforts emerges that 20-30\% of materials are topological\cite{bradlyn2017topological,TopoCatalogue,TopoCatalogue2,XGW_topoSearch}. Yet it appears that crystal symmetry may not hold up as the strict barrier it was thought to be, and via quasi-symmetries concepts, topology may reach even further than these estimates predict.

\clearpage

\noindent \textbf{Acknowledgements} 
We would like to acknowledge Dr. Joerg Harms for the assistance on graphic design. This work was funded by the European Research Council (ERC) under the European Union’s Horizon 2020 research and innovation programme (MiTopMat - grant agreement No. 715730). This project received funding by the Swiss National Science Foundation (Grants  No. PP00P2\_176789). C. X. Liu and L. H. Hu are supported by the Office of Naval Research (Grant No. N00014-18-1-2793) and Kaufman New Initiative research Grant No. KA2018-98553 of the Pittsburgh Foundation. K.M., and C.F. acknowledge the financial support from the European Research Council (ERC) Advanced Grant No. 742068 "TOP-MAT”; European Union’s Horizon 2020 research and innovation program (Grant Nos. 824123 and 766566)and Deutsche Forschungsgemeinschaft (DFG) through SFB 1143. Additionally, K. M. acknowledges Max Plank Society for the funding support under Max Plank–India partner group project. B.A.B. thanks funding from the European Research Council (ERC) under the European Union’s Horizon 2020 research and innovation programme (grant agreement no. 101020833). B.A.B. is also supported by the U.S. Department of Energy (Grant No. DE-SC0016239), and partially supported by the National Science Foundation (EAGER Grant No. DMR 1643312), a Simons Investigator grant (No. 404513), the Office of Naval Research (ONR Grant No. N00014-20-1-2303), the Packard Foundation, the Schmidt Fund for Innovative Research, the BSF Israel US foundation (Grant No. 2018226), the Gordon and Betty Moore Foundation through Grant No. GBMF8685 towards the Princeton theory program, and a Guggenheim Fellowship from the John Simon Guggenheim Memorial Foundation. B.A.B. and C. X. Liu is supported by the NSF-MERSEC (Grant No. 
MERSEC DMR 2011750). B.A.B. gratefully acknowledge financial support from the Schmidt DataX Fund at Princeton University made possible through a major gift from the Schmidt Futures Foundation. B.A.B. received additional support from the Max Planck Society. Further support was provided by the NSF-MRSEC No. DMR-1420541, BSF Israel US foundation No. 2018226, and the Princeton Global Network Funds.

\noindent \textbf{Author Contributions} C.C. and L.H. contributes equally to this work. Crystals were synthesized and characterized by K.M., C.S. and C.F.. The experiment design, FIB microstructuring and the magnetotransport measurements were performed by C.G., C.P., J.D., X.H. and P.J.W.M.. L.H., C.L. and B.A.B. developed and applied the general theoretical framework, and the analysis of experimental results has been done by C.G. C.P. and P.J.W.M.. Band structures were calculated by Y.S. F.F. and C.F.. All authors were involved in writing the paper.\\

\noindent \textbf{Competing Interests} 
The authors declare that they have no competing financial interests.

\section*{Methods}
\subsection*{Data Availability}
Data that support the findings of this study is deposited to Zendo with the access link: https://doi.
org/10.5281/zenodo.6336000.
\subsection*{Code Availability}
Matlab code used for this study is deposited to Zendo with the access link: https://doi.org/
10.5281/zenodo.6336013.

\clearpage

\clearpage
\clearpage
\clearpage

\clearpage

\bibliographystyle{Nature}

\clearpage

\end{document}


\renewcommand{\theequation}{S\arabic{equation}}
	\newcommand{\beginsupplement}{%
		\setcounter{table}{0}
		\renewcommand{\thetable}{S\arabic{table}}%
		\setcounter{figure}{0}
		\renewcommand{\thefigure}{S\arabic{figure}}%
	}
\title{Supplementary materials for "Quasi-symmetry protected topology in a semi-metal"}

\author{Chunyu Guo${}^{\ast}$}\affiliation{Laboratory of Quantum Materials (QMAT), Institute of Materials (IMX),\'{E}cole Polytechnique F\'{e}d\'{e}rale de Lausanne (EPFL), CH-1015 Lausanne, Switzerland}
\author{Lunhui Hu${}^{\ast}$}\affiliation{Department of Physics, The Pennsylvania State University, University Park, PA, USA}
\author{Carsten Putzke${}^{}$}\affiliation{Laboratory of Quantum Materials (QMAT), Institute of Materials (IMX),\'{E}cole Polytechnique F\'{e}d\'{e}rale de Lausanne (EPFL), CH-1015 Lausanne, Switzerland}\affiliation{Max Planck Institute for the Structure and Dynamics of Matter, 22761 Hamburg, Germany}
\author{Jonas Diaz${}^{}$}\affiliation{Laboratory of Quantum Materials (QMAT), Institute of Materials (IMX),\'{E}cole Polytechnique F\'{e}d\'{e}rale de Lausanne (EPFL), CH-1015 Lausanne, Switzerland}
\author{Xiangwei Huang${}^{}$}\affiliation{Laboratory of Quantum Materials (QMAT), Institute of Materials (IMX),\'{E}cole Polytechnique F\'{e}d\'{e}rale de Lausanne (EPFL), CH-1015 Lausanne, Switzerland}
\author{Kaustuv Manna${}^{}$}\affiliation{Max Planck Institute for Chemical Physics of Solids, 01187 Dresden, Germany}\affiliation{Department of Physics, Indian Institute of Technology Delhi, New Delhi 110016, India}
\author{Feng-Ren Fan${}^{}$}\affiliation{Max Planck Institute for Chemical Physics of Solids, 01187 Dresden, Germany}
\author{Chandra Shekhar${}^{}$}\affiliation{Max Planck Institute for Chemical Physics of Solids, 01187 Dresden, Germany}
\author{Yan Sun${}^{}$}\affiliation{Max Planck Institute for Chemical Physics of Solids, 01187 Dresden, Germany}
\author{Claudia Felser${}^{}$}\affiliation{Max Planck Institute for Chemical Physics of Solids, 01187 Dresden, Germany}
\author{Chaoxing Liu${}^{\dagger}$}\affiliation{Department of Physics, The Pennsylvania State University, University Park, PA, USA}\affiliation{Department of Physics, Princeton University, Princeton, New Jersey 08544, USA}
\author{B. Andrei Bernevig${}^{\dagger}$}\affiliation{Department of Physics, Princeton University, Princeton, New Jersey 08544, USA}\affiliation{Donostia International Physics Center, P. Manuel de Lardizabal 4, 20018 Donostia-San Sebastian, Spain}\affiliation{IKERBASQUE, Basque Foundation for Science, Bilbao, Spain}
\author{Philip J. W. Moll${}^{\dagger}$}\affiliation{Laboratory of Quantum Materials (QMAT), Institute of Materials (IMX),\'{E}cole Polytechnique F\'{e}d\'{e}rale de Lausanne (EPFL), CH-1015 Lausanne, Switzerland}
\date{\today}

\maketitle
\beginsupplement
	
\section{Experimental and theoretical methods}
\subsection{Single crystal growth}
\begin{figure}
	\includegraphics[width=0.95\columnwidth]{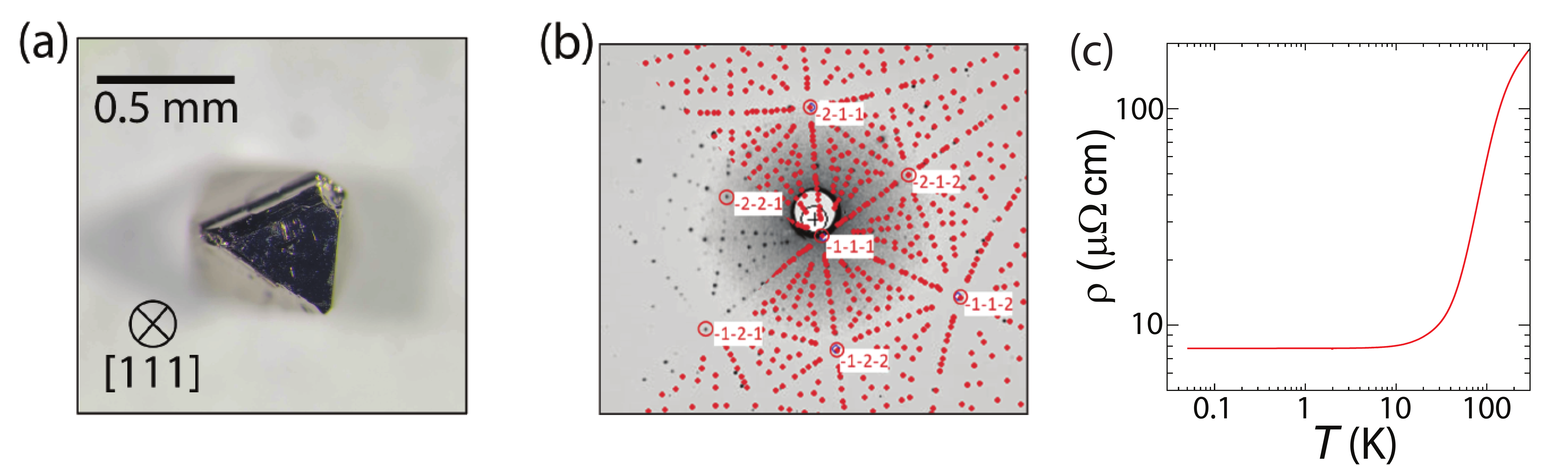}
	\caption{ (a) Optical microscope image of a CoSi single crystal sample. (b) Laue diffraction pattern of the grown CoSi single crystal, superimposed with a theoretically simulated one confirming high crystalline quality. (C) Temperature dependence of resistivity of device-1 measured with electrical current applied along [100] direction.}
	\label{Xray}
\end{figure}

CoSi single crystals were grown in Te-flux. The high purity starting materials Co (99.95\%, Alfa Aesar), Si (99.999\%, Chempur) and Te (99.9999\%, Alfa Aesar)  were mixed in the molar ratio of 1:1:20. All materials were kept in a cylindrical alumina crucible and sealed in a quartz tube. The entire assembly was heated to 1050 $^\circ$C at a rate of 100 $^\circ$C/h, and held there for for 15 h to ensure a homogeneous mixture. Successively, the sample was cooled to 700 $^\circ$C at a rate of 2 $^\circ$C/h, and extra Te-flux was removed by centrifugation at 700 $^\circ$C. High quality CoSi single crystals in the mm-range resulted from this growth protocol. These single crystals resemble octahedra, indicating a dominant growth along the [111] direction [Fig.~\ref{Xray} (a)]. The quality of the single-phase crystallinity was checked by Laue and single crystal X-Ray diffraction. The Laue diffraction pattern of CoSi matches well with the theoretically simulated one [Fig.~\ref{Xray} (b)], demonstrating high crystalline quality.

\subsection{Focused-ion-beam (FIB) microstructuring}
100 $\times$ 40 $\times$ 2 $\mu m$ thin slabs were cut out from a CoSi single crystal via a FEI Helios Plasma FIB using Xe-ions. The lamella was then transferred to a sapphire substrate ex-situ with a micro-manipulator and glued down with the red araldite epoxy. Afer that it was structured to the designed geometry with the Plasma FIB. 

\subsection{Details of ab-initio band structure calculation}
\begin{figure}
	\includegraphics[width=0.6\columnwidth]{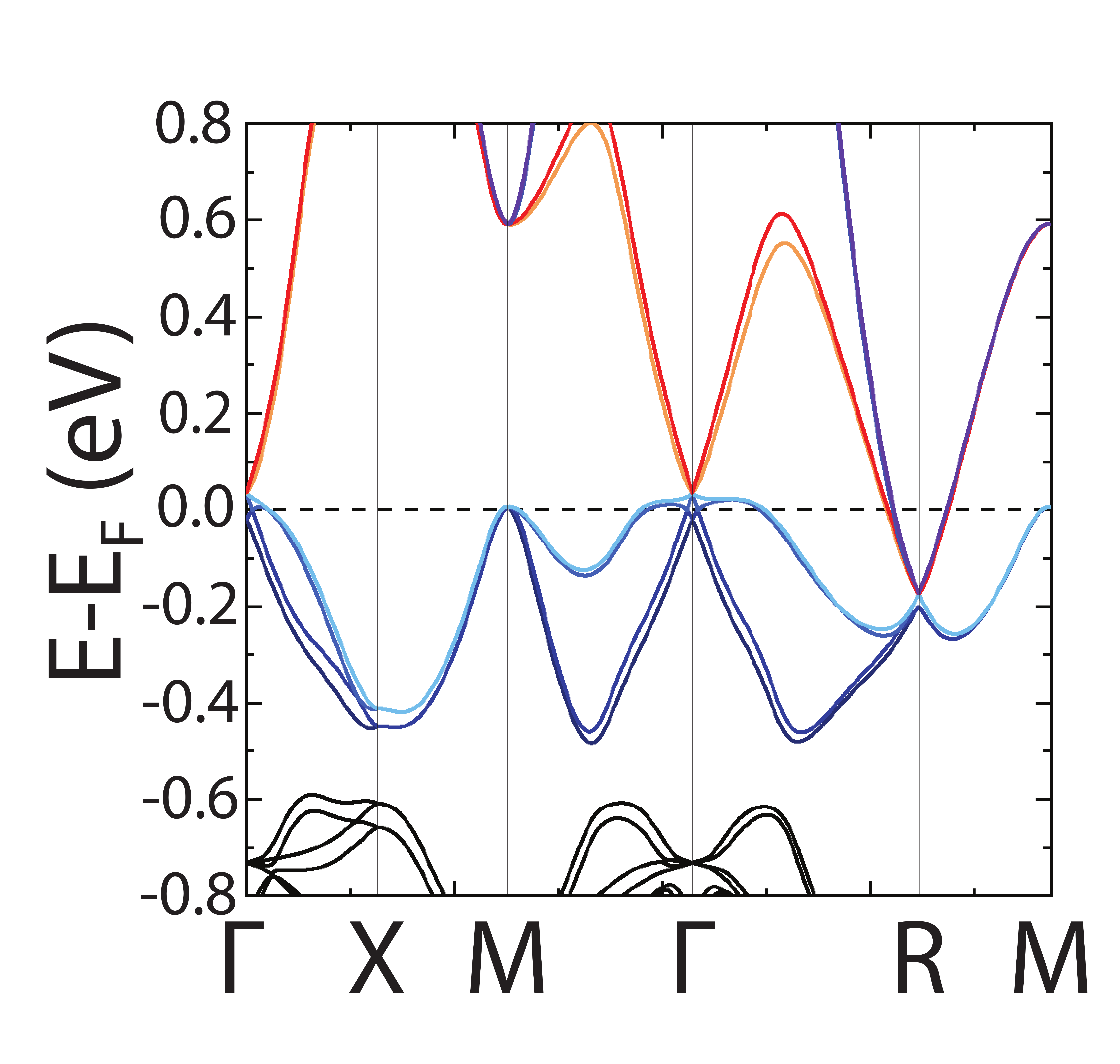}
	\caption{ Ab-initio-calculated band structure of CoSi.}
	\label{BS}
\end{figure}

Electronic band structures of CoSi were calculated by density functional theory (DFT) using the full-potential local-orbital code (FPLO) \cite{fplo} with a localized atomic basis and full potential treatment (Fig.~\ref{BS}). The exchange and correlation energies were considered in the generalized gradient approximation level \cite{gga}. To calculate the Fermi surface, we projected Bloch states onto high symmetric atomic-orbital-like Wannier functions and constructed tight-binding model Hamiltonian.

\section{Detailed analysis of Fermi surfaces at R-point}

\subsection{Fermi surface Identity}
\begin{figure}
	\includegraphics[width=1\columnwidth]{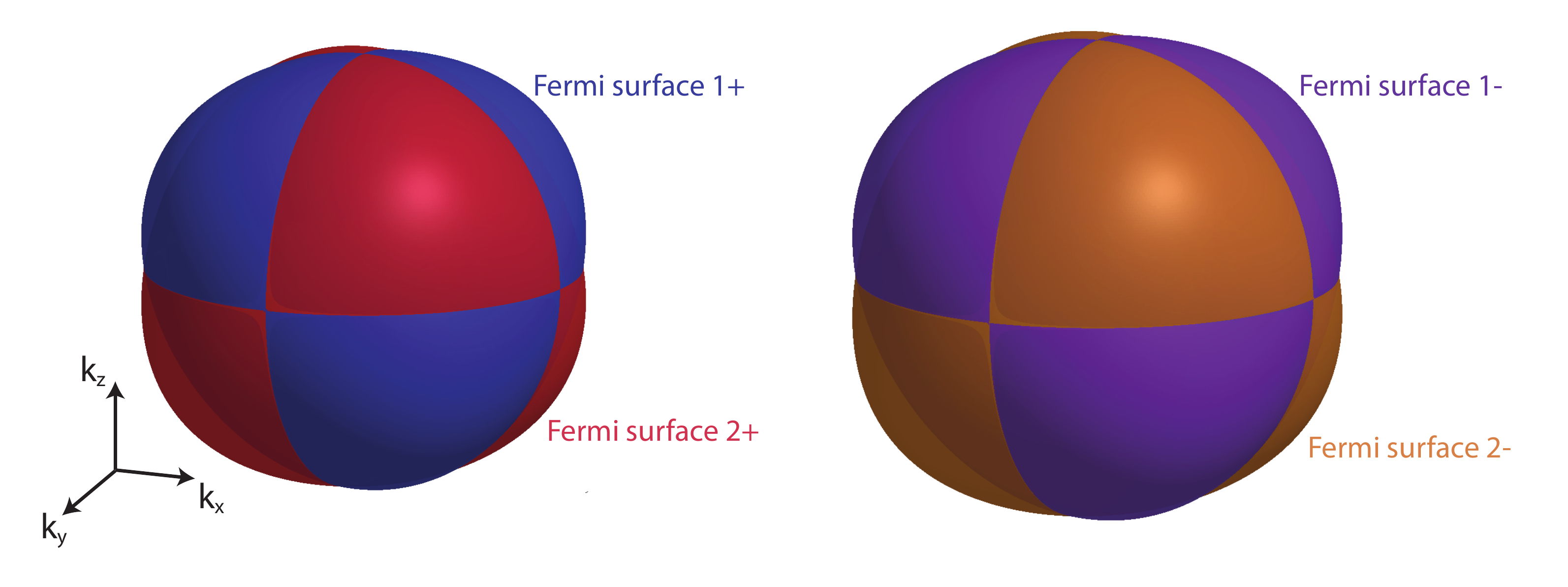}
	\caption{Fermi surface identification. Two pairs of Fermi surfaces are presented and labelled with their band characters, consistent with the identification in band structure calculations.}
	\label{FSID}
\end{figure}

Following the band character identification in band structure calculations [Fig. 2(d)] presented in the main manuscript, the four corresponding Fermi surfaces are also labelled as $1^+,1^-,2^+$ and $2^-$ (Fig.~\ref{FSID}). Here 1 and 2 denote the orbit character while +/- stand for the spin character. Note that the sizes of 1- and 2- Fermi surfaces are larger than their spin-opposite counterpart due to spin-orbit-coupling as discussed in Sec. \ref{SOCSec}. For the clarification we separate them into two pairs: $1^+$/$2^+$ and $1^-$/$2^-$. Due to crystalline symmetry, these Fermi surfaces intersect exactly at the Brillouin zone boundary.

\subsection{Slice-and-view of Fermi surfaces}
\begin{figure}
	\centering
	\includegraphics[width=0.77\columnwidth]{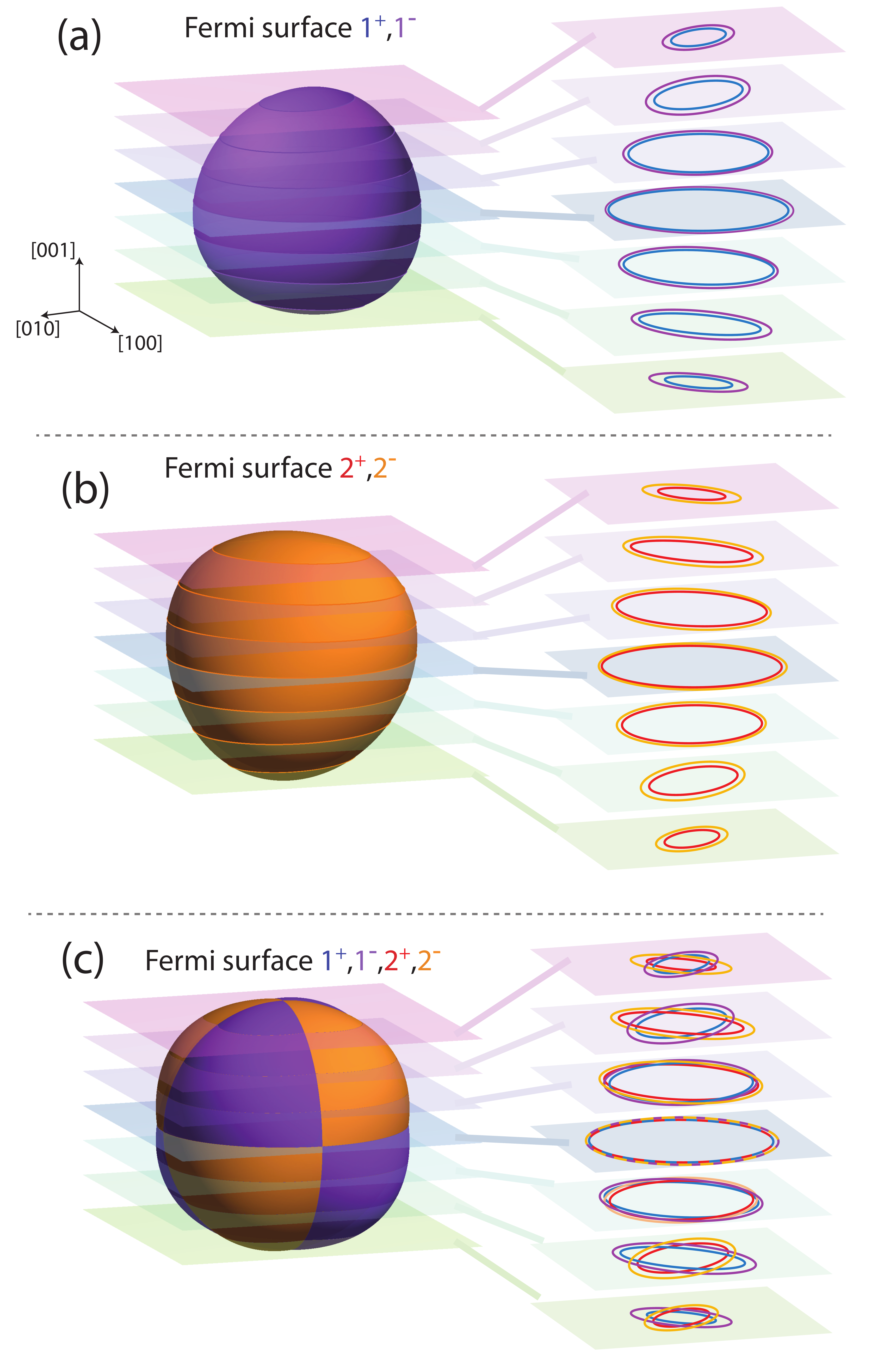}
	\caption{(a) and (b) present the slice-and-view of Fermi surfaces 1$^\pm$ and 2$^\pm$ respectively. As one can clearly see the sliced orbits of 1$^-$ or 2$^-$ are always larger than 1$^+$ or 2$^+$ at any level. This difference is due to spin splitting caused by spin-orbital coupling (SOC). (c) The summarized view of all four Fermi surfaces. It displays four orbits intersect with each other in a complex pattern which results in the degenerate points protected by either crystalline symmetry or quasi-symmetry.}
	\label{Slice-view}
\end{figure}

To further demonstrate the Fermi surface geometry, the slice-and-view of the four Fermi surfaces is displayed in Fig. \ref{Slice-view}. As slicing from the top to the bottom of the Fermi surfaces, the slice section changes from ellipsoidal to circular and then back to ellipsoidal. Note that the ellipsoids from the top and bottom slice are elongated along orthogonal directions. This orthogonality is generally true for all Fermi surfaces [Fig. \ref{Slice-view} (a) and (b)]. The combination of all sliced orbits, as displayed in Fig. \ref{Slice-view} (c), show both crossings at the high symmetry plane and crossings protected by quasi-symmetry.

\begin{figure}
	\includegraphics[width=0.95\columnwidth]{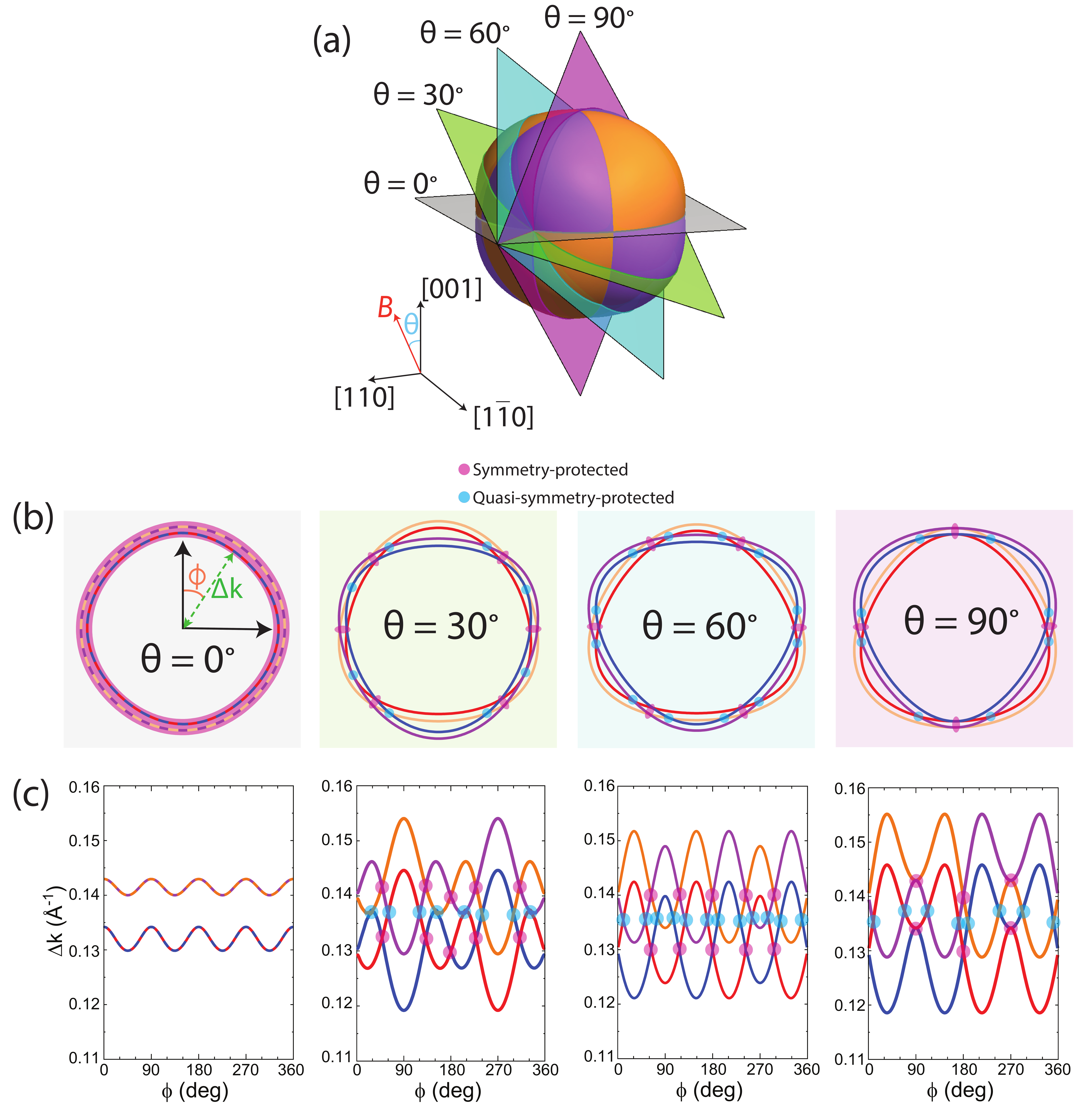}
	\caption{(a) Illustration of the rotation of magnetic field and the corresponding extremal cross section of Fermi surfaces. Here $\theta$ is defined as the angle between field direction and [001] direction, while the square planes stand for the planes perpendicular to the magnetic field. (b) Angle-dependent Fermi surface orbits. The crystalline symmetry and quasi-symmetry protected degeneracies are denoted with pink and blue circles respectively. (c) Momentum difference from R-point $\Delta k$ as a function of $\phi$, which is defined in (b), different types of degenerate points are colored following the color-code in (b).}
	\label{RotCut}
\end{figure}

The Fermi surface orbits with field applied along different directions are shown in Fig. \ref{RotCut} (b) which demonstrate the evolution of degeneracies at the Fermi surfaces. For all four field directions displayed, the Fermi surface orbits are intersecting at multiple k-points. Degeneracies occur at the Brilloiun zone boundary are protected by crystalline symmetry, while the others located at low symmetry k-points are formed due to quasi-symmetry. These two different types of degeneracies are denoted with different colors on Fig. \ref{RotCut} (b). The momentum difference from R-point $\Delta k$ as a function of angle $\phi$ with different applied field directions is presented in Fig. \ref{RotCut} (c) with $\phi$ defined in \ref{RotCut} (b). Two different types of degenerate points are marked with pink and blue circles respectively.

\subsection{Influence of spin-orbit-coupling}\label{SOCSec}
\begin{figure}
	\includegraphics[width=0.95\columnwidth]{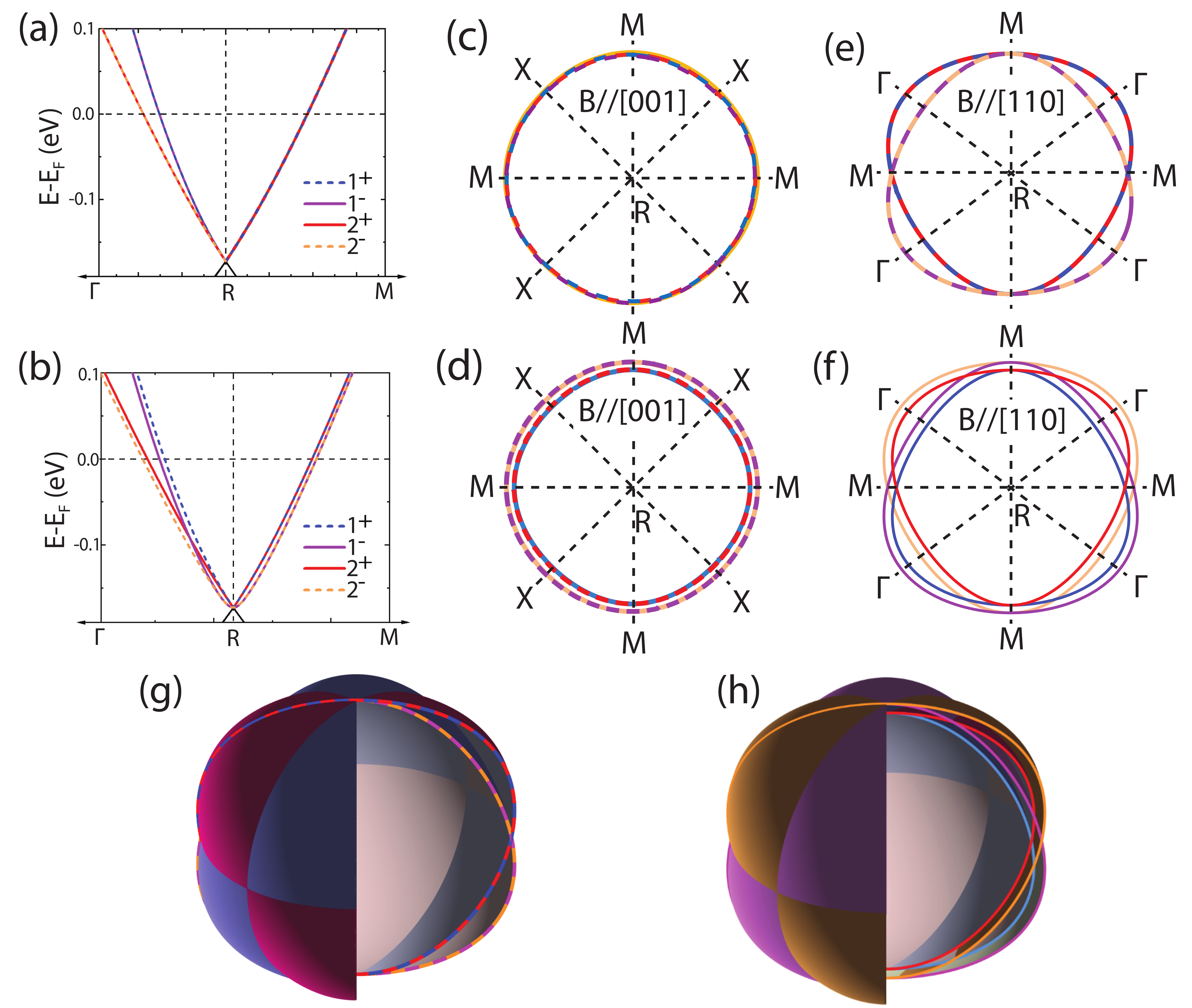}
		\centering
	\caption{(a) and (b) display the results of ab-initio band structure calculation around R-point without and with SOC respectively. When SOC is neglected, spin degeneracy is preserved and therefore all bands are at least two-fold degenerate. At the Brillouin zone boundary (i. e. R to M) all four bands are degenerate due to the protection of crystalline symmetry. Taking SOC into account, the spin degeneracy is therefore lifted and only at the Brillouin zone boundary the bands are fully two fold degenerate due to orbital degree of freedom. This spin splitting effect is also clearly demonstrated with the Fermi surface orbit presented in (c) to (f). At the high symmetry (100) plane, the orbits sit exactly at the zone boundary and therefore for (c) non-SOC case all four orbits are degenerate while for (d) SOC-included case they are doubly degenerate. Meanwhile for orbits sit at the (110) plane, the non-SOC scenario features two spin-degenerate orbits that show four-fold degeneracy only at the zone boundary. While for SOC-included case, the four non-degenerate orbits display a intersecting pattern which results in not only the similar degeneracy protected by crystalline symmetry but also quasi-symmetry protected degeneracies that are not located at the high symmetry directions.}
	\label{SOC}
\end{figure}

In this section, we discuss the influence of spin-orbit-coupling (SOC) on band structure. Without SOC, the spin dengeneracy is preserved and therefore all bands are at least two-fold degenerate throughout the Brilloiun zone. At the boundary of the Brilloiun zone, the crystalline symmetry enforces a four-fold band degeneracy, as revealed by the ab-initio band structure calculation [Fig. \ref{SOC} (a)]. If the SOC is included, spin degeneracy will be lifted therefore all bands are non-degenerate except either at the high symmetry planes or the quasi-symmetry protected planes [Fig. \ref{Slice-view} (b)]. This difference is exemplified for the Fermi surface orbits with different field directions. When the magnetic field is applied along [100] axis, the non-SOC orbits are four-fold degenerate [Fig. \ref{SOC} (c)] while for SOC-included case the orbits are doubly degenerate [Fig. \ref{SOC} (d)]. In the meantime for B//[110] the non-SOC orbits are doubly degenerate except at the high symmetry lines [Fig. \ref{SOC} (e)], while for SOC-included case the four orbits are almost non-degenerate [Fig. \ref{SOC} (f)] yet displays a complex intersecting pattern with degeneracies occur only at certain k-point. The three-dimensional (3D) view of the Fermi surfaces for both non-SOC and SOC-included cases are displayed in Fig. \ref{SOC} (g) and (h) respectively.

\section{Near-degeneracy due to quasi-symmetry}
\begin{figure}
	\includegraphics[width=0.95\columnwidth]{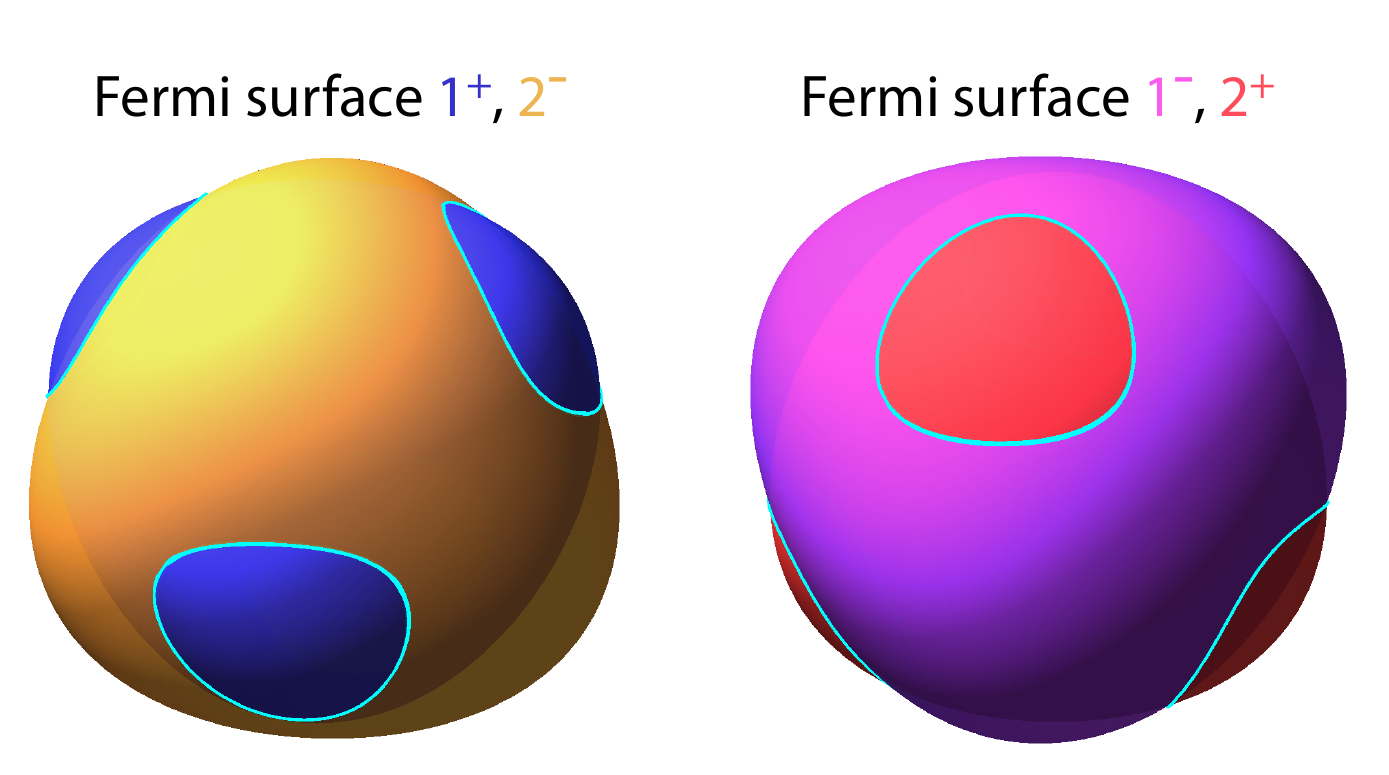}
	\caption{To clearly demonstrate the intersecting pattern of the Fermi surfaces, here we plot the Fermi surfaces 1$^+$/2$^-$ and 1$^-$/2$^+$, they intersect at the quasi-symmetry protected degenerate rings.}
	\label{FSring}
\end{figure}

\begin{figure}
	\includegraphics[width=0.95\columnwidth]{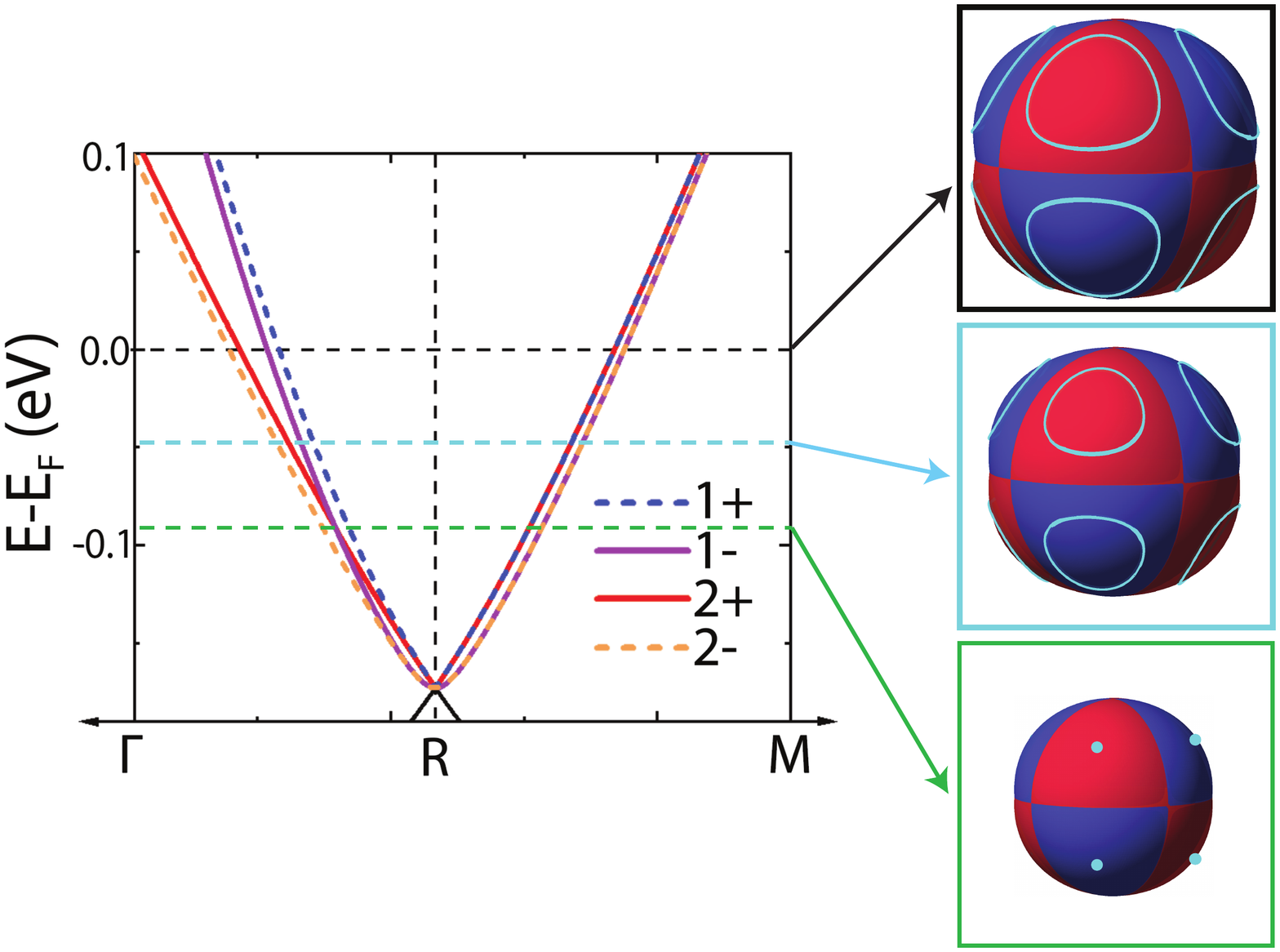}
	\caption{Fermi energy-dependence of Fermi surfaces 1$^+$ and 2$^+$. Except for the natural reduction of Fermi surface size, the degenerate rings also become smaller with lower Fermi energy. And when the Fermi level locates exactly at the type-II Weyl node, the rings shrink and reshape to eight Weyl points.}
	\label{FSringEF}
\end{figure}

\begin{figure}
	\includegraphics[width=0.95\columnwidth]{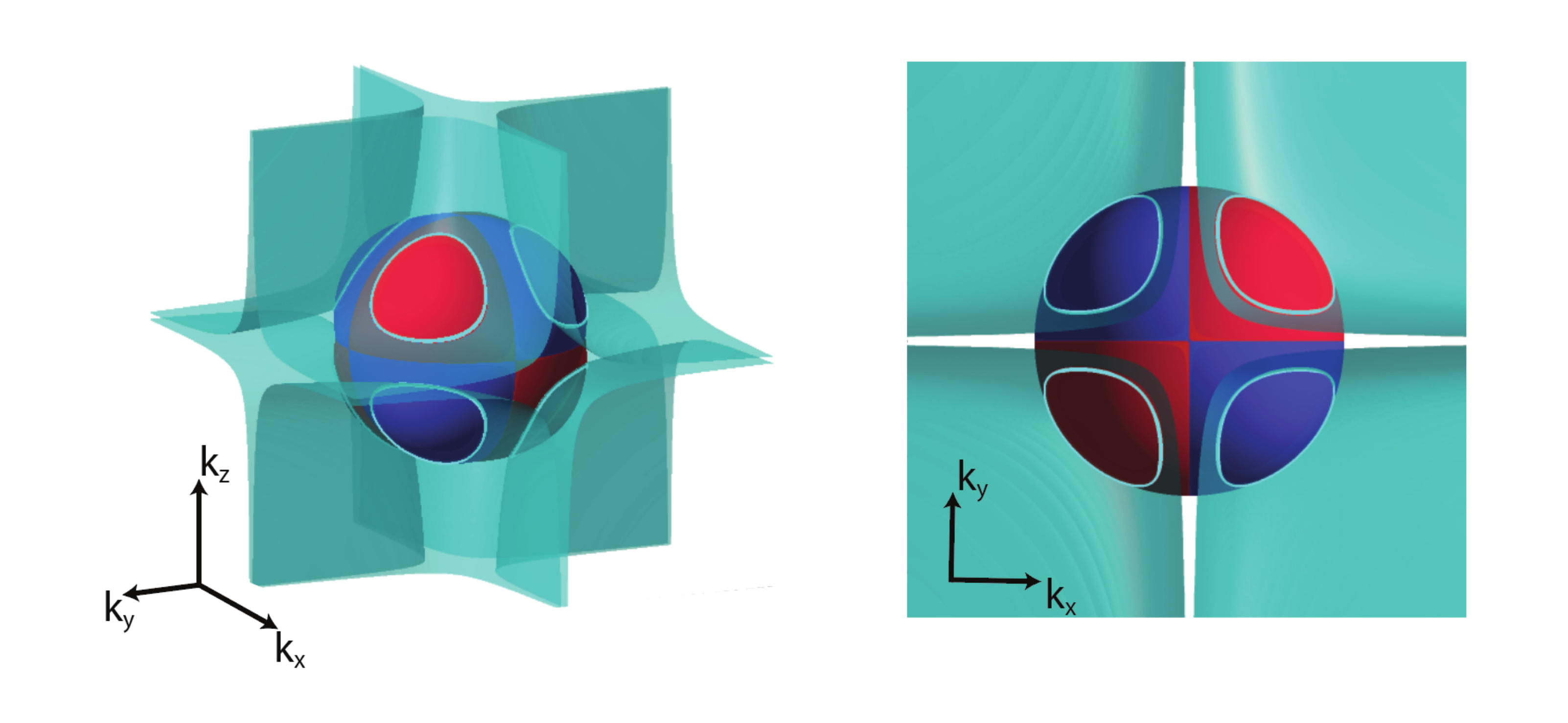}
	\caption{Fermi surfaces 1$^+$/2$^+$ and quasi-symmetry protected degenerate planes. They intersect exactly at the eight degenerate rings.}
	\label{PlaneFS}
\end{figure}

\begin{figure}
	\includegraphics[width=0.95\columnwidth]{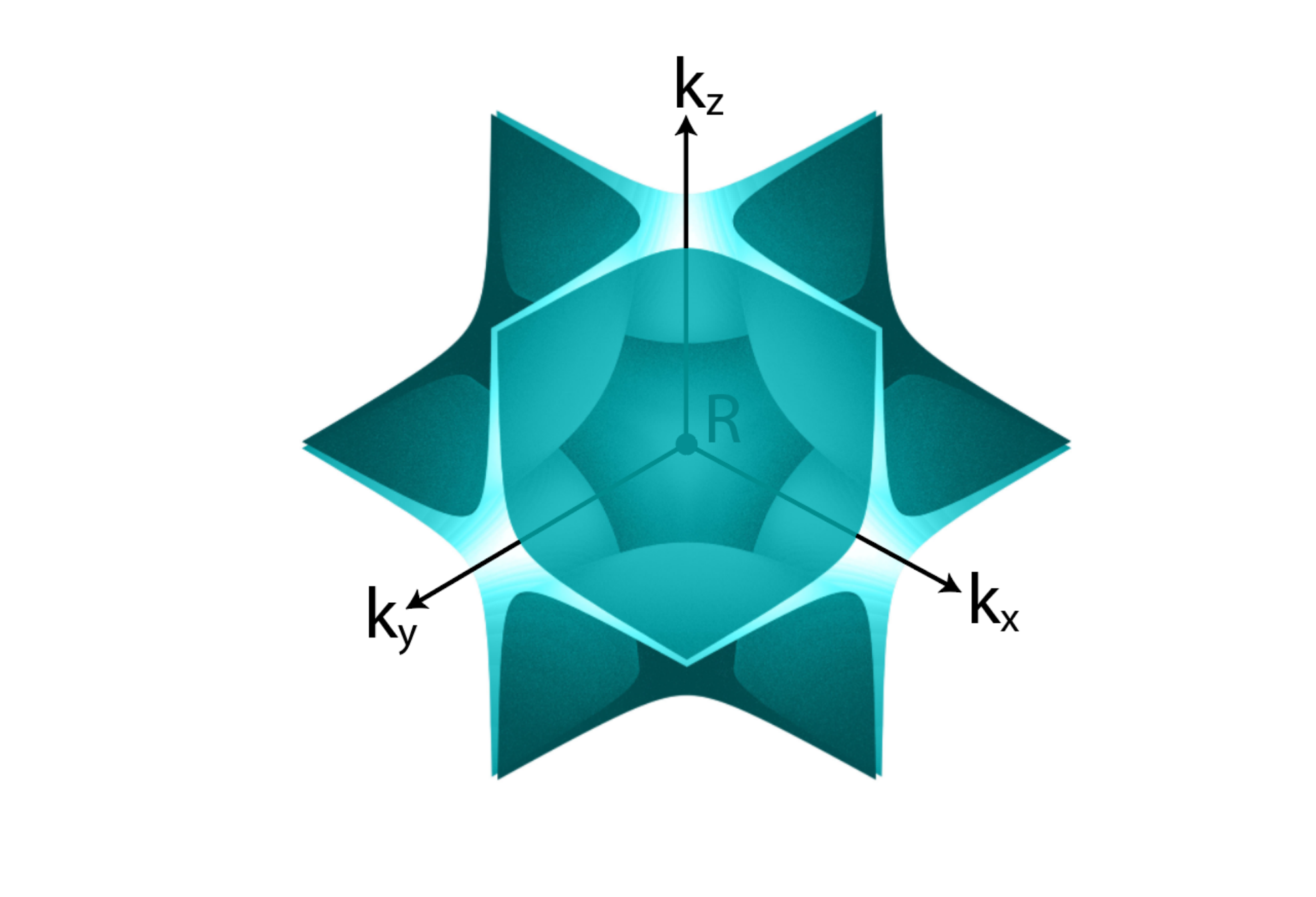}
	\caption{The eight quasi-symmetry protected degenerate planes.}
	\label{Plane}
\end{figure}

To display the near-degeneracy due to quasi-symmetries, Fermi surfaces with different orbital and spin characters, namely 1$^+$/2$^-$ and 1$^-$/2$^+$, are paired up (Fig. \ref{FSring}). The intersections between the Fermi surfaces form four rings for each pair and therefore eight rings in total. As the position of the degenerate rings strongly depend on the size and shape of the Fermi surfaces, by tuning the Fermi level of the system, the degenerate rings are reshaped and relocated in the Brilloiun zone (Fig. \ref{FSringEF}). When the Fermi level is tuned to the type-II Weyl point located along the R to $\Gamma$ line, the degenerate rings shrink to eight singular Weyl points at the Fermi surface and vanish with further reducing the Fermi level. 
By combining all degenerate rings at different Fermi energy, degenerate planes protected by the quasi-symmetry is constructed (Fig. \ref{PlaneFS}). The degenerate planes contain eight different planes which are all symmetrically related (Fig. \ref{Plane}).

\section{Shubnikov–de Haas oscillation measurements of CoSi}
\begin{figure}
	\includegraphics[width=0.95\columnwidth]{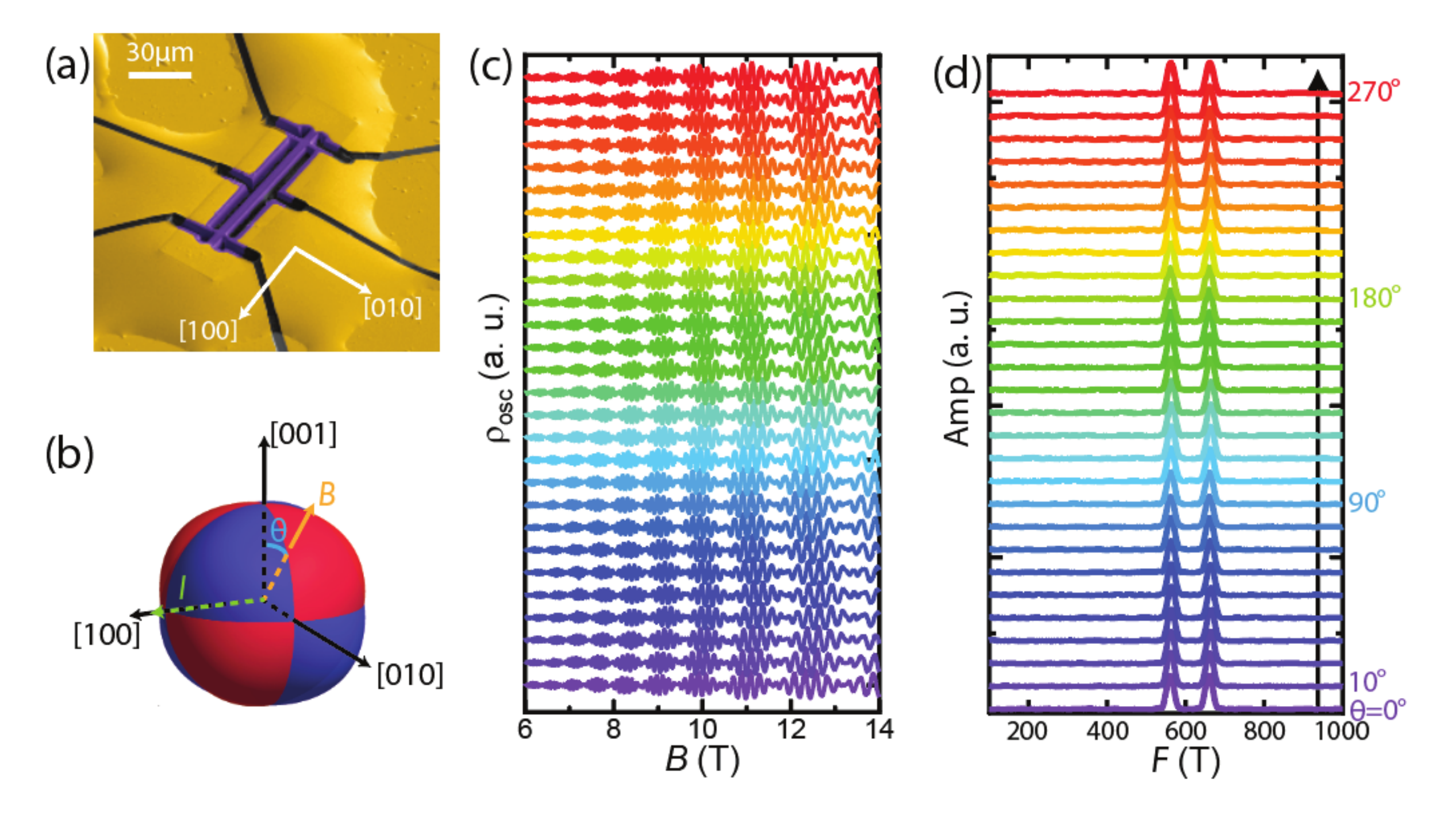}
	\caption{(a) Scanning electron microscope (SEM) image of device-1. A long bar of a 6 by 6 $\mu m^2$ cross section is fabricated with FIB. The golden part stands for the deposited gold thin film for obtaining low resistance ohmic contact and the CoSi crystalline part is colored purple. (b) Illustration of field and current orientation of Shubnikov-de-Hass oscillation measurements. $\theta$ is defined as the angle between the applied field and [001] axis. (c) SdH oscillations as a function of angle $\theta$. Here $\rho_{osc} = \Delta\rho/\rho_{BG}$ with $\Delta\rho$ the oscillating part of the magnetoresistance (MR) and $\rho_{BG}$ the MR background subtracted by a 3$^{rd}$-order polynomial function. (d) Fast-Fourier-transformation spectrum of SdH oscillations displayed in (c). The two main peaks correspond to two pairs of Fermi surface orbits with same cross section areas.}
	\label{L1}
\end{figure}

\begin{figure}
	\includegraphics[width=0.95\columnwidth]{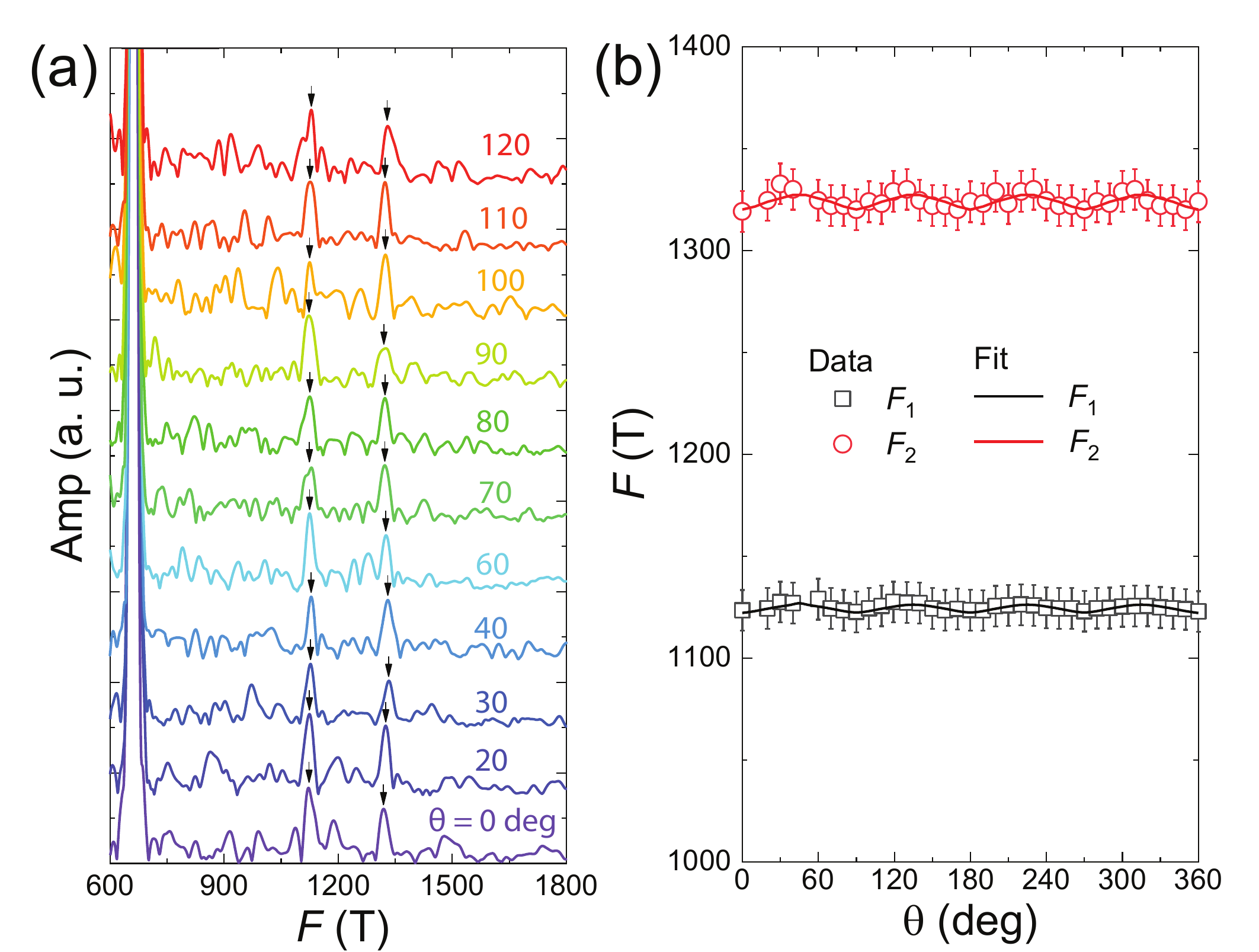}
	\caption{Analysis of angle-dependent frequencies corresponding to the second harmonic oscillations. Despite of the larger error bar, the angular dependence of the extracted frequencies can still be nicely described by the quasi-symmetry-protected breakdown orbit scenario. These results are consistent with the analysis of first harmonics presented in Fig. 3.}
	\label{2ndH}
\end{figure}

Four microstructure devices are fabricated by focused-ion-beam technique in order to perform Shubnikov–de Haas (SdH) oscillation measurements with field applied along different directions. For device-1, a microbar along [100] axis with a cross section area of 6 by 6 $\mu m^2$ is fabricated [Fig. \ref{L1} (a)], and the quantum oscillations are measured with magentic field rotate within the (100) plane [Fig. \ref{L1} (b)]. This ensures the magnetic field direction is always perpendicular to the current direction which eliminates the possible influence of longitudinal magnetotransport. The SdH oscillations are measured with a 10-degree-step rotation starting from [001] axis [Fig. \ref{L1} (c)], and the corresponding angle-dependent fast-Fourier-transformation (FFT) spectrum is displayed in Fig. \ref{L1} (d). In order to demonstrate the consistency of our analysis, we have also analyzed the angular dependence of second harmonic oscillation frequencies which can be nicely described by the theoretical prediction as well (Fig. \ref{2ndH}).

\begin{figure}
	\includegraphics[width=0.95\columnwidth]{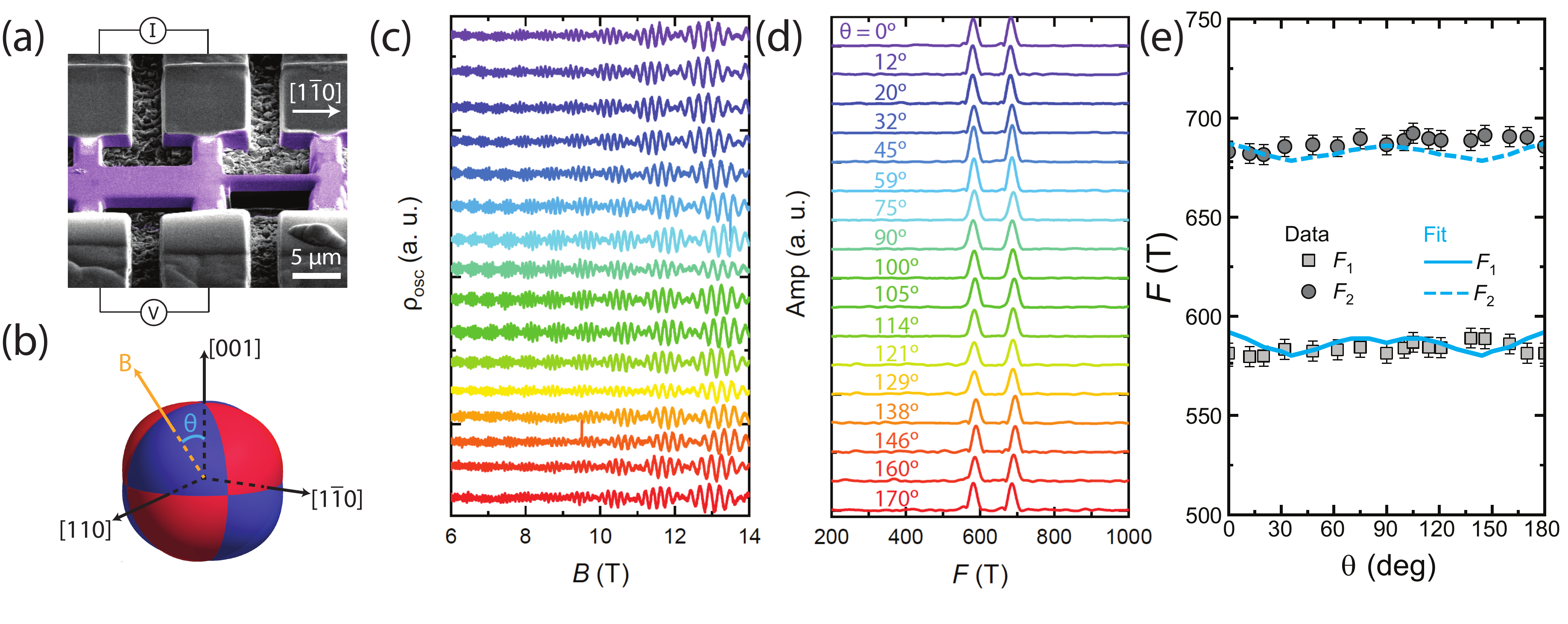}
	\caption{(a) SEM image of device-2. Here the gold thin film is covered with FIB-deposition of Carbon thin film (grey) for protection of ion beam tail during microstructure fabrication. The current is applied along [1$\bar{1}$0] axis. (b) Demonstration of field and current orientation of SdH measurements. For device-2 the magnetic field is applied within the (1$\bar{1}$0) plane, and $\theta$ is defined as the angle between the applied field and [001] axis. (c) Angular dependence of SdH oscillations $\rho_{osc}$. (d) FFT spectrum of SdH oscillations. Again two and only two mean frequencies are observed. (e) Angular dependence of SdH oscillation frequencies and theoretical prediction which successfully reproduced the experimental results.}
	\label{PML1}
\end{figure}

To further investigate the Fermiology of CoSi, another micro-bar along [1$\overline{1}$0] axis is fabricated [Fig. \ref{PML1} (a)]. The magnetic field is therefore rotated from [001] to [110] axis [Fig. \ref{PML1} (b)]. Angle-dependent SdH oscillations and corresponding FFT spectrum are displayed in Fig. \ref{PML1} (c) and (d) respectively. The angular dependence of oscillation frequency is summarized in Fig. \ref{PML1} (e). Similarly, the experimental results match well with the theoretical prediction, which further demonstrates the importance of quasi-symmetry in CoSi.

\subsection{Determination of cyclotron mass}
\begin{figure}
	\includegraphics[width=0.95\columnwidth]{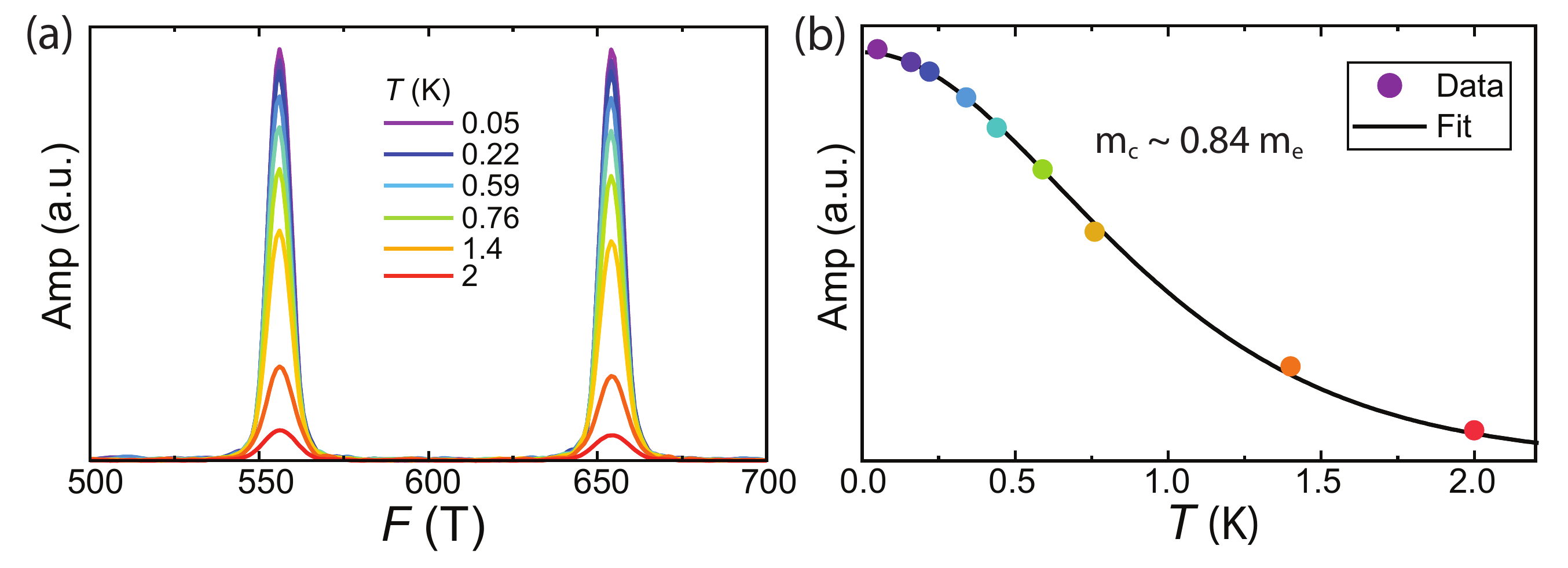}
	\caption{ (a) Fast-Fourier-transformation spectrum of the SdH oscillations presented in Fig. 3(a) with the field window of 3 to 14 T. Two main peaks can be clearly observed. The suppression of peak amplitude with increasing temperature is due to the thermal damping effect. (b) Lifshitz-Kosevich fit to the temperature dependence of cyclotron mass. The fitting yields a cyclotron mass $m_c \sim 0.84~m_e$, comparable to the previously reported values\cite{CoSi_JS,CoSi_Plef}.}
	\label{mstar}
\end{figure}

To accurately determine the corresponding cyclotron mass of the Fermi surfaces, we have performed SdH oscillation measurements of device-1 down to 50~mK [Fig. 3 (a)] with both magnetic field and current applied along [100] direction. The temperature dependence of the oscillations follows well the Lifshitz-Kosevich form, leading to a low cyclotron effective mass of $m_c\approx 0.84~m_e$ (Fig. \ref{mstar}) which is consistent with the previous report\cite{CoSi_JS,CoSi_Plef}.

\subsection{Analysis of quantum oscillation spectrum under tensile strain along [111] axis}
\begin{figure}
	\includegraphics[width=0.58\columnwidth]{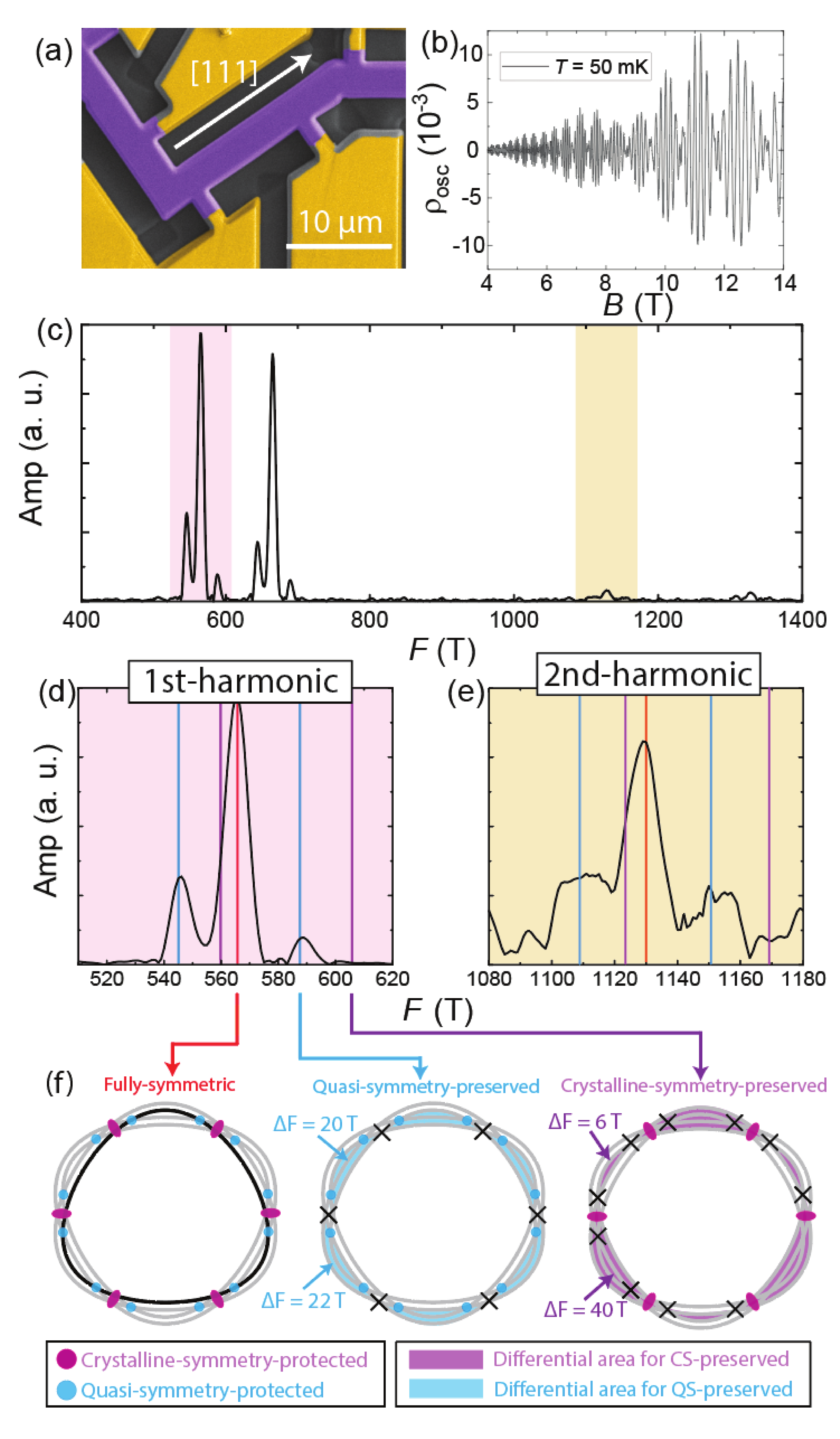}
	\caption{(a) Scanning electron microscope image of CoSi microdevice with a short bar along [111] direction (device-4). This configuration generates a tensile strain along [111] which breaks all $C_2$ rotation symmetries. (b) SdH oscillations with field and current applied along [111] axis at $T = 50$ mK. (c) FFT spectrum of SdH oscillations. (e) Enlarged view of satellite peaks correspond to the 1$^{st}$ and 2$^{nd}$ harmonic oscillations. The red, purple and blue vertical lines correspond to the FFT spectrum produced by the fully symmetric, crystalline-symmetry-preserved and quasi-symmetry-preserved scenarios respectively. (f) Corresponding Landau orbits for three different scenarios. Here the colored area illustrates the orbital area difference compared to the fully-symmetric case. Only the quasi-symmetry-preserved scenario reproduces FFT peaks that match well with the experimental data.}
	\label{111}
\end{figure}

In the main manuscript, we explained how tensile strain breaks crystalline symmetry of CoSi which leads to the observation of additional quantum oscillation with different frequencies. To further demonstrate the relation between strain-induced additional oscillation frequencies and crystalline symmetry breaking, here we presented detailed analysis of SdH oscillations of device-3 measured at $T = 50$ mK with a tensile strain along [111] direction(Fig. \ref{111}). Similar to the tensile strain approximately along [110] axis, additional satellite peaks around the main frequencies are also clearly resolved. By calculating the additional orbital area generated by breaking either crystalline or quasi-symmetry, we confirmed that the experimental results can only be described by the quasi-symmetry-preserved scenario which, on the other hand, further demonstrates the stability of quasi-symmetry against strain-induced crystalline symmetry breaking.

\subsection{Gaussian-type multi-peak fit of FFT spectrum}
\begin{figure}
	\includegraphics[width=0.9\columnwidth]{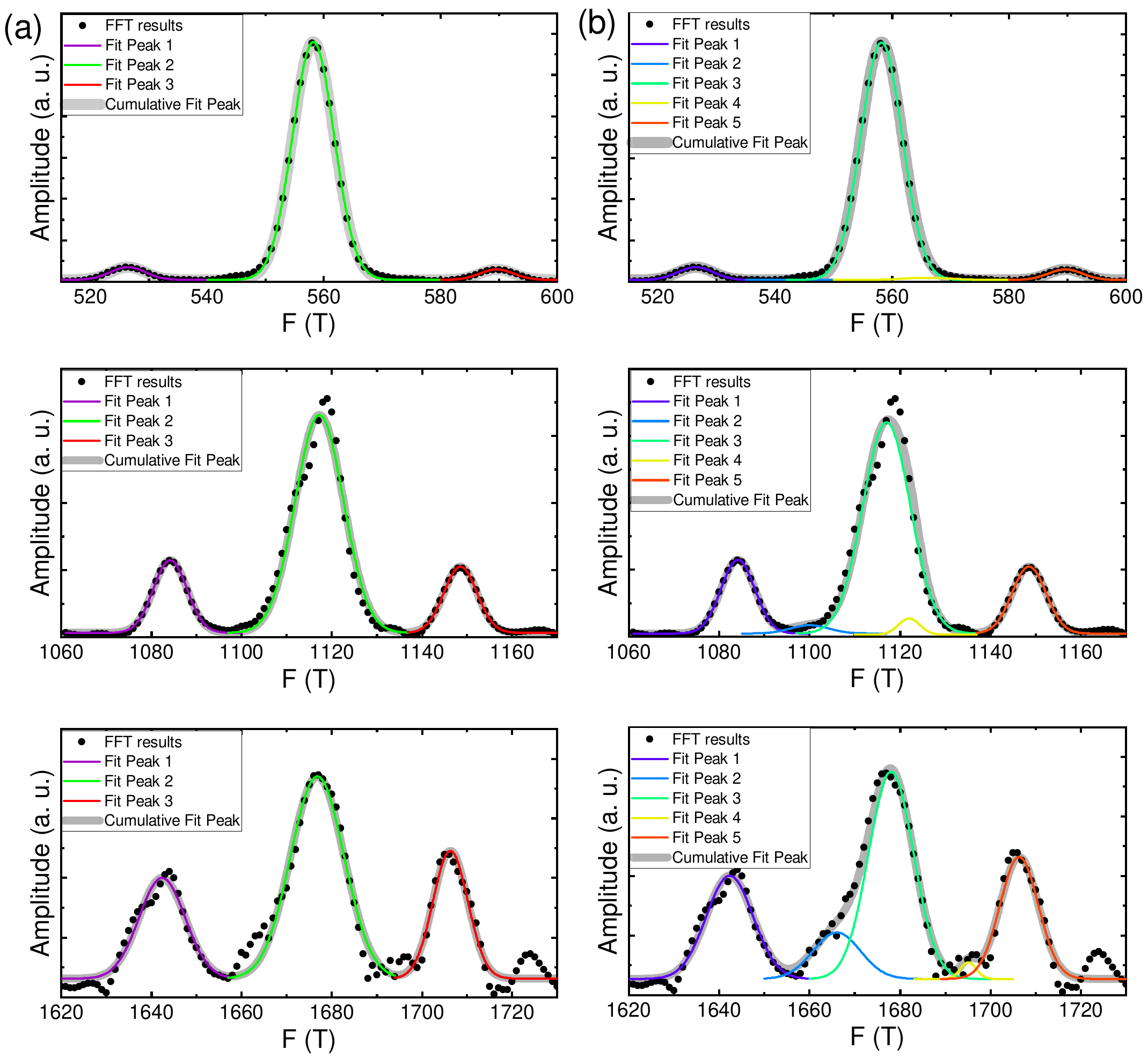}
	\caption{3-peak (a) and 5-peak (b) Guassian fitting of the FFT spectrum of quantum oscillations measured with field and tensile strain applied along [110] axis at $T = $ 50~mK.}
	\label{Gaus}
\end{figure}

In order to accurately determine the position and relative size of the satellite peaks induced by tensile strain along [110], we performed both 3-peak and 5-peak Gaussian fit to the the FFT spectrum presented in Fig.4 (e). 3-peak Gaussian fit [Fig. \ref{Gaus} (a)] describes the experimental spectrum reasonably well which clearly demonstrates the existence of two satellite peaks. The frequency difference between the main and satellite peaks always stay around 32~T for all principle frequencies and their higher harmonics. Slight discrepancy between the spectrum of third harmonic oscillations and the 3-peak Gaussian fit is mainly due to its much reduced amplitude which is comparable to the noise floor of the FFT analysis. 5-peak Gaussian fit for both first and second harmonic spectrum is almost identical to the 3-peak fit since the two additional peaks are almost negligible. While for the third harmonic spectrum, although the fitting quality is improved with including the two additional peaks, they do not match with the theoretical prediction of crystalline-symmetry-preserved scenario. These results rigorously demonstrate that strain-induced crystalline symmetry breaking is the origin of the satellite peaks in the FFT spectrum. To note that since these additional frequencies rely on the magnetic breakdown tunnelling at the symmetry breaking point, their amplitudes are much smaller compared to the principle ones. Therefore it is necessary to perform the measurement down to $T$ = 50 mK as to magnify the quantum oscillation amplitude as much as possible. In the mean time, the breakdown pattern may also be related to the slight misalignment of strain direction from [110] which we discussed in details in Sec.~\ref{TheoryStrain}.

\begin{figure}
	\includegraphics[width=0.9\columnwidth]{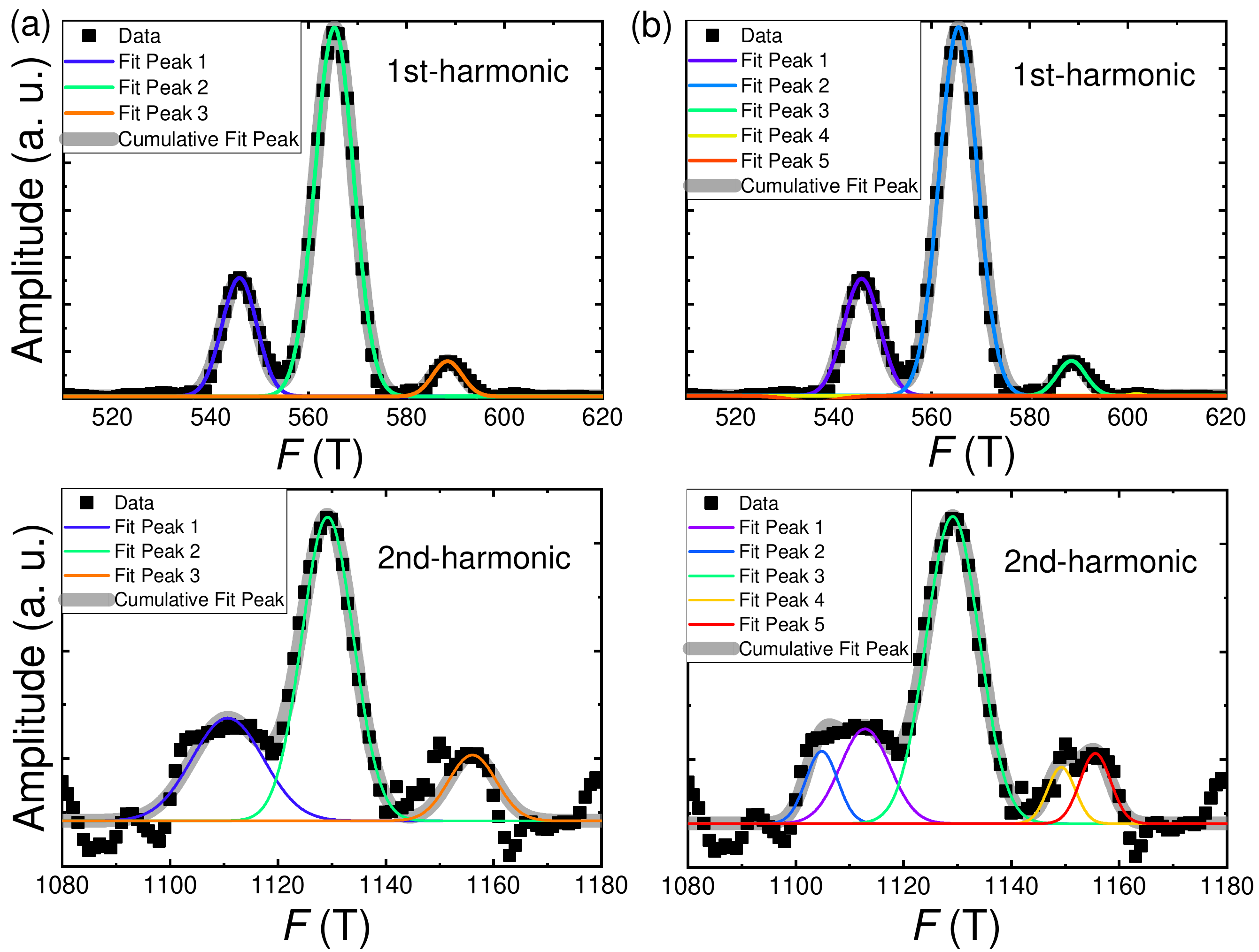}
	\caption{3-peak (a) and 5-peak (b) Guassian fitting of the FFT spectrum of quantum oscillations measured with field and tensile strain applied along [111] axis at $T = $ 50~mK.}
	\label{111G}
\end{figure}

In comparison, for strain along [111] direction the first harmonic FFT spectrum can also be well described by a 3-peak Gaussian fit (Fig. \ref{111G}). Here the amplitude difference between the two different satellite peaks is possibly due to the slight strain imhomogenity along the microstructure. For the second harmonic spectrum since the satellite peak height is again close to the FFT noise floor, the 3-peak Gaussian fit displays limited fitting quality which is improved by including two additional peaks in the 5-peak Gaussian fit. Similar to the [110] case, since the included additional peaks are distinct from the expected ones for crystalline-symmetry-preserved case, they are most likely due to the limited resolution of FFT analysis.

\section{Calculation of magnetic breakdown field and corresponding cyclotron mass}
\subsection{Calculation of magnetic breakdown field}

Theoretically, the magnetic breakdown probability is defined as\cite{MB}:
\begin{equation}
    P = e^{-H_0/H}
\end{equation}
where $H_0$ is the magnetic breakdown field given by:
\begin{equation}
    H_0=\frac{\pi}{4 \hbar e} ~ \frac{\Delta^2}{v_\parallel v_\bot}
\end{equation}
here $v_\parallel$ and $v_\bot$ stand for the Fermi velocity along two in-plane directions perpendicular to the magnetic field. For example, with field applied along [110] axis, $v_\parallel$ is the Fermi velocity along [001] axis while $v_\bot$ stands for Fermi velocity along [1$\overline{1}$0] direction. From band structure calculation we obtained: 
\begin{equation}
    v_\parallel~=~2.11 \times 10^5~m/s,~~v_\bot~=~2.64 \times 10^5~m/s
\end{equation}
Meanwhile the breakdown gap $\Delta \approx~$2 meV as presented in Fig. \ref{MBgap} (d). Therefore the magnetic breakdown field is estimated to be around 0.11 T. Such a small magnetic breakdown field supports the validity of our quasi-symmetry model. For other field orientations, the breakdown gap remains smaller than 2~meV (Fig. \ref{MBgap}) and the corresponding breakdown field $H_0$ at any arbitrary field direction is always smaller than 0.11 T. This means at the lowest field limit ($B_c\approx$ 3 T) where the quantum oscillation starts to become observable, the breakdown transmission possibility is about 96.4\%. This explains the clean quantum oscillation spectrum we observed (Fig. 3). This small breakdown gap also implies the possibility of zero-field topological application above $T$ = 20 K as the thermal broadening renders the breakdown gap transparent. 

\subsection{Calculation of corresponding cyclotron mass of the breakdown orbits}
\begin{figure}
	\includegraphics[width=0.95\columnwidth]{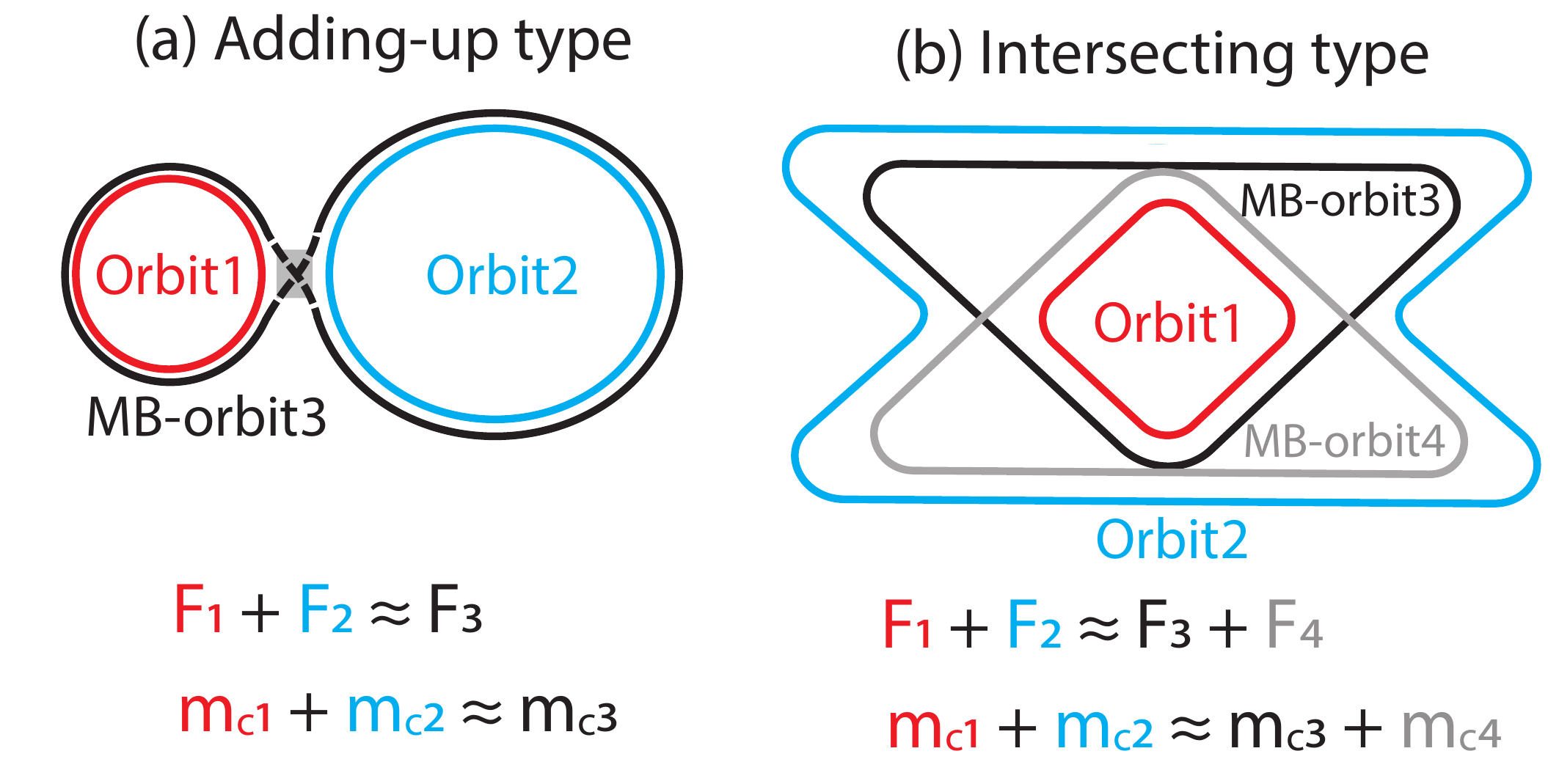}
	\caption{Illustration of different magnetic breakdown scenarios. For the adding-up type, according to semi-classical theory, the oscillation frequency and cyclotron mass of the breakdown orbit can be simply calculated by summing up the values from the original orbits. While for the intersecting type since there exists more than one breakdown orbits the situation is more subtle.}
	\label{MBMC}
\end{figure}

\begin{figure}
	\includegraphics[width=0.95\columnwidth]{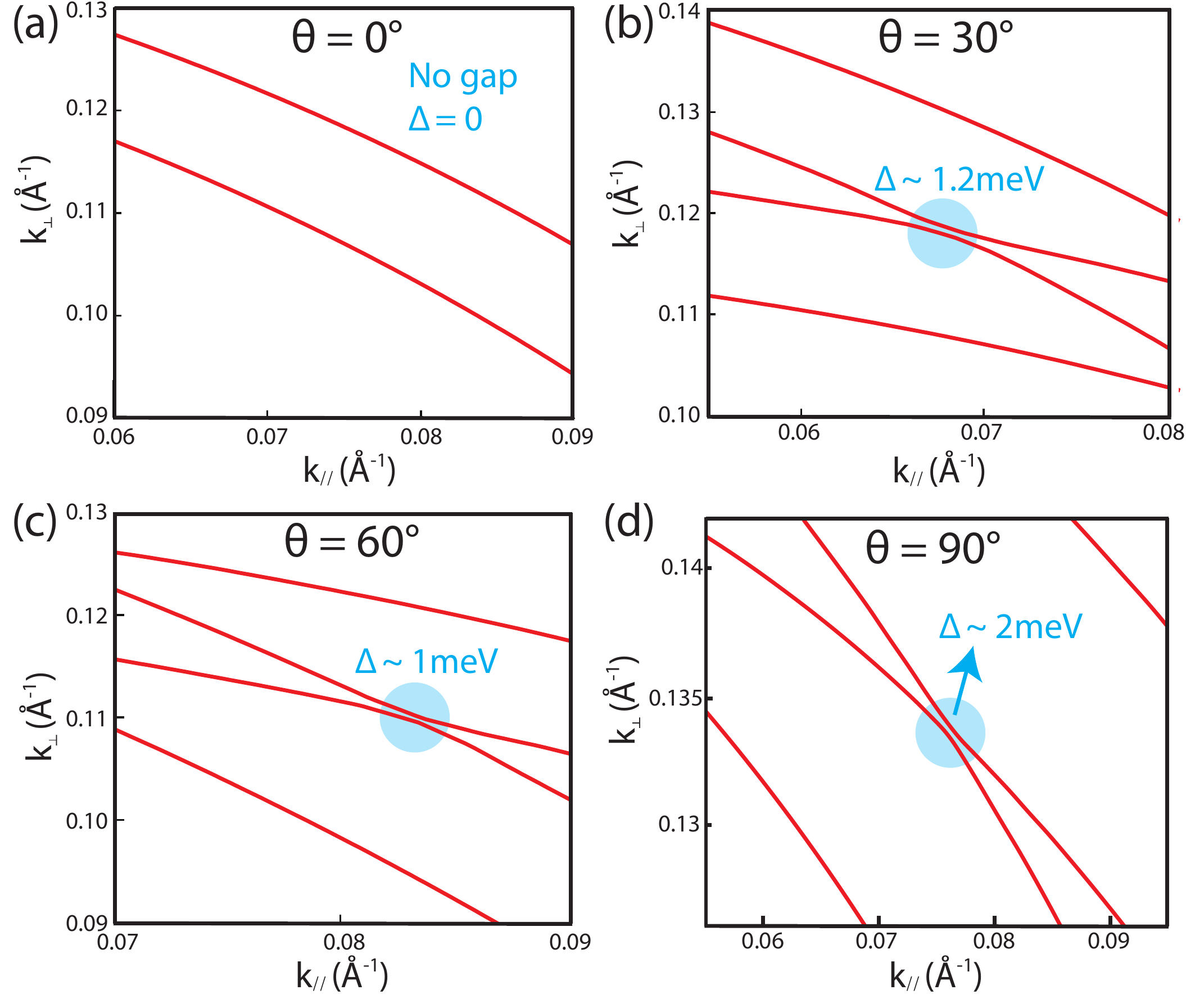}
	\caption{Fermi surface orbits with field applied along different directions obtained from DFT band structure calculations. Here $\theta$ is defined the same as in Fig. S3(a). The largest estimated breakdown gap $\Delta \approx $ 2 meV occurs when field is applied along [110] axis ($\theta = 90^\circ$).}
	\label{MBgap}
\end{figure}

Based on DFT calculations one can easily obtain the expected cyclotron mass $m_c \approx 0.78~m_e$ for all four pockets with field applied along [100] direction, which is consistent with the experimental result of $m_c \approx 0.84~m_e$. Note that in semi-classical theory\cite{SMMB}, if the breakdown orbit consists with two original orbits as shown in Fig. \ref{MBMC}(a), the cyclotron mass of the breakdown orbit can be estimated by directly adding the masses of individual pockets. However this simple scenario is not applicable for CoSi where the intersection of different original orbits results in two instead of only one breakdown orbits. Therefore based on semi-classical theory one can deduce that the sum of cyclotron mass of the two original/breakdown orbits should be similar. This means the original and breakdown orbit should have similar cyclotron mass value $m_c \approx 0.78 m_e$ which is consistent with the experimental results. 
We can also directly calculate the cyclotron mass of the breakdown orbits in CoSi assuming the quasi-symmetry protected degeneracy is gapless. The calculation yields a similar cyclotron mass value around 0.77 $m_e$, which again matches well with the experimental results.

\section{Generality of quasi-symmetry}
\begin{figure}
	\includegraphics[width=0.8\columnwidth]{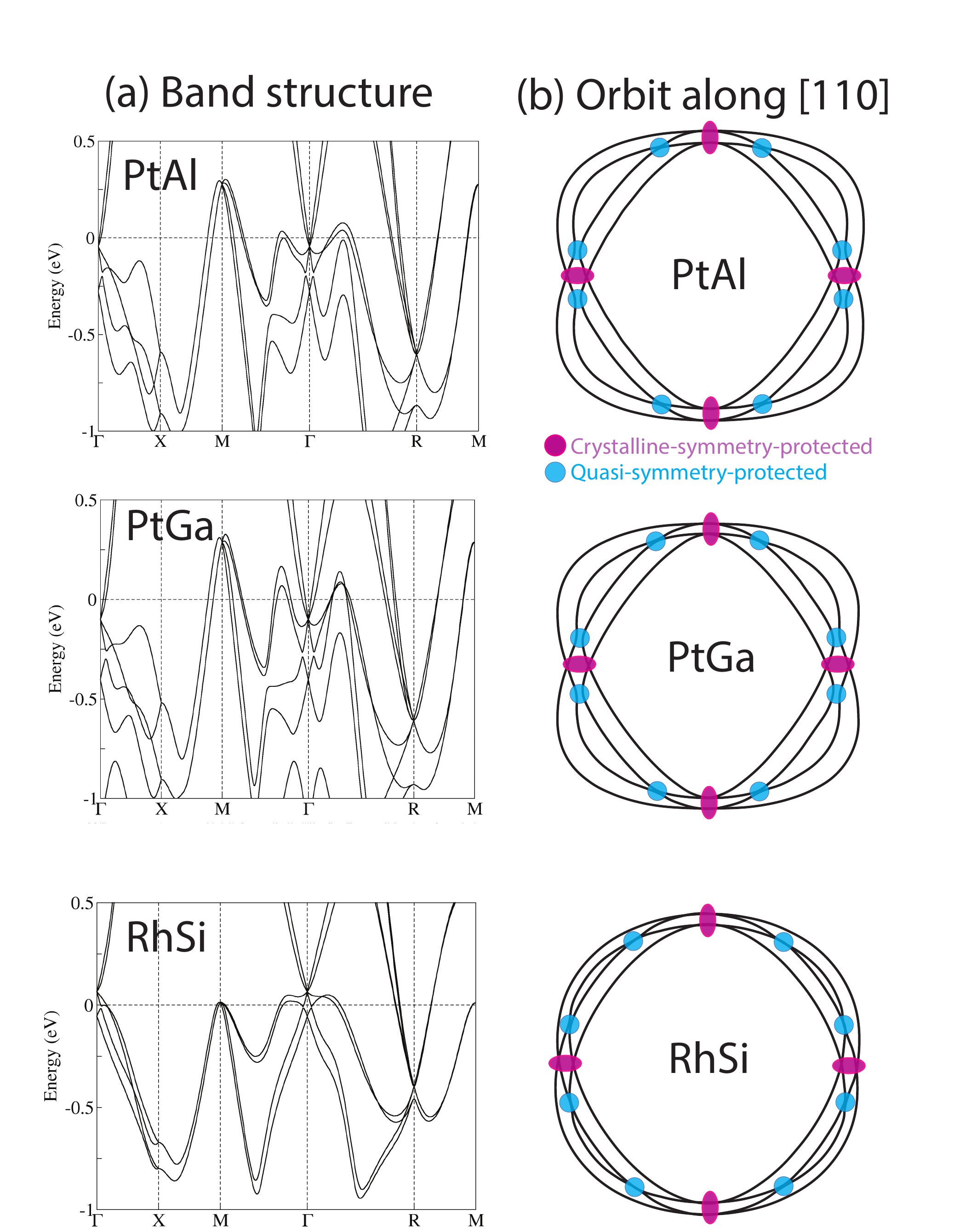}
	\caption{(a) Band structure calculation of PtAl, PtGa and RhSi. The similarity of electronic structure among these materials are expected as they share the crystal structure. (b) Fermi surface orbits with field applied along [110] axis for all three materials. The crystalline symmetry and quasi-symmetry-protected degeneracies are denoted with purple and blue circles respectively.}
	\label{gene}
\end{figure}

As the quasi-symmetry originates from the ${\bf k\cdot p}$-type expansion of effective model, we would expect it is a general property that exists in materials with the same crystal structure. To demonstrate this point, we studied the electronic band structure of three different compounds, PtAl, PtGa and RhSi, which share the same crystal structure as CoSi. We notice that the band structure of RhSi is almost identical to that of CoSi with an electron pocket around R point and a hole pocket around $\Gamma$ point, while for PtAl and PtGa, the electron pockets exist for both $\Gamma$ and R points and the additional hole pockets appear around the M points. The electron Fermi pockets around the R point are all similar and described by the same effective model for all these compounds and as shown in Fig. \ref{gene}, these Fermi surfaces in the [110] plane display both crystalline-symmetry-protected exact degeneracies and quasi-symmetry-protected near degeneracies, similar to CoSi. Since the quasi-symmetry is approximate, a small gap is expected for the near degeneraceis, as labelled by blue dots in Fig. \ref{gene}, and the sizes of these gaps vary at the Fermi surfaces for different materials. We find this gap is extremely small also for RhSi, but a bit larger for PtAl and PtGa, which depends on the material details. Nevertheless, our calculations here demonstrate that the scenario of quasi-symmetry can generally be applied to all these compounds. 

\clearpage

\section{Theoretical modelling}

\subsection{Space group 198 of CoSi}
The chiral crystal CoSi has a space group 198 (SG198), which can be generated by

\begin{align}\label{eq-little-SP198-cxl-operators}
S_{2x} = \{ C_{2x} | \tfrac{1}{2} \tfrac{1}{2} 0 \},
C_{3} = \{ C_{3,(111)} | 0 0 0 \},
\end{align}

in addition to the translation sub-group\cite{new_fermions}.
Hereafter, the Seitz notation is taken for the non-symmorphic symmetry operations, i.e., a point group operation $\mathcal{O}$ followed by a translation $\mathbf{v}=v_i\mathbf{t}_i$, labeled as $ \hat{\mathcal{O}} = \{ \mathcal{O} | \mathbf{v} \} \text{ or } \hat{\mathcal{O}} = \{ \mathcal{O} | v_1v_2v_3\}$, with $\mathbf{t}_i$ ($i=1,2,3$) representing three basis vectors for a Bravais lattice in three dimensions [see Fig.~2(a) in the main text]
The reciprocal space lattice vectors are generated by $\mathbf{g}_i$, where $\mathbf{g}_i\cdot \mathbf{t}_j=2\pi \delta_{ij}$.
The rules for multiplication and inversion are defined as

\begin{align}
\{ \mathcal{O}_2 | \mathbf{v}_2 \} \{ \mathcal{O}_1 | \mathbf{v}_1 \} &= \{ \mathcal{O}_2\mathcal{O}_1 | \mathbf{v}_2 + \mathcal{O}_2 \mathbf{v}_1 \}, \\
\{\mathcal{O}\vert \mathbf{v}  \}^{-1} &= \{ \mathcal{O}^{-1}\vert -\mathcal{O}^{-1}\mathbf{v} \}.
\end{align}

The $S_{2y} = \{ C_{2y} | 0 \tfrac{1}{2} \tfrac{1}{2}  \}$ symmetry can be generated by

\begin{align}
 S_{2y} = C_3^{-1} S_{2x} C_3,
\end{align}

where we take the convention for $C_{3,(111)}: \; (x,y,z)\to(y,z,x)$, leading to $C_{2y} = C_{3,(111)}^{-1} C_{2x}$ $ C_{3,(111)} $.
Likewise, the $S_{2z}$ symmetry can be given by the combination of $S_{2x}$, $S_{2y}$ and the translation operator $E_\mathbf{v}=\{E|\mathbf{v}\}$,

\begin{align}
S_{2x} S_{2y} = \{E|00\bar{1}\} S_{2z} \triangleq E_{00\bar{1}}S_{2z}.
\end{align}

Similar to the MnSi in Ref.~[\onlinecite{MnSi_Plef}], the symmetry-enforced nodal planes (high symmetry planes) also exist for CoSi.
On these high-symmetry planes, the two-fold degeneracies are protected by the combined anti-unitary symmetries,  $S_{2x}\mathcal{T}$, $S_{2y}\mathcal{T}$ and $S_{2z}\mathcal{T}$, where $\mathcal{T}$ is the time-reversal operator. For spinless fermions, $\mathcal{T}=\mathcal{K}$  with the complex conjugate $\mathcal{K}$, while for the spin-1/2 fermions, $\mathcal{T}=is_y \mathcal{K}$ with $s_y$ the Pauli matrix acting on the spin subspace.
These operations transform the Hamiltonian as

\begin{align}
\left(S_{2x}\mathcal{T}\right)^\dagger \mathcal{H}(k_x,k_y,k_z) \left(S_{2x}\mathcal{T}\right) &= \mathcal{H}(-k_x,k_y,k_z), \\
\left(S_{2y}\mathcal{T}\right)^\dagger \mathcal{H}(k_x,k_y,k_z) \left(S_{2y}\mathcal{T}\right) &= \mathcal{H}(k_x,-k_y,k_z), \\
\left(S_{2z}\mathcal{T}\right)^\dagger \mathcal{H}(k_x,k_y,k_z) \left(S_{2z}\mathcal{T}\right) &= \mathcal{H}(k_x,k_y,-k_z).
\end{align}

It indicates that $[S_{2x}\mathcal{T}, \mathcal{H}(k_x=0,\pi,k_y,k_z)]=0$ for the $k_x=0,\pi$ planes.
Likewise, the $S_{2y}\mathcal{T}$-invariant ($S_{2z}\mathcal{T}$-invariant) planes are $k_y=0,\pi$ ($k_z=0,\pi$).
Moreover, we act the operators $S_{2x}\mathcal{T}$, $S_{2y}\mathcal{T}$ and $S_{2z}\mathcal{T}$ on the Bloch wave functions at $\mathbf{k}$ and find
$(S_{2x}\mathcal{T})^2 = (S_{2y}\mathcal{T})^2= (S_{2z}\mathcal{T})^2=-1$ at the $k_i=\pi$ plane.
Thus, $S_{2x}\mathcal{T}$, $S_{2y}\mathcal{T}$ and $S_{2z}\mathcal{T}$ behave as the time reversal operator for spinful fermions, and the two-fold degeneracy on the $k_{x,y,z}=\pi$-plane can be guaranteed by these anti-unitary symmetries, similar to the Kramers' theorem, as depicted in Fig.~S4(c).

\subsection{The effective model Hamiltonian around the $R$-point}
In this section, we construct the effective $\mathbf{k}\cdot\mathbf{p}$ Hamiltonian around the $R$-point ($\mathbf{k}_R=(\pi,\pi,\pi)$), and fit parameters for the $\mathbf{k}\cdot\mathbf{p}$ bands with the density-functional-theory (DFT) bands.

We first consider the commutation relations of the symmetry operators $S_{2x}$, $S_{2y}$ and $\mathcal{T}$ in SG198 for spinless fermions at $R$, given by

\begin{align}
\{S_{2x},S_{2y}\}=0 ,\; [\mathcal{T},S_{2x}] = [\mathcal{T},S_{2y}]=0 \text{ and } (S_{2x}\mathcal{T})^2 = (S_{2y}\mathcal{T})^2 = -1.
\end{align}

For a common eigen-state $\vert \Psi\rangle$ of $S_{2x}$ with eigenvalue $\lambda$ and the Hamiltonian, the $S_{2x}$-eigenvalues of the states $S_{2y}\vert \Psi\rangle$, $\mathcal{T}\vert \Psi\rangle$ and $S_{2y}\mathcal{T}\vert \Psi\rangle$ are given in the following table,

\begin{center}
	\begin{tabular}{|c|c|c|c|c|}
		\hline
		& $\vert \Psi\rangle$ & $S_{2y}\vert \Psi\rangle$ & $\mathcal{T}\vert \Psi\rangle$ & $S_{2y}\mathcal{T}\vert \Psi\rangle$ \\
		\hline
		$S_{2x}$ & $\lambda$ & $-\lambda$ & $\lambda^\ast = -\lambda$ & $-\lambda^\ast=\lambda$ \\
		\hline
	\end{tabular}
\end{center}

from which we find $\vert \Psi\rangle, S_{2y}\mathcal{T}\vert \Psi\rangle$ and $S_{2y}\vert \Psi\rangle, \mathcal{T}\vert \Psi\rangle$ carry opposite $S_{2x}$-eigenvalues. Here we have used the fact that $\lambda$ is a purely imaginary number due to $S_{2x}^2=-1$. Furthermore, since $( S_{2y} \mathcal{T})^2=-1$ and $S_{2y} \mathcal{T}$ is an anti-unitary symmetry operator, $\langle \Psi \vert S_{2y}\mathcal{T}\vert \Psi \rangle=0$, which means the states $\vert \Psi\rangle$ and $S_{2y}\mathcal{T}\vert \Psi\rangle$ ($S_{2y}\vert \Psi\rangle$ and $\mathcal{T}\vert \Psi\rangle$) are orthogonal to each other. Thus, $\{\vert \Psi\rangle, S_{2y}\vert \Psi\rangle, \mathcal{T}\vert \Psi\rangle,S_{2y}\mathcal{T}\vert \Psi\rangle \}$ are four independent degenerate eigen-states of the Hamiltonian. This proves all the eigen-states of the Hamiltonian at the R-point are four-fold degenerate (without spin), and thus supports the discussions in the main text above Eq.~(1).

Next we will construct the effective model on the four-fold degenerate states based on the symmetry argument. Let us choose $G_{1}, G_2$ and $G_3$ to be the matrix representations for the $S_{2x}$, $S_{2y}$, and $C_3$, respectively. Without spin, these matrix representations need to satisfy the relations

\begin{align}\label{eq-cxl-g1g2g3-relation}
G_3^3=1, G^2_1=G^2_2=-1, \;
G_1G_2=-G_2G_1,\;
G_3^{-1}G_1G_3 = G_2, \;
G_3^{-1}G_2G_3 = -G_1G_2.
\end{align}

Moreover, all of them commute with time-reversal symmetry, $[G_1, \mathcal{T}]=[G_2, \mathcal{T}]=[G_3, \mathcal{T}]=0$ for $\mathcal{T}=K$. Thus, one can construct the 4D irreducible representation matrices for $G_{1}, G_2$ and $G_3$ as

\begin{align}
G_3  = \begin{pmatrix}
1 & 0  & 0  &  0 \\
0 & 0  & -1  & 0 \\
0 & 0  & 0  & 1  \\
0 & -1  & 0  & 0
\end{pmatrix} ,
G_1  = \begin{pmatrix}
0 & 0  & 1  & 0 \\
0 & 0  & 0  & 1 \\
-1 & 0  & 0  & 0  \\
0 & -1  & 0  & 0
\end{pmatrix} =i\sigma_y\tau_0  ,
G_2  = \begin{pmatrix}
0 & 0  & 0  & 1 \\
0 & 0  & -1  & 0 \\
0 & 1  & 0  & 0  \\
-1 & 0  & 0  & 0
\end{pmatrix}=i\sigma_x\tau_y .
\end{align}

Here we have chosen two sets of Pauli matrices $\sigma$ and $\tau$ to rewrite the four by four matrices.

Based on the above three transformation matrices for $G_{1}, G_2$ and $G_3$ and time reversal operator $\mathcal{T}=\mathcal{K}$, we can construct the effective Hamiltonian around the R-point as (without spin)

\begin{align}\label{sm-eq-R-model-Ham-nonsoc}
\mathcal{H}_{non-soc}(\mathbf{k}) &= \mathcal{H}_{0}(\mathbf{k}) + \mathcal{H}_{k^2}(\mathbf{k}) ,
\end{align}

where the the leading order Hamiltonian reads

\begin{align}
\mathcal{H}_{0}(\mathbf{k}) &= C_0 \sigma_0\tau_0 + A_1( k_x \sigma_y\tau_0 +  k_y \sigma_x\tau_y - k_z \sigma_z\tau_y),
\end{align}

and the $k^2$-order Hamiltonian is 

\begin{align}
\begin{split}
\mathcal{H}_{k^2}(\mathbf{k})  &= B_1 k^2 \sigma_0\tau_0 + C_1(k_xk_y\sigma_z\tau_0 + k_yk_z\sigma_0\tau_z + k_xk_z\sigma_z\tau_z )
+C_2(k_xk_y\sigma_x\tau_x \\  &-k_yk_z\sigma_0\tau_x +  k_xk_z\sigma_x\tau_0 )  + C_3(k_xk_y\sigma_x\tau_z -k_yk_z\sigma_y\tau_y -k_xk_z\sigma_z\tau_x ),
\end{split}
\end{align}

with $C_0, A_1, B_1, C_1, C_2, C_3$ material-dependent parameters and $k=\sqrt{k_x^2+k_y^2+k_z^2}$.

Next we will include spin-orbit coupling into the Hamiltonian. To do that, we note that spin is a pseudo-vector and behaves exactly the same as a vector due to the lack of inversion, mirror or other roto-inversion symmetries for a chiral crystal. Correspondingly, we just need to replace the momentum $\mathbf{k}$ by the Pauli matrices $\mathbf{s}$ of the electron spin to get the effective spin-orbit coupling, which is written as

\begin{align}
\mathcal{H}_{soc} &= \lambda_0 ( s_x \sigma_y\tau_0 + s_y \sigma_x\tau_y - s_z \sigma_z\tau_y ),
\end{align}

up to the first order in spin, with the spin-orbit coupling parameter $\lambda_0$. The full Hamiltonian is given by

\begin{align}\label{sm-eq-R-model-Ham}
\mathcal{H}_{R}(\mathbf{k}) &= \mathcal{H}_{non-soc} \otimes s_0 +  \mathcal{H}_{soc}.
\end{align}

For $\lambda_0=0$, one can check that the Hamiltonian $\mathcal{H}_{R}(\mathbf{k})$ has a 8-fold degeneracy at $R$-point ($\mathbf{k}=0$). Once the spin-orbit coupling is turned on, the 8-fold degeneracy at $R$-point is split into a 6-fold (energy $\lambda_0$) and a 2-fold (energy $-3\lambda_0$).
From the DFT calculation [see Fig. 2(c) in the main text], we find that the 6-fold degenerated states have higher energy than that of the 2-fold states, which implies $\lambda_0>0$.

\begin{figure}[!htbp]
	\centering
	\includegraphics[width=0.95\linewidth]{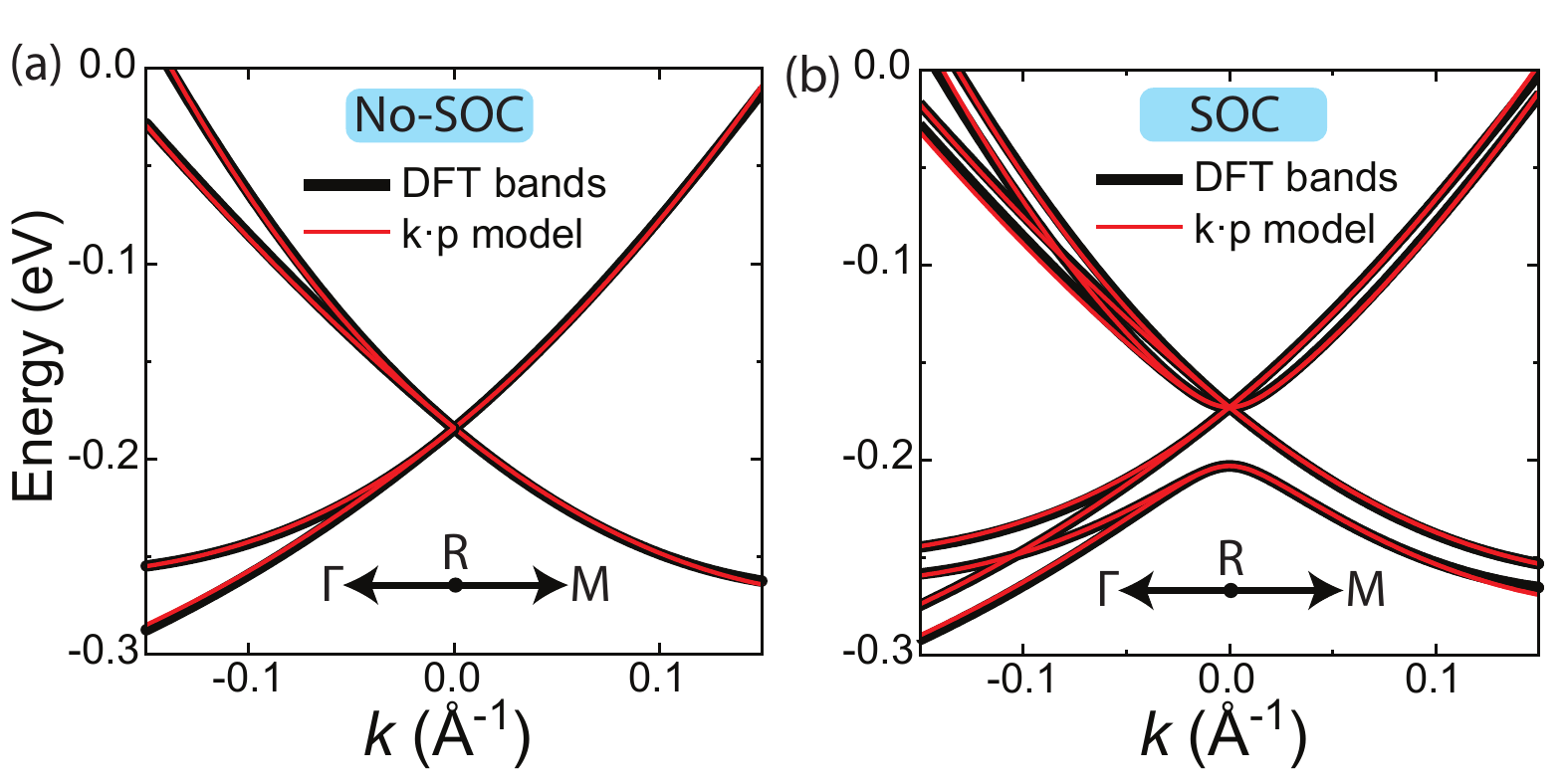}
	\caption{DFT-calculated Band structures (black) and $k \cdot p$ model fitting (red) around the $R$-point. (a) The fitting along the $R-M$ line generates parameters $C_0, A_1, B_1$ for the $\mathbf{k}\cdot\mathbf{p}$ Hamiltonian without SOC. The remaining parameters $C_1,C_2, C_3$ are obtained for the $\mathbf{k}\cdot\mathbf{p}$ Hamiltonian without SOC along the $R-\Gamma$ line. (b) Fitting the strength of SOC $\lambda_0$ for the $\mathbf{k}\cdot\mathbf{p}$ Hamiltonian with SOC (the $R$-model) along the $\Gamma-R-M$ line.}	
	\label{sm-theory-fig1}
\end{figure}

The spin-independent Hamiltonian $\mathcal{H}_{0}(\mathbf{k})$ is isotropic with the full rotation symmetry. To see that, we could define the emergent angular momentum operators as

\begin{align}
 L_x = \frac{1}{2}\sigma_y\tau_0, \; L_y = \frac{1}{2}\sigma_x\tau_y, \; L_z = -\frac{1}{2}\sigma_z\tau_y,
\end{align}

which satisfies the commutation relation $[L_i,L_j] = i\epsilon_{ijk} L_k$ with Levi-Civita symbol $\epsilon_{ijk}$ and $i=x,y,z$.
Therefore, $\mathcal{H}_{0}(\mathbf{k})$ and $\mathcal{H}_{soc}$ can be re-written as

\begin{align}
\mathcal{H}_{0}(\mathbf{k}) &= C_0\sigma_0\tau_0 + 2A_1 (\mathbf{k}\cdot\mathbf{L}), \\
\mathcal{H}_{soc} &= 2\lambda_0 (\mathbf{s}\cdot \mathbf{L}),
\end{align}

which is shown in the Eq.~(1) in the main text. $\mathcal{H}_{k^2}(\mathbf{k})$ breaks the full rotational symmetry.
Moreover, we extract the parameters for the $R$-model in Eq.~\eqref{sm-eq-R-model-Ham} by fitting the DFT bands along high symmetry lines, shown in Fig.~\ref{sm-theory-fig1}. The parameters are

\begin{align}\label{sm-eq-fitting-parameters}
\begin{split}
C_0=-0.18  \; eV,  \lambda_0 = 0.0075 \; eV,
B_1=2.123 \; eV\cdot\AA^2,
A_1=0.853 \; eV\cdot\AA, \\
C_1=-0.042 \; eV\cdot\AA^2,
C_2=0.546 \; eV\cdot\AA^2,
C_3=3.345 \; eV\cdot\AA^2.
\end{split}
\end{align} 

With this set of parameters, the bands near the $R$-point from the DFT are well reproduced by the effective model, shown in Fig.~\ref{sm-theory-fig1} (b).

\subsection{The perturbation theory around the Fermi energy}\label{sm-subsection-perturbation-theory}
\begin{figure}[!htbp]
	\centering
	\includegraphics[width=0.85\linewidth]{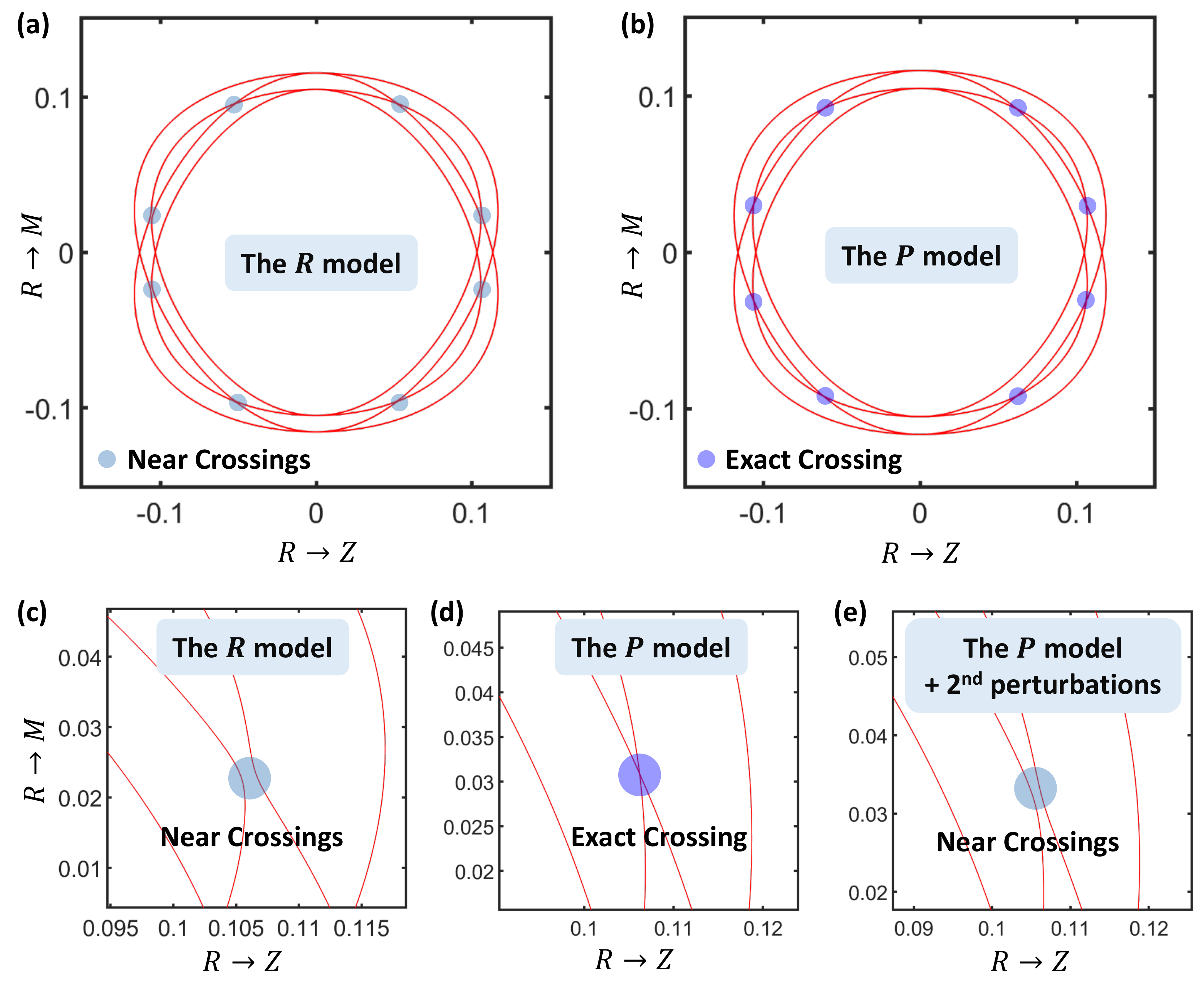}
	\caption{Fermi surfaces at $E_f=-0.06$ eV in the $\Gamma-R-M$ plane.
		(a) Fermi surfaces generated from the $R$-model, eight near crossings are found and marked by light-blue solid circles.  
		As a comparison, the Fermi surfaces for the $P$-model is shown in (b), which display eight exact crossings (marked by purpose solid circles).
		The zoomed plots for the near/exact crossings are shown in (c),(d) and (e). 
		(c) R-model calculations indicate a very tiny gap for the near crossings.
		(d) The exact crossings in P-model with only first order perturbation is shown for a comparison, of which the two-fold degeneracy is protected by the quasi-symmetry. 
		(e) After adding the second-order perturbation corrections to the $P$-model, the quasi-symmetry is broken and thus the exact crossings are gapped. In all figures, the unit for energy is eV and that for momentum is $\AA^{-1}$.
	}	
	\label{sm-theory-fig2}
\end{figure}

According to the DFT calculation as well as the results from the effective $R$ model in Eq.~\eqref{sm-eq-R-model-Ham}, we find the near crossings which are consistent with the experimental results. A tiny gap ($\sim 1$ meV) for the near crossings is shown in Fig.~\ref{sm-theory-fig2} (a). To understand the origin of this tiny gap, we apply the perturbation theory to the model Hamiltonian \eqref{sm-eq-R-model-Ham}. 
We treat $\mathcal{H}_0(\mathbf{k})\otimes s_0$ as the unperturbed Hamiltonian and $\mathcal{H}_{perb}=\mathcal{H}_{soc} + \mathcal{H}_{k^2}(\mathbf{k})$ as the perturbation Hamiltonian.

We first solve the eigen-problem of the Hamiltonian $\mathcal{H}_0(\mathbf{k})\otimes s_0$ and due to its full rotation symmetry, we choose the spherical coordinate for the momentum ${\bf k}=(k_x,k_y,k_z) = k(\sin\theta\cos\phi, \sin\theta\sin\phi, \cos\theta)$. The eigen-energies of $\mathcal{H}_0(\mathbf{k})\otimes s_0$ have two branches $E_{\pm}=C_0\pm A_1k$ for which each branch has four-fold degeneracy.
The four degenerate eigen-wave functions of the positive energy branch ($E_+$) are given by

\begin{align}\label{sm-eq-four-basis}
\begin{split}
\vert \Psi_{A+\uparrow}(\theta,\phi) \rangle &= \vert \Psi_{A+}(\theta,\phi) \rangle \otimes (1,0)^T , \\
\vert \Psi_{B+\uparrow}(\theta,\phi) \rangle &= \vert \Psi_{B+}(\theta,\phi) \rangle \otimes (1,0)^T ,\\
\vert \Psi_{A+\downarrow}(\theta,\phi) \rangle &= \vert \Psi_{A+}(\theta,\phi) \rangle\otimes  (0,1)^T , \\
\vert \Psi_{B+\downarrow}(\theta,\phi) \rangle &=  \vert \Psi_{B+}(\theta,\phi)\rangle \otimes (0,1)^T ,
\end{split}
\end{align}

where the spin-independent components are

\begin{align}
\vert \Psi_{A+}(\theta,\phi) \rangle &= \tfrac{1}{\sqrt{2}} \left( \cos\theta\cos\phi - i\sin\phi, -\cos\theta\sin\phi-i\cos\phi, 0 , \sin\theta\right)^T , \\
\vert \Psi_{B+}(\theta,\phi) \rangle &= \tfrac{1}{\sqrt{2}} \left( -i\sin\theta\cos\phi, i\sin\theta\sin\phi,1,i\cos\theta\right) ^T,
\end{align}

here we use $\{A+,B+\}$ to label the two bands and $\{\uparrow,\downarrow\}$ for the electron spin.

Then, we can project the angular momentum operator $\mathbf{L}$ into the eigenstate subspace and find

\begin{align}
\langle \Psi_{A+} (\theta,\phi)\vert \mathbf{L} \vert \Psi_{A+}(\theta,\phi)\rangle = \langle \Psi_{B+} (\theta,\phi)\vert \mathbf{L} \vert \Psi_{B+}(\theta,\phi)\rangle = \frac{\mathbf{k}}{2k}, \\
\langle \Psi_{A+} (\theta,\phi)\vert \mathbf{L} \vert \Psi_{B+}(\theta,\phi)\rangle = \langle \Psi_{B+} (\theta,\phi)\vert \mathbf{L} \vert \Psi_{A+}(\theta,\phi)\rangle = 0,
\end{align}

which means the emergent angular momentum operator $\mathbf{L}$ is projected into the momentum direction. Thus up to the first-order perturbation, the spin-orbit coupling term becomes the form of $\frac{\lambda_0}{k}(\mathbf{k}\cdot\mathbf{s})\otimes I_{2\times2}$ after projecting into the eigen-state subspace,

\begin{align}
\vert \Psi_{upper}\rangle =\{\vert \Psi_{A+\uparrow}(\theta,\phi) \rangle, \vert \Psi_{B+\uparrow}(\theta,\phi) \rangle, \vert \Psi_{A+\downarrow}(\theta,\phi) \rangle, \vert \Psi_{B+\downarrow}(\theta,\phi) \rangle \},
\end{align}

which supports the discussions below Eq.~(1) in the main text.
The identity matrix $I_{2\times2}$ indicates that the first-order perturbation Hamiltonian of SOC does not couple different bands. 

Next, we can apply the perturbation theory to consider the perturbation Hamiltonian $\mathcal{H}_{perb}$ in the subspace
$\vert \Psi_{upper}\rangle$ order by order. Up to the first-order perturbation, the perturbed Hamiltonian $\mathcal{H}_{P}^{eff(1)}$ reads

\begin{align}\label{eq-eff-ham-p-model}
\begin{split}
\mathcal{H}_{P}^{eff(1)} &= (C_0+B_1k^2+A_1k)s_0\omega_0 + \lambda_0 \left( \lambda_x s_x + \lambda_y s_y + \lambda_z s_z \right)  \omega_0 \\
&+  \tilde{C}k^2 s_0  \left( d_x \omega_x + d_y\omega_y + d_z\omega_z \right).
\end{split}
\end{align}

where $\tilde{C}=C_1-C_2+C_3$, and $\omega_{x,y,z}$ are Pauli matrices for the $\{A+,B+\}$ band subspace [$+$ means the upper four bands in Eq.~\eqref{sm-eq-four-basis}].
The coefficient $\lambda_{x,y,z}$ are defined as

\begin{align}\label{sm-eq-soc-p-model-2}
(\lambda_x,\lambda_y,\lambda_z)&=(\sin\theta\cos\phi,\sin\theta\sin\phi,\cos\theta)= \tfrac{\mathbf{k}}{\vert \mathbf{k}\vert}, 
\end{align}

and $d_{x,y,z}$ are given by

\begin{align}\label{sm-eq-dxyz-pmodel}
\begin{split}
d_x &= \cos\theta\sin^2\theta \sin\phi\cos\phi ( \cos\phi + \sin\phi ), \\
d_y &= \cos\theta\sin^2\theta \sin\phi\cos\phi ( \sin\theta +\cos\theta (\cos\phi - \sin\phi) ), \\
d_z &= \sin^2\theta\sin\phi\cos\phi ( \cos^2\theta + \sin\theta\cos\theta (-\cos\phi+\sin\phi) ).
\end{split}
\end{align}

The perturbation theory is valid only for $A_1k \gg \lambda_0$ and $A_1k  \gg \tfrac{\sqrt{3}}{4} \tilde{C}k^2$. With the parameters used in this work [see Eq.~\eqref{sm-eq-fitting-parameters}], we can estimate the valid range of the momentum as

\begin{align}
k_{min} < k < k_{max} \quad \Leftarrow\quad  k_{min} = \frac{\lambda_0}{A_1}\approx 0.01 \; \AA^{-1} \text{ and } k_{max} = \frac{4A_1}{\sqrt{3}\tilde{C}}\approx 0.7 \; \AA^{-1}.
\end{align}

The eigen-energies of the effective Hamiltonian \eqref{eq-eff-ham-p-model} are given by

\begin{align}\label{eq-eig-value-p-model}
E_{\alpha\beta}(k,\theta,\phi) = C_0+B_1k^2+A_1k +\alpha\lambda_0 +\beta\tfrac{\sqrt{3}}{4}\tilde{C}k^2 \vert \sin2\phi\sin2\theta\sin\theta\vert,
\end{align}

where $\alpha=\pm$ and $\beta=\pm$. These energy dispersions show the exact crossings in Fig.~\ref{sm-theory-fig2}(b).
Note that $ \sin2\phi\sin2\theta\sin\theta = 4k_xk_yk_z/k^3$, indicating the perturbation correction of $\mathcal{H}_{k^2}$ breaks the full rotational symmetry down to three-fold rotational symmetry.

Next we consider the second-order perturbation corrections, which open a tiny gap for the emergent nodal lines [see Fig.~\ref{sm-theory-fig2}(c)].
For the second-order perturbation, besides the eigen-state basis set $\vert \Psi_{upper}\rangle$ in Eq.~\eqref{sm-eq-four-basis}, we also need to take four negative energy ($E_-$) eigen-states

\begin{align}\label{sm-eq-lower-four-bands}
\vert \Psi_{lower}\rangle= \left\{ \vert \Psi_{A-\uparrow}(\theta,\phi) \rangle,  \vert \Psi_{B-\uparrow}(\theta,\phi) \rangle,  \vert \Psi_{A-\downarrow}(\theta,\phi) \rangle, \vert \Psi_{B-\downarrow}(\theta,\phi) \rangle  \right\},
\end{align}

with the explicit expressions 

\begin{align}
\begin{split}
\vert \Psi_{A-\uparrow}(\theta,\phi) \rangle &= \tfrac{1}{\sqrt{2}} \left( \cos\theta\cos\phi + i\sin\phi, -\cos\theta\sin\phi+i\cos\phi, 0 , \sin\theta,0,0,0,0 \right)^T , \\
\vert \Psi_{B-\uparrow}(\theta,\phi) \rangle &= \tfrac{1}{\sqrt{2}} \left( i\sin\theta\cos\phi, -i\sin\theta\sin\phi,1,-i\cos\theta, 0,0,0,0 \right) ,\\
\vert \Psi_{A-\downarrow}(\theta,\phi) \rangle &= \tfrac{1}{\sqrt{2}} \left(0,0,0,0, \cos\theta\cos\phi + i\sin\phi, -\cos\theta\sin\phi+i\cos\phi, 0 , \sin\theta \right)^T , \\
\vert \Psi_{B-\downarrow}(\theta,\phi) \rangle &=  \tfrac{1}{\sqrt{2}} \left(0,0,0,0 , i\sin\theta\cos\phi, -i\sin\theta\sin\phi,1,-i\cos\theta\right) .
\end{split}
\end{align}\\
The second-order perturbed Hamiltonian is given by 

\begin{align}\label{sm-eq-2nd-perturbation}
\Delta\mathcal{H}_{P}^{eff(2)} = \frac{1}{\Delta E} \left( \langle \Psi_{upper} \vert (\mathcal{H}_{soc} + \mathcal{H}_{k^2}(\mathbf{k})) \hat{P}_{lower} ( \mathcal{H}_{soc} + \mathcal{H}_{k^2}(\mathbf{k}) ) \vert \Psi_{upper}\rangle  \right),
\end{align}

where $\Delta E=E_+-E_-=2A_1k$ is the energy difference and the projection operator $\hat{P}_{lower}=\vert \Psi_{lower} \rangle \langle \Psi_{lower}\vert$ onto the lower four bands in Eq.~\eqref{sm-eq-lower-four-bands}.
To simplify the problem, we only consider the mixed terms of $\mathcal{H}_{soc}$ and $\mathcal{H}_{k^2}(\mathbf{k})$, given by

\begin{align}\label{sm-eq-ham-p-eff2}
\Delta\mathcal{H}_{P}^{eff(2)} = \frac{1}{\Delta E} \left( \langle \Psi_{upper} \vert \mathcal{H}_{soc}  \vert \Psi_{lower} \rangle \langle \Psi_{lower} \vert \mathcal{H}_{k^2}(\mathbf{k}) \vert \Psi_{upper} \rangle  \right) + \text{h.c.},
\end{align}

The matrix elements of $\Delta\mathcal{H}_P^{eff(2)}$ are,

\begin{align}
\left \lbrack \Delta\mathcal{H}_P^{eff(2)}\right\rbrack _{1,1} &=
2 \sin ^2(\theta ) \sin (\phi ) \cos (\phi ) \left(\sin ^3(\theta ) \sin (\phi )+\sin (\theta ) \cos ^2(\theta ) \cos (\phi )-\cos ^3(\theta )\right), \\
\left \lbrack \Delta\mathcal{H}_P^{eff(2)}\right\rbrack _{1,2} &=
\sin ^2(\theta ) \cos (\phi ) (\sin ^2(\theta ) \sin (2 \phi )+2 i \cos (\theta ) \sin (\phi ) (\cos (\theta ) (\cos (\theta ) \cos (\phi ) \nonumber \\
&+\sin (\theta ))+\sin (\phi ) (\sin ^2(\theta )+i \cos (\theta ) ) ) ),  \\
\left \lbrack \Delta\mathcal{H}_P^{eff(2)}\right\rbrack _{1,3} &=
\tfrac{1}{4} e^{-i \phi } \sin (\theta ) \cos (\theta ) (\sin (2 \phi ) (\sin (3 \theta ) \sin (\phi )+4 \cos ^3(\theta ))-8 \sin (\theta ) \cos ^2(\theta ) \nonumber \\ & \sin (\phi ) \cos ^2(\phi ) \nonumber 
+8 i \cos (\theta ) \sin ^2(\phi )+(-\tfrac{3}{2}+6 i) \sin (\theta ) \cos (\phi )+(\tfrac{3}{2}+2 i) \sin (\theta ) \\ & \cos (3 \phi )), \\
~
\left \lbrack \Delta\mathcal{H}_P^{eff(2)}\right\rbrack _{1,4} &= e^{-i \phi } \sin (\theta ) ((\cos (\theta )-1) (\cos ^2(\theta ) \cos (\phi )\nonumber+i \sin (\theta ) \sin (\phi ) \cos (\phi ) (-\cos ^2(\theta )  \nonumber\\ &+\sin (\theta ) (\cos (\theta )+i) \cos (\phi )) +\sin (\theta ) \cos (\theta ) \sin ^2(\phi ) (-1-i \sin (\theta ) \cos (\phi )))\nonumber \\ &-2 i \cos ^2(\tfrac{\theta }{2}) (-i \cos ^2(\theta ) \cos (\phi ) \cos (2 \phi )+\sin (\phi ) \cos (\phi ) ((\cos ^3(\theta )+\cos (\theta )) \nonumber \\
& \cos (\phi )-i \sin ^2(\theta ) \cos (\phi )+\sin (\theta ) \cos ^2(\theta )) +\sin (\theta ) \cos (\theta ) \sin ^2(\phi ) (\sin (\theta ) \nonumber \\&\cos (\phi )+i)))
\end{align}
and
\begin{align}
\left \lbrack \Delta\mathcal{H}_P^{eff(2)}\right\rbrack _{2,2} &=
-\tfrac{1}{4} \sin ^5(\theta ) \sin (\phi ) \sin (2 \phi ) \nonumber \\&~~~~\left(\cot ^2(\theta ) (-4 \cot (\theta ) \csc (\phi )+4 \cot (\phi )-1)+\csc ^2(\theta )+3\right), \\
\left \lbrack \Delta\mathcal{H}_P^{eff(2)}\right\rbrack _{2,3} &=
\tfrac{1}{2} e^{-i \phi } \sin (\theta ) \cos (\theta ) (-4 \sin (\phi ) (\sin (\theta ) \sin (\phi )\nonumber +(2+2 i) \sin ^2(\frac{\theta }{2}) \cos ^2(\phi ))+\nonumber \\
&\cos ^2(\theta ) \sin (2 \phi ) (-4 i \sin ^2(\tfrac{\theta }{2}) \cos (\phi )+2 i \sin (\theta )+\csc (\phi ))+\cos (\theta ) \csc (\phi ) \nonumber \\
&(-(2+2 i) \sin ^2(\tfrac{\theta }{2}) \sin ^2(2 \phi ) +\sin ^2(\tfrac{\theta }{2}) \sin (4 \phi )+4 i \sin ^2(\theta ) \sin ^3(\phi ) \cos (\phi )\nonumber \\
&+\sin (2 \phi ))), \\
\left \lbrack\Delta\mathcal{H}_P^{eff(2)}\right\rbrack _{2,4} &=
-\tfrac{1}{8} e^{-i \phi } \sin (\theta ) \cos (\theta ) (8 \cos ^3(\theta ) \sin (2 \phi )-16 \sin (\theta ) \cos ^2(\theta ) \sin (\phi ) \cos ^2(\phi )\nonumber \\&+16 i \cos (\theta ) \sin ^2(\phi ) +\cos (\phi ) (4 \sin (3 \theta ) \sin ^2(\phi )+(-3+12 i) \sin (\theta ))\nonumber \\
&+(3+4 i) \sin (\theta ) \cos (3 \phi )),
\end{align}
and
\begin{align}
\left \lbrack \Delta\mathcal{H}_P^{eff(2)}\right\rbrack _{3,3} &=
-\tfrac{1}{4} \sin ^5(\theta ) \sin (\phi ) \sin (2 \phi ) \nonumber \\
&\left(\cot ^2(\theta ) (-4 \cot (\theta ) \csc (\phi )+4 \cot (\phi )-1)+\csc ^2(\theta )+3\right), \\
\left \lbrack \Delta\mathcal{H}_P^{eff(2)}\right\rbrack _{3,4} &=
-i \sin ^2(\theta ) \sin (\phi ) \cos (\phi ) (2 \cos ^3(\theta ) \cos (\phi )+2 \cos ^2(\theta ) (\sin (\theta )+i \sin (\phi )) \nonumber \\
&+\sin (\theta ) (\sin (2 \theta ) \sin (\phi )-2 i \sin (\theta ) \cos (\phi ))),
\end{align}
and
\begin{align}
\left \lbrack \Delta\mathcal{H}_P^{eff(2)}\right\rbrack _{4,4} &=  2 \sin ^2(\theta ) \sin (\phi ) \cos (\phi ) (\sin ^3(\theta ) \sin (\phi )+\sin (\theta ) \cos ^2(\theta ) \cos (\phi )-\cos ^3(\theta )),
\end{align}
And the other parts are related by conjugation $\left \lbrack \Delta\mathcal{H}_P^{eff(2)}\right\rbrack _{i,j} =  \left \lbrack\Delta\mathcal{H}_P^{eff(2)}\right\rbrack _{j,i}^\ast$.
The second-order perturbation corrections provides the spin-flipping terms in the effective model and thus generates a tiny gap for the emergent nodal lines on the Fermi surfaces, as shown in Fig.~\ref{sm-theory-fig2} (c).

\vspace{12pt}

\begin{figure}[!htbp]
	\centering
	\includegraphics[width=0.85\linewidth]{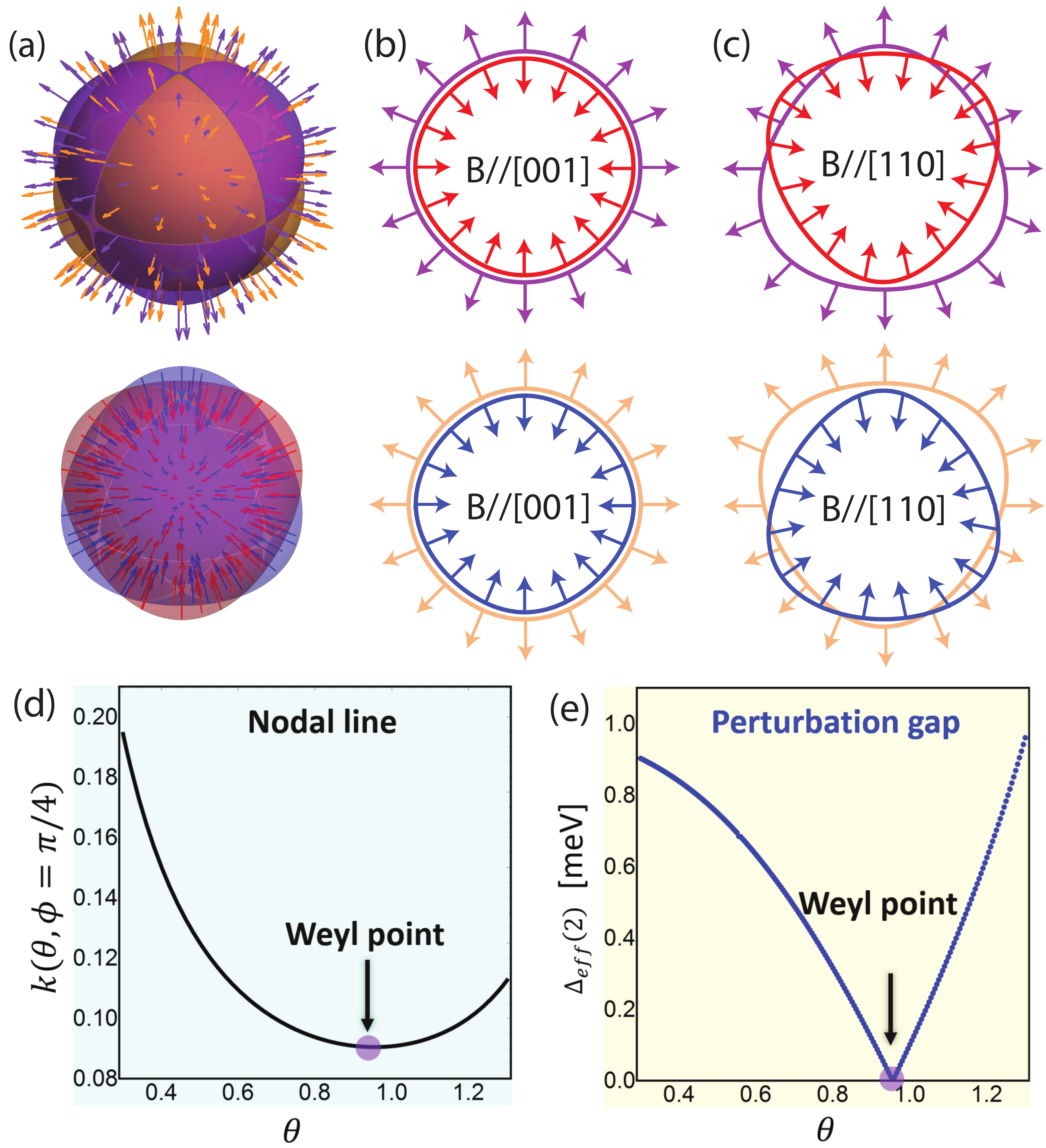}
	\caption{(a) The spin texture (colored arrow) for all four Fermi surfaces calculated from the $P$-model. As discussed in the main manuscript, since the spin of the eigen-states is parallel or anti-parallel to the momentum, Fermi surfaces with different spin character(+/-) have opposite spin-polarization at the same momentum.  (b) and (c) present the spin texture of Fermi surface orbits with field applied along [001] and [110] direction respectively.
	(d) Nodal line on the $\Gamma-R-M$ plane with $\theta\in\{0,\pi/2\}$ and $\phi=\pi/4$ obtained by solving $d_z(\mathbf{k})=0$ in Eq.~\eqref{sm-eq-heff2-two-band}. 
	(e) Energy gap on the nodal line caused by the second-order perturbation corrections, $\delta_{eff(2)}=2\sqrt{(\delta d_x)^2+(\delta d_y)^2}$, in unit of meV. 
	This gap is resulted from the second-order perturbation which involves the spin-flipping terms. The momentum $k$ is in unit of \AA$^{-1}$.
	}	
	\label{sm-theory-fig4}
\end{figure}

To simplify the above discussion, we can further project the above four-band $P$ model to the effective model that only consists of the two bands forming the nodal plane. To do that, we consider the eigen-states of the model Hamiltonian~\eqref{eq-eff-ham-p-model} and consider two eigen-states $ \{\vert \alpha=+,\beta=-\rangle , \vert \alpha=-,\beta=+ \rangle\}$ that give the eigen-energies $E_{\alpha=+,\beta=-}$ and $E_{\alpha=-,\beta=+}$ in Eq. \ref{eq-eig-value-p-model}. Within these two eigen-states, the effective Hamiltonian $\mathcal{H}_{P}^{eff(1)}$ can be reduced to

\begin{align}\label{sm-eq-heff2-two-band}
\mathcal{H}_{eff} = \epsilon_0 + d_z(\mathbf{k}) \sigma_z,
\end{align}

where $\epsilon_0=C_0+B_1k^2+A_1k $ and $d_z(\mathbf{k}) =\lambda_0  - \tfrac{\sqrt{3}}{4}\tilde{C}k^2 \vert \sin2\phi\sin2\theta\sin\theta\vert=\lambda_0 - \sqrt{3}\tilde{C} \frac{\vert k_xk_yk_z\vert}{k} $.
This Hamiltonian is the exactly two-band model discussed in the main text. 
Due to the SOC with $\mathbf{k}\cdot\mathbf{s}$ in the $P$-model~\eqref{eq-eig-value-p-model}, the spin texture is hexagonal on each Fermi surfaces ($\langle \mathbf{s}\rangle \sim \alpha\mathbf{k}$ with $\alpha=\pm$).
Moreover, we plot the spin texture on $E_{\alpha=+1,\beta=-1}(k,\theta,\phi) =\epsilon_0+d_z(\mathbf{k})=E_f$ in Fig.~\ref{sm-theory-fig4} (a), and that for $E_{\alpha=-1,\beta=+1}(k,\theta,\phi) =\epsilon_0-d_z(\mathbf{k})=E_f$ in Fig.~\ref{sm-theory-fig4} (b).
It indicates that the crossings between these two Fermi surfaces have opposite spin-polarization.
Therefore, according to the low-energy effective Hamiltonian, we can conclude that the spin-conservation plays the role of the quasi-symmetry.

Now we discuss the breaking of the quasi-symmetry by second-order perturbation corrections. To see that,
we can further project $\mathcal{H}_{P}^{eff(2)}$ into the subspace of $ \{\vert \alpha=+,\beta=-\rangle , \vert \alpha=-,\beta=+ \rangle\}$ and the effective Hamiltonian will include two more terms as

\begin{align}
\mathcal{H}_{eff(2)} = \delta d_x(\mathbf{k}) \sigma_x+\delta d_y(\mathbf{k}) \sigma_y, 
\end{align}

which are due to the spin-flipping terms, breaking the quasi-symmetry.
As a result, the co-dimension of the nodal plane becomes 3 instead of 1, explaining the gap opening.
While the detailed expressions for $\delta d_x$ and $\delta d_y$ are complex, we give an estimate of the typical magnitude of the gap $\Delta_{eff(2)}=2\sqrt{(\delta d_x)^2+(\delta d_y)^2}\sim 0.8 $ meV for $k_F=0.13$ \AA$^{-1}$, being the same order with the DFT estimations [see FIG.~S11].
By solving $d_z(\mathbf{k})=0$ in Eq.~\eqref{sm-eq-heff2-two-band}, we can find the nodal line on the $\Gamma-R-M$ plane ($\phi=\pi/4$) that is shown in  Fig.~\ref{sm-theory-fig4} (c).
Furthermore, we calculate the energy gap ($\sqrt{(\delta d_x(\mathbf{k}))^2+(\delta d_y(\mathbf{k}))^2}$) for the nodal line on the $\Gamma-R-M$ plane, shown in Fig.~\ref{sm-theory-fig4} (d).
Please note that the gap vanishes for the type-II Weyl point which locates at $\theta = \arcsin(\sqrt{2/3})\sim0.96$ (along the $\Gamma-R$ line), due to the $C_3$ protection (the two Weyl states have different $C_3$-eigenvalues).

\subsection{The Berry curvature distributions}
\begin{figure}[!htbp]
	\centering
	\includegraphics[width=0.95\linewidth]{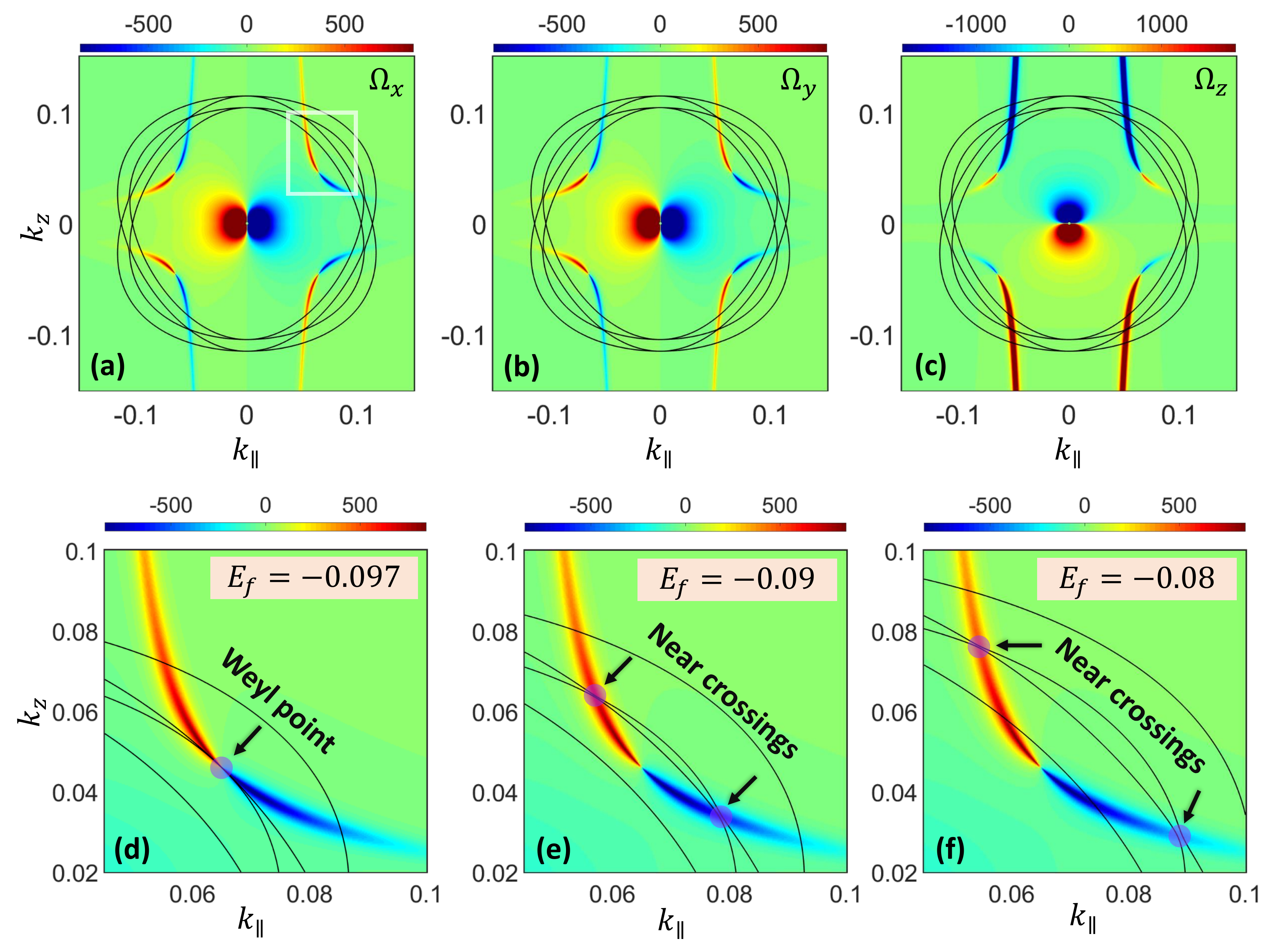}
	\caption{The Berry curvature distributions around the near crossings on the $\Gamma-R-M$ plane.
		In (a),(b) and (c), the three component of the Berry curvature ($\Omega_{x,y,z}$) are shown for the the $1^{-}$-band. The black lines stand for the Fermi surface contour with $E_f=-0.06$ eV.
		The zoomed plot for $\Omega_x$ in (a)  is shown in (d),(e) and (f), where we compare the Fermi surface flow as the distributions of $\Omega_x$. 
	    The Fermi energies are $E_f=-0.097$ eV, $E_f=-0.09$ eV, $E_f=-0.08$ eV for (d), (e), and (f), respectively.	
		The singularity is at the type-II Weyl points, as well as that for the 6-fold degeneracy at the $R$-point. 
		And the large Berry curvature almost sits at the near crossings (marked by the dark-red solid circles) on the Fermi surface.
		In all figures, the unit for energy is eV and for momentum it is $\AA^{-1}$.
	}	
	\label{sm-theory-fig3}
\end{figure}

In this section, we show the topologically physical consequence for the emergent nodal lines (near crossing), characterized by the Berry curvature distributions in the momentum space.
We calculate the Berry curvature based on the $R$-model in the $\Gamma-R-M$ plane.
Here, we take a standard formula to compute the gauge-invariant Berry curvature

\begin{align}
\mathbf{\Omega}_n(\mathbf{k}) = -\text{Im} \sum_{m\neq n} \frac{\langle n\vert \bm{\nabla} \mathcal{H}_{R}(\mathbf{k}) \vert m\rangle \times \langle m\vert \bm{\nabla} \mathcal{H}_{R}(\mathbf{k}) \vert n\rangle }{(E_m-E_n)^2},
\end{align}

where the velocity operators are given by $\mathbf{\hat{V}}=\bm{\nabla} \mathcal{H}_{R}(\mathbf{k}) \triangleq (\hat{V}_x,\hat{V}_y,\hat{V}_z) $.
Similar to the analysis in Re.~[\onlinecite{MnSi_Plef}] (see Eq.~(37) in the Supplementary Materials for the Nature paper of MnSi), we find that

\begin{align}
&S_{2x}\mathcal{T}: (k_x,k_y,k_z)\to (-k_x,k_y,k_z) \text{ and } \bm{\Omega}(k_x,k_y,k_z) = \bm{\Omega}(-k_x,k_y,k_z) \begin{pmatrix}
-1 & 0 & 0 \\
0 & +1 & 0 \\
0 & 0 & +1
\end{pmatrix} ,  \\
&S_{2y}\mathcal{T}: (k_x,k_y,k_z)\to (k_x,-k_y,k_z) \text{ and } \bm{\Omega}(k_x,k_y,k_z) = \bm{\Omega}(k_x,-k_y,k_z) \begin{pmatrix}
+1 & 0 & 0 \\
0 & -1 & 0 \\
0 & 0 & +1
\end{pmatrix} , \\
&S_{2z}\mathcal{T}: (k_x,k_y,k_z)\to (k_x,k_y,-k_z) \text{ and } \bm{\Omega}(k_x,k_y,k_z) = \bm{\Omega}(k_x,k_y,-k_z) \begin{pmatrix}
+1 & 0 & 0 \\
0 & +1 & 0 \\
0 & 0 & -1
\end{pmatrix} ,
\end{align}

where we take the vector for Berry curvature as $\bm{\Omega}_n(k_x,k_y,k_z) = (\Omega_{n,yz},\Omega_{n,zx},\Omega_{n,xy}) $ for the $n$-th band.
The calculated results are shown in Fig.~\ref{sm-theory-fig3} on the $\Gamma-R-M$ plane, confirming the above symmetry requirements.
Here we present the Berry curvature for the $1^{-}$-band (see the band notation in Fig.~2 (d) in the main text).
Namely, this band is just the second upper-band from the $\mathbf{k}\cdot\mathbf{p}$-bands, shown in Fig.~\ref{sm-theory-fig1} (c).
The three components are shown in Fig.~\ref{sm-theory-fig3} (a),(b) and (c), where the four Fermi surfaces are shown with $E_f=-0.06$ eV.
Moreover, the Berry curvature distribution coincides with the emergent nodal lines, illustrated in Fig.~\ref{sm-theory-fig3} (d),(e) and (f).
At the $R$-point or Weyl points, the $U(1)$ Berry curvature is not well defined due to degeneracy, which corresponds to the singularity.
In Fig.~\ref{sm-theory-fig3} (a), (b) and (c), the singularity at the $R$-points is due to the 6-fold degeneracy.
And the type-II Weyl points are shown in Fig.~\ref{sm-theory-fig3} (d) for the singularity (marked by the dark-red solid circle).

\subsection{The strain effects on the crystalline symmetries and quasi-symmetries}\label{TheoryStrain}
In this subsection, we study the strain effect on the electronic bands, which is characterized by the strain tensor 
\begin{align}
u_{ij} = \frac{1}{2} \left( \partial_{x_i}u_j + \partial_{x_j}u_i \right)
\end{align}
where $u_i$ is the displacement at $\mathbf{x}$.
And $u_{ij}$ transfers as $k_ik_j$ under point group symmetry operators. 
Thus, the leading order correction to the 8-band R-model in Eq.~\eqref{sm-eq-R-model-Ham} due to the strain-induced Hamiltonian is generally given by
\begin{align}\label{eq-ham0-strain}
\begin{split}
\mathcal{H}_{strain} &= D_0(u_{xx}+u_{yy}+u_{zz}) + D_1(u_{xy}\sigma_z\tau_0 + u_{yz}\sigma_0\tau_z + u_{xz}\sigma_z\tau_z ) \\
&+D_2(u_{xy}\sigma_x\tau_x -u_{yz}\sigma_0\tau_x + u_{xz}\sigma_x\tau_0 ) + D_3(u_{xy}\sigma_x\tau_z -u_{yz}\sigma_y\tau_y -u_{xz}\sigma_z\tau_x ).
\end{split}
\end{align}
It is invariant under both all the symmetry generators of the space group 198 and the time-reversal symmetry operator $\mathcal{T}=is_y\mathcal{K}$ with $\mathcal{K}$ the complex conjugate.

Here, we first describe the calculation of the strain tensor resulting from an uniaxial stress of magnitude $P$ along an arbitrary direction.
Analysis begins by adopting a coordinate system $ (x',y',z')$ in which the $ x'$ axis is parallel to the stress direction. This system is related to the coordinate system $ (x,y,z)$ of the primary crystallographic axes of the semiconductor by a rotation
\begin{align}
U(\theta,\phi)=\left(
\begin{array}{ccc}
\cos (\theta ) \cos (\phi ) & \cos (\theta ) \sin (\phi ) & -\sin (\theta ) \\
-\sin (\phi ) & \cos (\phi ) & 0 \\
\sin (\theta ) \cos (\phi ) & \sin (\theta ) \sin (\phi ) & \cos (\theta ) \\
\end{array}
\right)
\end{align}
where $ \theta $ and $ \phi $ are the polar and azimuthal angles of the stress direction relative to the crystallographic coordinate system. In the primed coordinate system, the stress tensor has only one non-zero component, $ \sigma'_{zz} = P$. The stress tensor in the crystallographic system can be calculated from
\begin{align}
\displaystyle \sigma_{ij} = U_{\alpha i} U_{\beta j} \sigma'_{\alpha \beta}\ .
\end{align}
If uniaxial stress is applied along one of the directions [100], [110], and [111], the related stress tensors in the principal system become:
\begin{align}
\sigma_{[100]} = \begin{pmatrix}
P & 0 & 0 \\ 
0 & 0 & 0 \\ 
0 & 0 & 0
\end{pmatrix} , 
\sigma_{[110]} = \begin{pmatrix}
P/2 & P/2 & 0 \\ 
P/2 & P/2 & 0 \\ 
0 & 0 & 0
\end{pmatrix} , 
\sigma_{[111]} = \begin{pmatrix}
P/3 & P/3 & P/3 \\ 
P/3 & P/3 & P/3 \\ 
P/3 & P/3 & P/3
\end{pmatrix} , 
\end{align}

Then, we study with the (110) strain, whose lowest order strain Hamiltonian reads,
\begin{align}
\begin{split}
\mathcal{H}_{(110)strain}&= D_0(u_{xx}+u_{yy}) + D_1(u_{xy}\sigma_z\tau_0  ) +D_2(u_{xy}\sigma_x\tau_x) + D_3(u_{xy}\sigma_x\tau_z).
\end{split}
\end{align}
In the following, the $u_{xx}$ and $u_{yy}$ are absorbed into the Fermi energy, thus only $u_{xy}$-terms will be considered.
Let us check the symmetry breaking,
the $S_{2x}=i\sigma_y\tau_0$ and $S_{2y}=i\sigma_x\tau_y$ are broken, while the $S_{2z}=i\sigma_z\tau_y$ is still preserved because of
\begin{align}
[S_{2z}, \mathcal{H}_{(110)strain}] =0.
\end{align}
It indicates that the strain-induced gaps only appear for two high-symmetry planes: $k_x=0$ or $k_y=0$; while the $k_z=0$ plane still has twofold degeneracy.
Moreover, let us also check the $R$-point with $k_x=k_y=k_z=0$, only $\mathcal{T}$ and $S_{2z}$ are preserved, all the other symmetries are broken, then, each state is twofold degenerate. 

To confirm the above symmetry analysis, we plot the bands without/with strain effect in Fig.~\ref{StrainBandsGaps}, where we use the strain parameters: $D_1=0.003$, $D_2=0.001$, and $D_3=0.002$ in unit of eV. 
From Fig.~\ref{StrainBandsGaps} (a) and (b), we can see the bands along $k_z=0$ line are still degenerate. However, in (c) and (d), the bands along $k_x,k_y=0$ are no longer degenerate due to strain effect (crystalline symmetry breaking). In (e), the Weyl point is no longer along the $\Gamma-R$ line since the $C_3$ rotation is broken. 
Based on the bands in the $\Gamma-R-M$ plane, we define two gaps as
\begin{align}
    \Delta_{cs} &= \left\vert E_{1^+}(0,0,0.2) - E_{1^-}(0,0,0.2) \right\vert, \\
    \Delta_{qs} &= \text{min}\left\vert E_{1^-}(k,k,k_z) - E_{2^+}(k,k,k_z) \right\vert.
\end{align}
where the crystalline-symmetry gap $ \Delta_{cs} $ represents the breaking of $S_{2x}$ and the quasi-symmetry gap $\Delta_{qs}$ is for the qausi-symmetry gap (near degeneracy).
As shown in (f), we conclude that the quasi-symmetry is almost unaffected by the strain. 

Next, we further consider the the (11$\delta$) and (111) strain effects, whose Hamiltonian can be represented as
\begin{align}
\mathcal{H}_{(11\delta)strain}&=  D_1(u_{xy}\sigma_z\tau_0  ) +D_2(u_{xy}\sigma_x\tau_x) + D_3(u_{xy}\sigma_x\tau_z) + D_4(u_{xy}\sigma_0\tau_z), \\
\mathcal{H}_{(111)strain}&=  D_1(u_{xy}\sigma_z\tau_0 + u_{yz}\sigma_0\tau_z + u_{xz}\sigma_z\tau_z ).
\end{align}
The additional terms due to (11$\delta$) strain might be in principle smaller than those parameters used for (110) strain. Notice that the Hamiltonian for (11$\delta$) strain has been simplified, so only a $D_4$ term is added, compared with the Hamiltonian of (110) strain. 
To illustrate clearly the breaking of $S_{2z}$ rotation by $D_4$ term, we take $D_4$ as the same order as $D_{1}$.

To schematically demonstrate these strain effects on the electronic Fermi surfaces, we plot the Fermi surfaces perpendicular to (100), (1$\bar{1}$0) and (111) axes in Fig.~\ref{StrainFermiSurfaces}.
As expected, once the crystalline symmetry is easily broken by the strain, the crystalline symmetry-protected twofold degeneracy is spitted. 
Thus, we may conclude that, the breaking of crystalline symmetry due to strain effect, more magnetic orbits can be observed due to trivial magnetic breakdown.
It is consistent with the experimental observations of quantum oscillation measurement.

By using perturbation theory discussed in Sec.~\ref{sm-subsection-perturbation-theory}, we discuss the strain effect on the robustness of the quasi-symmetry in CoSi.
To the first-order perturbation, we find the $P$-model around $R$-point that is modified as,
\begin{align}\label{sm-eq-new-p-model}
\mathcal{H}_P = \mathcal{H}_{\text{soc},+}^{eff(1)} + \mathcal{H}_{k^2,+}^{eff(1)} + \mathcal{H}_{strain} ,
\end{align}
where each part reads
\begin{align}
\begin{split}
\mathcal{H}_{\text{soc},+}^{eff(1)} &= \lambda_0 \left( \lambda_x s_x + \lambda_y s_y + \lambda_z s_z \right)  \omega_0, \\
\mathcal{H}_{k^2,+}^{eff(1)} &=  s_0  \left( d_x \omega_x + d_y\omega_y + d_z\omega_z \right),\\
\mathcal{H}_{strain} &=  s_0 \left( d_x' \omega_x + d_y'\omega_y + d_z'\omega_z \right),
\end{split}
\end{align}
where the coefficients are defined as
\begin{align}
\begin{split}
(\lambda_x,\lambda_y,\lambda_z)&=(\sin\theta\cos\phi,\sin\theta\sin\phi,\cos\theta)= \tfrac{\mathbf{k}}{\vert \mathbf{k}\vert}. \\
d_x &= \tfrac{\tilde{C}k^2}{4} \sin\theta\sin(2\theta) \sin (2\phi) (\cos\phi + \sin\phi) , \\
d_y &= \tfrac{\tilde{C}k^2}{4}\sin\theta\sin(2\theta) \sin (2\phi) ( \sin\theta +\cos\theta (\cos\phi - \sin\phi) ), \\
d_z &= \tfrac{\tilde{C}k^2}{4}\sin\theta\sin(2\theta) \sin (2\phi) ( \cos\theta - \sin\theta (\cos\phi-\sin\phi) ). \\
d_x' &= u_{\text{xy}}\cos (\theta )  \left(-d_2 \sin (\phi )+d_3 \cos (\phi ) \right), \\
d_y' &= u_{\text{xy}}\cos (\theta )  \left(-d_2  \cos (\theta ) \cos (\phi ) - d_3  \cos (\theta ) \sin (\phi ) - d_1 \sin (\theta )\right), \\
d_z' &=u_{\text{xy}}  \cos (\theta )  \left(\sin (\theta ) \left(d_3 \sin (\phi )+d_2 \cos (\phi )\right)+d_1 \cos (\theta )\right).
\end{split}	
\end{align}
As a result, we find that the new $P$-model in Eq.~\eqref{sm-eq-new-p-model} with corrections from the strain effect is still a stabilizer code Hamiltonian, the quasi-symmetry $\mathcal{M}_{eff}$ is still preserved. 
Thus, the dispersion consists two parts,
\begin{align}
E_{\alpha,\beta} = \alpha \lambda_0 + \beta \sqrt{ (d_x+d_x')^2 +(d_y+d_y')^2  + (d_z+d_z')^2},
\end{align}
where the second part is give by
\begin{align}\label{eq-second-part-eng}
\frac{1}{2} \sqrt{3 \tilde{C}^2 k_r^4 \sin ^4(\theta ) \cos ^2(\theta ) \sin^2(2 \phi ) + \tilde{C}\tilde{d} u_{\text{xy}} k_r^2 \sin ^2(2 \theta )  \sin (2 \phi )+4 \left(d_1^2+d_2^2+d_3^2\right) \cos ^2(\theta ) u_{\text{xy}}^2}
\end{align}
here $\tilde{C}=C_1-C_2+C_3$ and $\tilde{d}=d_1-d_2+d_3$.
Please note that 
\begin{align}
\sin ^2(\theta ) \cos(\theta ) \sin(2 \phi ) \times \cos(\theta ) = \frac{1}{4} \sin ^2(2 \theta )  \sin (2 \phi ),
\end{align}
therefore, Eq.~\eqref{eq-second-part-eng} becomes
\begin{align}
\frac{1}{2}\sqrt{ \left(\tilde{C} \tilde{k}  + d_1u_{\text{xy}}\cos(\theta )  \right)^2  + \left(\tilde{C}\tilde{k}  + d_3u_{\text{xy}}\cos(\theta )  \right)^2 + \left(\tilde{C}\tilde{k}  - d_2u_{\text{xy}}\cos(\theta )  \right)^2        }
\end{align}
where we have defined $\tilde{k}= k_r^2 \sin^2(\theta ) \cos(\theta ) \sin(2 \phi ) = 2k_x k_y k_z/k_r$.
The constraint equation for the nodal plane is
\begin{align}
\lambda_0 = \frac{1}{2}\sqrt{ \left(\tilde{C} \tilde{k}  + d_1u_{\text{xy}}\cos(\theta )  \right)^2  + \left(\tilde{C}\tilde{k}  + d_3u_{\text{xy}}\cos(\theta )  \right)^2 + \left(\tilde{C}\tilde{k}  - d_2u_{\text{xy}}\cos(\theta )  \right)^2     } .
\end{align}
The general solution is plotted in Fig.~\ref{StrainNewNodalPlanes}. 
Recall that, in the $d_1=d_2=d_3=0$ limit (no strain), the above equation is $\lambda_0 = \sqrt{3}\tilde{C} \vert k_x k_y k_z\vert /k_r $, which is reduced back to our previous results. 

Next, let us consider a simple case with by setting $d_2=d_3=0$, the equation becomes
\begin{align}\label{eq-new-nodal-plane-equ}
\lambda_0 = \frac{1}{2}\sqrt{ \left(\tilde{C} \tilde{k}  + d_1 \cos(\theta )  \right)^2  + 2\left(\tilde{C}\tilde{k} \right)^2     }
 =  \frac{\tilde{C}}{2}\left\vert \frac{k_z}{k_r} \right\vert\times \sqrt{ (2k_xk_y + d_1)^2 + 8 k_x^2k_y^2 }
\end{align}
where $u_{xy}=1$ is used for simplicity. 
In principle, it could also leads to the nodal plane solution, which indicates that the quasi-symmetry protected nodal plane are robust against strain effect.
And Eq.~\eqref{eq-new-nodal-plane-equ} gives rise to the following results,
\begin{itemize}
\item[1.)] $k_z=0$, each band has two-fold degeneracy. Thus, the are only two Fermi surfaces. Since $C_{2z}$ is still preserved for the $(110)$ strain.
\item[2.)] $k_x=0$ or $k_y=0$, the crystalline symmetry protected twofold degeneracy is broken, and the gap is about 
 \begin{align}
  \Delta = \sqrt{d_1^2+d_2^2+d_3^2} u_{xy} \vert \cos\theta\vert
 \end{align}
\item[3.)] If Eq.~\eqref{eq-new-nodal-plane-equ} has physical solutions, it means there is quasi-symmetry protected nodal plane.
First, let us analytically understand Eq.~\eqref{eq-new-nodal-plane-equ}. By fixing $k_z\neq0$, we get the equation in the $k_x-k_y$ plane,
\begin{align}
(2k_xk_y + d_1)^2 + 8 k_x^2k_y^2 = ( f_\lambda(k_x,k_y,k_z) )^2
\end{align}
where $  f_\lambda(k_x,k_y,k_z) = \frac{2\lambda_0}{\tilde{C}} \left\vert \frac{k_r}{k_z} \right\vert>0$. 
For $d_1=0$, nodal plane locates at $k_xk_y\neq0$.
When $d_1>0$, and it is increased, at the critical value $ \frac{2\lambda_0}{\tilde{C}}$, two nodal planes with $k_xk_y>0$ touch with each other at $k_x=k_y=0$, forming a single nodal plane.
We further increase $d_1$, then, the location of nodal planes depends on the value of $k_z$. 
\end{itemize}

As a brief conclusion, the quasi-symmetry is a symmetry of the lowest Hamiltonian, but not symmetry of the crystal. It explains that the strain effect does not affect the quasi-symmetry, which is consistent with the experiments.

\clearpage

\clearpage

\begin{figure}
	\includegraphics[width=1\columnwidth]{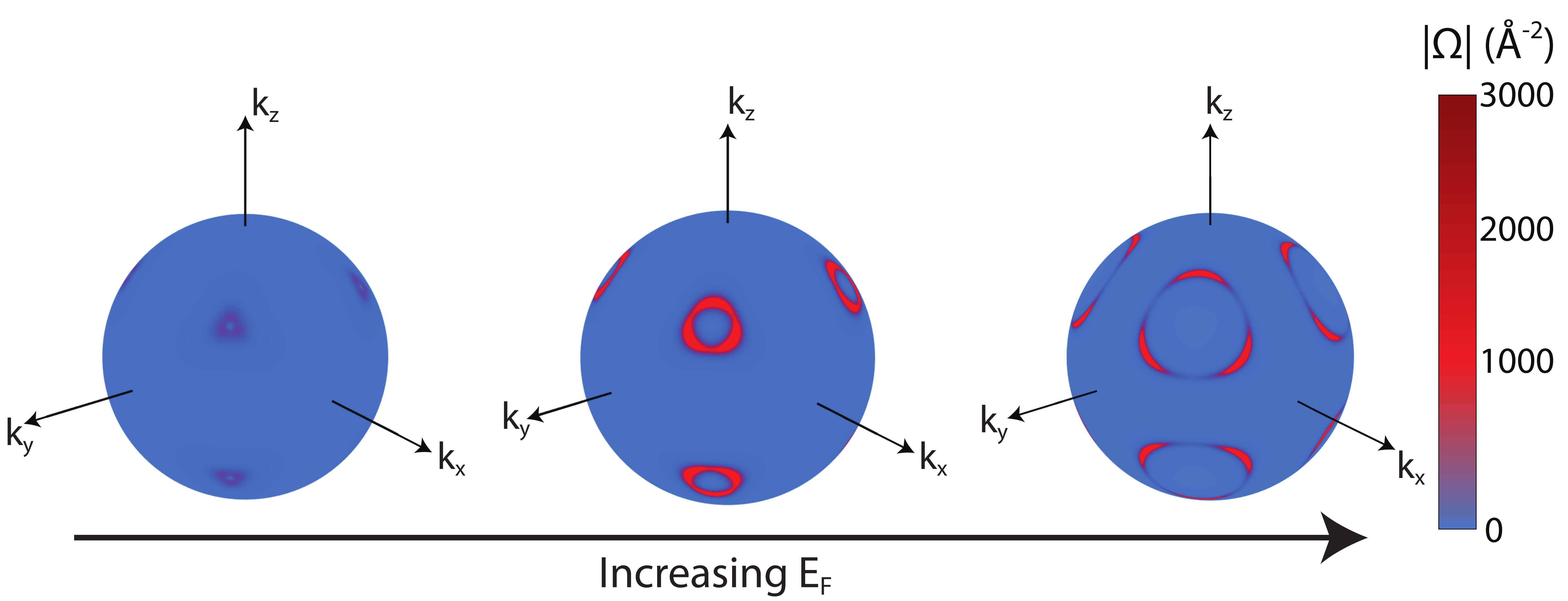}
	\caption{Energy-dependent Berry curvature in momentum space for the $1^{-}$-band. The large Berry curvature is a direct result of near-degeneracy due to quasi-symmetry.}
	\label{BC3D}
\end{figure}

\begin{figure}
	\centering
	\includegraphics[width=1\columnwidth]{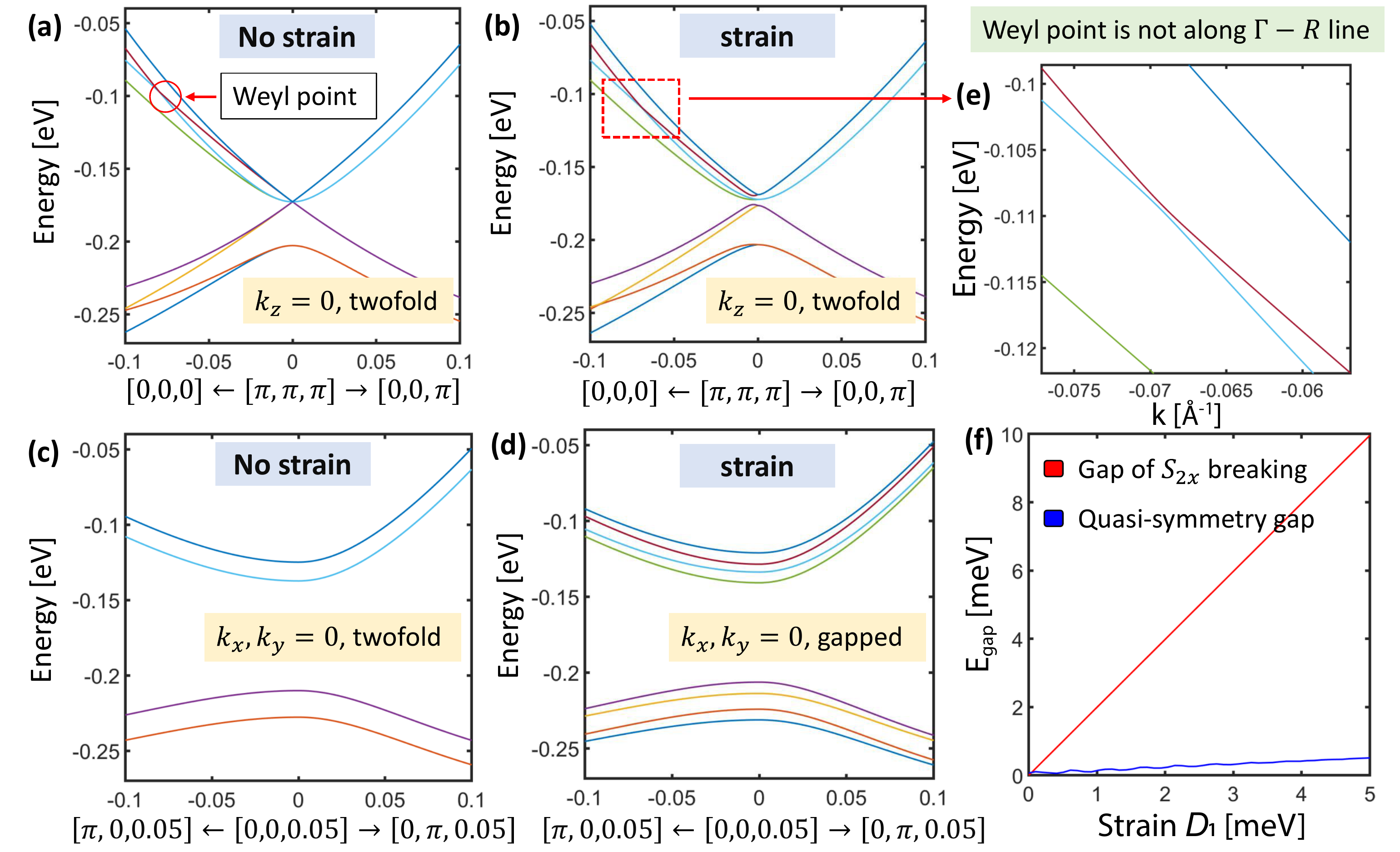}
	\caption{The bands without/with strain effect. To model the [110] strain that breaks both $S_{2x}$ and $S_{2y}$ screw rotations, we use $D_1=0.003$, $D_2=0.001$, and $D_3=0.002$ in unit of eV. However, the $S_{2z}$ rotation is preserved for the [110] strain. In (a) without strain and (b) with strain, we can see the bands along $k_z=0$ line are still degenerate. However, in (c) and (d), the bands along $k_x,k_y=0$ are no longer degenerate due to strain effect (crystalline symmetry breaking). In (e), the Weyl point is no longer along the $\Gamma-R$ line since the $C_3$ rotation is broken. In (f), the band gap due to the breaking of $S_{2x}$ and the quasi-symmetry gap are shown for the bands in the $\Gamma-R-M$ plane, which indicates that the quasi-symmetry is almost unaffected by the strain. In this schematic plot, we set $D_2=D_3=0$ and $D_1\neq0$ for [110] strain. At zero strain ($D_1=0$), the gap for the $S_{2x}$ breaking is exactly zero; while the quasi-symmetry gap is small $\sim 0.1$ meV. }
	\label{StrainBandsGaps}
\end{figure}

\begin{figure}
	\centering
	\includegraphics[width=1\columnwidth]{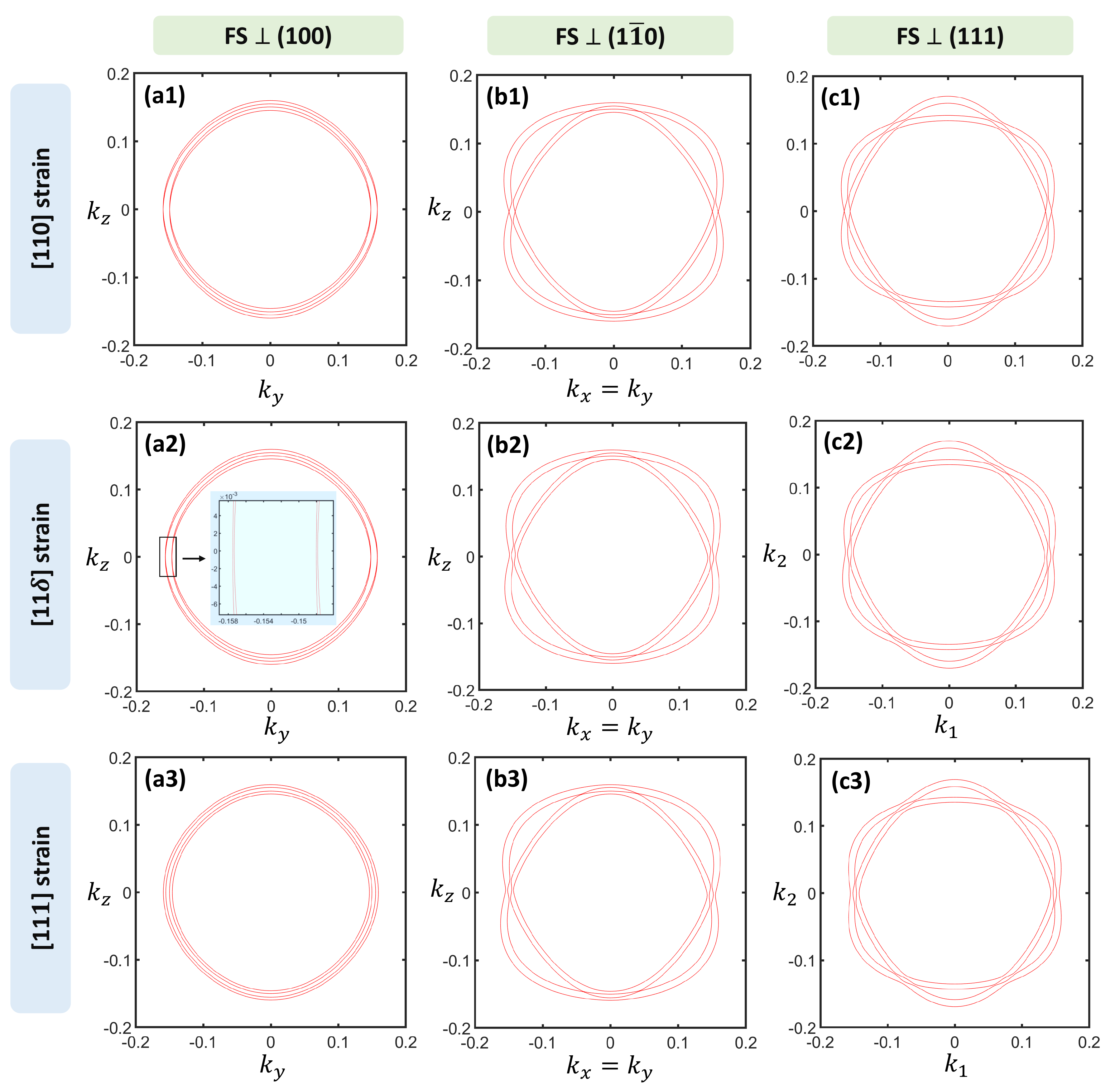}
	\caption{The Fermi surfaces in planes perpendicular to the (100), (1$\bar{1}$0) and (111) axes for three types of strain effect: [110] strain (breaks $S_{2x}$ and $S_{2y}$), [11$\delta$] strain (breaks all crystalline symmetries) and [111] strain (breaks $S_{2x}$, $S_{2x}$ and $S_{2z}$), respectively. In (a1, b1, c1), the Fermi surfaces for the [110] strain ($D_1=0.003$, $D_2=0.001$, and $D_3=0.002$ in unit of eV). In (a2, b2, c2), the Fermi surfaces for the [11$\delta$] strain ($D_1=0.003$, $D_2=0.001$, $D_3=0.002$ and $D_4=0.004$ in unit of eV). In (a3, b3, c3), the Fermi surface for the [111] strain ($D_1=0.003$ in unit of eV). Note that a possible $D_4$ as the same order as other parameters is assumed for the illustration of the Fermi surface plot where $S_{2z}$ rotation is also broken so that the degeneracy at $k_z=0$ is broken.   }
	\label{StrainFermiSurfaces}
\end{figure}

\begin{figure}
	\centering
	\includegraphics[width=1\columnwidth]{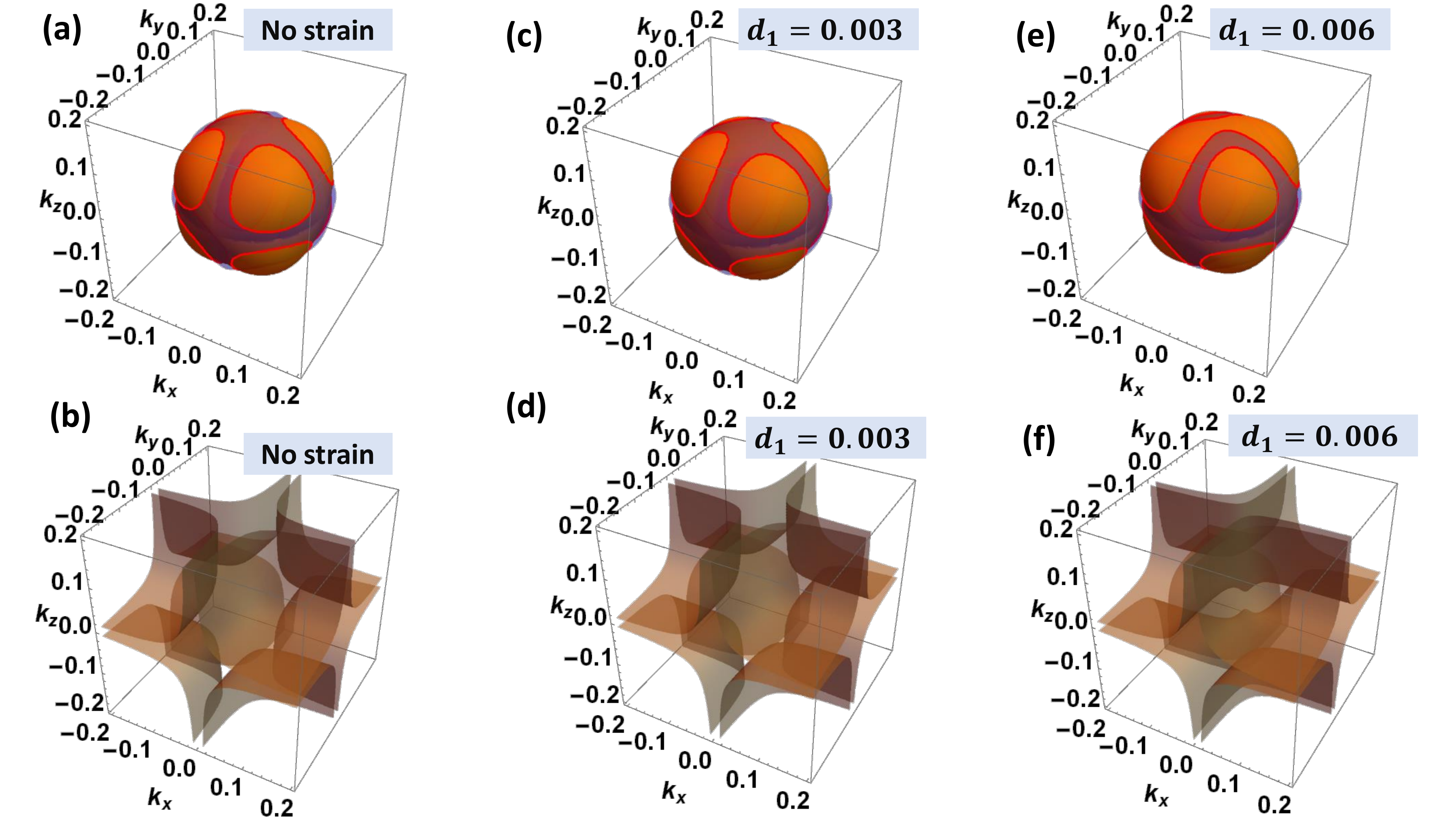}
	\caption{The quasi-symmetry protected nodal planes with/without strain effect. In (a, c, e), the two Fermi surfaces of the P-model show the intersecting features (see the red lines). In (b, d, f), the quasi-symmetry protected nodal planes are found for both without and with strain effects.  }
	\label{StrainNewNodalPlanes}
\end{figure}